\documentclass{article}
    \usepackage{latexsym}
\usepackage{amsmath}
\usepackage{mathtools}
\usepackage{amsfonts,amssymb,amsthm}
\usepackage{xcolor}
\usepackage{amsmath}
\usepackage{rotating}
\usepackage{booktabs}
\usepackage{makecell}
\usepackage[bottom]{footmisc}
\usepackage{tikz}

\def\init{\setcounter{equation}{0}}

\newtheorem{theoreme}{Theorem }[section]
\newtheorem{proposition}[theoreme]{Proposition}
\newtheorem{lemma}[theoreme]{Lemma}
\newtheorem{definition}[theoreme]{Definition}
\newtheorem{corollary}[theoreme]{Corollary}
\newtheorem{remark}[theoreme]{Remark}

\newcommand{\beq}{\begin{equation}}
\newcommand{\eeq}{\end{equation}}

\newcounter{smallarabics}
\newenvironment{arabicenumerate}
{\begin{list}{{\normalfont\textrm{(\arabic{smallarabics})}}}
  {\usecounter{smallarabics}\setlength{\itemindent}{0cm}
   \setlength{\leftmargin}{5ex}\setlength{\labelwidth}{4ex}
   \setlength{\topsep}{0.75\parsep}\setlength{\partopsep}{0ex}
   \setlength{\itemsep}{0ex}}}
{\end{list}}
\newcounter{smallroman}

\def\bel{\begin{lemma}}
\def\eel{\end{lemma}}
\def\bet{\begin{theoreme}}
\def\eet{\end{theoreme}}
\def\bed{\begin{definition}}
\def\eed{\end{definition}}
\def\bep{\begin{proposition}}
\def\eep{\end{proposition}}
\def\ben{\begin{arabicenumerate}}
\def\een{\end{arabicenumerate}}

\def\rr{{\mathbb R}}
\def\zz{{\mathbb Z}}
\def\cc{{\mathbb C}}
\def\nn{{\mathbb N}}

\def\hh{{\mathbb H}}
\def\SS{{\mathbb S}}

\def\ii{{\bf i}}

\def\Re{{\rm Re}}
\def\Im{{\rm Im}}

\newcommand\p{\mathrm{p}}
\newcommand\hol{\mathrm{hol}}
\newcommand\un{\mathrm{un}}

\newcommand\qq{\mathbb{Q}}
\newcommand\pp{\mathbb{P}}

\def\s{{\rm s}}

\def\h{{\rm h}}
\def\p{{\rm p}}
\newcommand\dS{\mathrm{dS}}

\def\red{}

\def\i{{\rm i}}

\def\Tr{{\rm Tr}}

\def\e{{\rm e}}
\def\d{{\rm d}}

\def\w{{\rm w}}

\def\D{{\mathrm D}}
\def\N{{\mathrm N}}
\def\cG{{\cal G}}

\def\P{{\cal P}}
\newcommand\cP{\mathcal{P}}

\def\F{{\cal F}}
\def\cF{{\cal F}}
\def\cA{{\cal A}}
\def\cH{{\cal H}}
\def\cX{{\cal X}}
\def\cK{{\cal K}}

\newcommand\cQ{\mathcal{Q}}
\newcommand\cL{\mathcal{L}}

\def\W{{\cal W}}
\def\cW{{\mathcal W}}
\newcommand\cD{\mathcal{D}}

\renewcommand\bar{\overline}

\def\12{\frac{1}{2}}

\def\qed{\hfill$\Box$\medskip}
\def\proof{\noindent{\bf Proof.}\ \ }

\def\trho{{\tilde \rho}}

\newcommand\sa{\mathrm{sa}}

\DeclarePairedDelimiter{\norm}{\lVert}{\rVert}

\begin{document}
\title{Exactly  solvable Schr\"odinger operators \\ related to the
  hypergeometric equation}
\author{
	Jan Derezi\'{n}ski*\\
	\small Department of Mathematical Methods in Physics,\\
	\small Faculty of Physics, University of Warsaw, \\
	\small Pasteura 5, 02-093 Warszawa, Poland\\
	\small email: \texttt{jan.derezinski@fuw.edu.pl}\\
	\and
        Pedram Karimi\\
	\small Faculty of Physics, University of Warsaw, \\
	\small Pasteura 5, 02-093 Warszawa, Poland\\
	\small email: \texttt{pedram.karimi@fuw.edu.pl}\\
}

\maketitle

\begin{abstract}

	We study one-dimensional Schr\"odinger operators defined as closed operators that are exactly solvable in terms of the Gauss hypergeometric function. We allow the potentials to be complex.
  These operators fall into three groups. The first group can be reduced to the Gegenbauer equation, up to an affine transformation, a special case of the hypergeometric equation. The two other groups, which we call {\em hypergeometric of the first}, resp. {\em second kind}, can be reduced to the general Gauss hypergeometric equation. Each of the group is subdivided in three families, acting on the Hilbert space $L^2]-1,1[,$ $L^2(\rr_+)$ resp. $L^2(\rr)$. Motivated by geometric applications of these families, we call them {\em spherical}, {\em hyperbolic}, resp. {\em deSitterian}. All these families are known from applications in Quantum Mechanics: e.g. spherical hypergeometric Schr\"odinger operators of the first kind are often called {\em trigonometric P\"oschl-Teller Hamiltonians}.
  For operators belonging to each family we compute their spectrum and determine their Green function (the integral kernel of their resolvent). We also describe transmutation identities that relate these Green functions. These identities interchange spectral parameters with coupling constants across different operator families. Finally, we describe how these operators arise from separation of variables of (pseudo-)Laplacians on symmetric manifolds.
  Our paper can be viewed as a sequel to \cite{DL}, where closed realizations of one-dimensional Schr\"odinger operators solvable in terms Kummer's confluent equation were studied.
\vspace{2em}
\\
\textbf{Mathematics Subject Classification MSC2020:} 33C05; 34L40; 47A10; 81Q80		

\end{abstract}

\tableofcontents
\section{Introduction}

{\em One-dimensional Schr\"odinger operators} are operators of the form
\begin{equation}
	\label{oper} L := -\partial_x^2 + V(x),
\end{equation}
where $V(x)$ is the {\em potential}, which in this paper is allowed to be complex-valued.
Our paper is devoted to several families of operators of the form
\eqref{oper}, interpreted as {\em closed operators} on $L^2(]a,b[)$
for appropriate $-\infty\leq a<b\leq+\infty$, which can be reduced
to the {\em (Gauss) hypergeometric equation}
\beq
\big(z(1-z)\partial_z^2+(c-(a+b+1)z)\partial_z-ab\big)f(z)=0,\label{hyperge1}\eeq
and whose Green functions can be expressed
in terms of the {\em (Gauss) hypergeometric function}.

We will also consider operators of the form
\eqref{oper}
that can be reduced to the {\em Gegenbauer equation}.
\beq
\Big((1-w^2)\partial_w^2-2(1+\alpha )w\partial_w
+\lambda ^2-\big(\alpha +\tfrac{1}{2}\big)^2\Big)g(w)=0.\label{gege01}\eeq
The Gegenbauer equation 
 is up to an affine transformation a special
case of the hypergeometric equation.
Its special property is the mirror symmetry.

Our paper can be viewed as the sequel to \cite{DL}, where one of the
authors (JD) together with Jinyeop Lee studied a similar problem for
the {\em confluent equation}.  We will mostly use the same terminology
and methods. We try to make the present paper reasonably
self-contained, however the reader is  encouraged to consult
\cite{DL}, especially concerning the general theory of closed
realizations of operators of the form \eqref{oper}.

In the remaining part of the introduction we give a summary of
  the results of our paper. In the later section these results will be
  discussed in detail.

  \subsection{3$\times$3 families of hypergeometric Hamiltonians}

Abusing the terminology, for the sake of brevity, we will use the term {\em Hamiltonian} for 
one-dimensional Schr\"odinger operators.
We study three categories of Hamiltonians:
\begin{enumerate}
	\item[(1)] Those reducible to the Gegenbauer
          equation; they will depend on a single
          complex parameter, and can be viewed as a subclass of
          hypergeometric Hamiltonians, both of the first and second kind.
	\item[(2)] Those reducible to the hypergeometric equation 
by the substitution $z=\sin^2\frac{r}{2}$ (or similar); they will be 
called {\em hypergeometric of the first kind};  they will depend on 
two complex parameters.
\item[(3)]
Those reducible to the hypergergeometric equation 
by the substitution $z=\frac1{1+\e^{2r}}$ (or similar); they will be 
called {\em hypergeometric of the second kind};  they will depend on 
two complex parameters.
\end{enumerate}

Within each category we will consider 3 families, which differ by the
choice of the interval $]a,b[$. This interval can be viewed as a subset of
the complex plane. We will always assume that the endpoints are
singular points of the equation. For each of
$3\times 3=9$ cases, for a  set of parameters with a
nonempty interior
the operator $L$, defined originally on $C_\mathrm{c}^\infty]a,b[$,
possesses a unique closed realization in  the sense of
$L^2]a,b[$. This realization depends
 holomorphically on parameters, and extends to a holomorphic
family of closed operators on a larger domain. We will call it the
{\em basic family} of closed realizations of $L$. For some
special ranges of parameters there exist
other closed realizations of $L$ with {\em mixed boundary conditions}---we will not 
consider them in this paper.

For each  operator $L_\bullet$  from those families we will find its
spectrum, denoted $\sigma(L_\bullet)$.
In all cases with real potentials, these operators will be
  self-adjoint (so that $\sigma(L_\bullet)\subset \rr$).
More generally,  the resolvent set (the complement of the
spectrum) of these operators  will be nonempty. For $z$ in the resolvent set we will find the
resolvent, that is $(L_\bullet-z)^{-1}$, which we will usually denote
$\frac{1}{L_\bullet-z}$.
The {\em Green function of $L_\bullet-z$}, that is integral kernel of the
resolvent $\frac1{L_\bullet-z}$, will be denoted $\frac1{(L_\bullet-z)}(x,y)$, with
$x,y\in]a,b[$.
We will find expressions of Green functions of Hamiltonians from all
3$\times$3  families in terms of the Gamma function and the Gauss hypergeometric function.

These $9$ families were discovered in the early days of Quantum
Mechanics by physicists trying to find exactly solvable models for
various quantum systems. In the literature they are usually named
after the researchers who discovered them. Instead of the traditional
names, we will prefer to use different names:
spherical, hyperbolic
and deSitterian.
 In the spherical case 
$]a,b[$ is $]-1,1[,$ in the hyperbolic case it is 
$]0,+\infty[=:\rr_+$, and  in the deSitterian case it is $\rr$.

Our names indicate a
major geometric application of these families:
 spherical, hyperbolic, resp. deSitterian  Gegenbauer Hamiltonians appear when we
separate variables for the (pseudo-)Laplacian on the sphere,  on
the hyperbolic space,
resp.  on the deSitter space. 

\subsection{Review of 9 families}

Let us briefly review  the 9 families described in the paper, referring the
reader to the main text for precise statements.
We will write $\cA_\un$ for the domain of uniqueness, that is, the set of
parameters (a subset of $\cc$ or $\cc\times\cc$) for which there exists
a unique closed realization of a given differential expression. This
realization in all cases depends holomorphically 
on  its parameters, and extends to a larger domain, denoted
$\cA_\hol$.
$\cA_\sa$ will indicate the set of parameters for which the operator
is self-adjoint. Note that the operators are essentially self-adjoint
on $C_\mathrm{c}^\infty]a,b[$ if and only if  the parameter belongs to
$\cA_\sa\cap\cA_\un$.

{\red Note that we always use the notion of an analytic family of closed
  operators  in the sense of Kato \cite{Kato,DW1}. In Appendix
  \ref{Holomorphic families of operators} we give a concise explanation of
  this concept.}

\begin{enumerate}\item Gegenbauer Hamiltonians
  \begin{enumerate}
\item Spherical Gegenbauer Hamiltonian, $L^2]0,\pi[$:
\beq
  \begin{aligned}
L_\alpha^\s&:=-\partial_r^2+
\left(\alpha^2-\frac{1}{4}\right)\frac{1}{\sin^2r},
\label{hami1-}
            \\ 
\cA_\un&=\{\Re\,\alpha\geq1\},\quad \cA_\hol=\{\Re\,\alpha>-1\},\quad\cA_\sa=]-1,+\infty[.
  \end{aligned}
\eeq
\item
Hyperbolic  Gegenbauer Hamiltonian, $L^2(\rr_+)$:
\beq
\begin{aligned}\label{hami2-}
  L_{\alpha}^\h&:
  =-\partial_r^2+
  \left(\alpha^2-\frac{1}{4}\right)\frac{1}{\sinh^2 r},
       \\ \cA_\un&=\{\Re\,\alpha\geq1\},\quad \cA_\hol=\{\Re\,\alpha>-1\},\quad\cA_\sa=]-1,+\infty[.
             \end{aligned}
             \eeq
\item DeSitterian Gegenbauer Hamiltonian, $L^2(\rr)$:
\beq
  \begin{aligned}\label{hami3-}
L_{\alpha}^\dS&:=-\partial_r^2
-\left(\alpha^2-\frac{1}{4}\right)\frac{1}{\cosh^2 r},
         \\ \cA_\un&= \cA_\hol=\cc,\quad\cA_\sa=\rr.
             \end{aligned}
\eeq
\end{enumerate}
\item Hypergeometric Hamiltonians of the first kind:
  \begin{enumerate}
  \item Spherical hypergeometric Hamiltonian of the first kind,
    $L^2]0,\pi[$:
    \beq
  \begin{aligned}\label{posch1-}
L_{\alpha,\beta}^\s&:=-\partial_r^2+
\left(\alpha^2-\frac{1}{4}\right)\frac{1}{4\sin^2\frac{r}{2}}+
                        \left(\beta^2-\frac{1}{4}\right)\frac{1}{4\cos^2 \frac{r}{2}},\\
          \cA_\un&=\{\Re\,\alpha,\Re\,\beta\geq1\},\quad \cA_\hol=\{\Re\,\alpha,\Re\,\beta>-1\},\quad\cA_\sa=]-1,+\infty[\times]-1,+\infty[.
             \end{aligned}
\eeq

\item
Hyperbolic  hypergeometric Hamiltonian of the first kind,
$L^2(]0,\infty[)$:
\beq
\begin{aligned}\label{posch2-}
       L_{\alpha,\beta}^\h&:=-\partial_r^2+
\left(\alpha^2-\frac{1}{4}\right)\frac{1}{4\sinh^2\frac{r}{2}}-
\left(\beta^2-\frac{1}{4}\right)\frac{1}{4\cosh^2 \frac{r}{2}}
,\\     \cA_\un&=\{\Re\,\alpha\geq1\}\times\cc,\quad \cA_\hol=\{\Re\,\alpha>-1\}\times\cc ,\quad\cA_\sa=]-1,+\infty[\times\rr.
\end{aligned}
\eeq
\item DeSitterian hypergeometric Hamiltonian of the first kind, $L^2(\rr)$,
\beq
  \begin{aligned}
    \label{scarf-}L_{\alpha,\beta}^\dS&
:=-\partial_r^2-
\left(\alpha^2-\frac{1}{4}\right)\frac{1}{\cosh^2 r}\left(\frac12+\frac{\i\sinh
     r}{2}\right){-}
\left(\beta^2-\frac{1}{4}\right)\frac{1}{\cosh^2 r}\left(\frac12-\frac{\i\sinh
     r}{2}\right), \\      \cA_\un&= \cA_\hol=\cc\times\cc ,\quad\cA_\sa=\{\alpha=\bar\beta\}.
\end{aligned}
\eeq
\end{enumerate}
\item Hypergeometric Hamiltonians of the second kind:
  \begin{enumerate}
  \item Spherical hypergeometric Hamiltonian of the second kind, $L^2]0,\pi[$:
  \beq
\begin{aligned}\label{second1-}
K_{\tau,\mu}^\s&:=
-\partial_u^2+\tau\frac{\cos u}{\sin  u}+
\left(\frac{\mu^2}{4}-\frac14\right) 
\frac{1}{\sin^2
  u},\\
        \cA_\un&=\cc\times\{\Re\,\mu\geq2\}, \quad \cA_\hol=\cc\times\{\Re\,\mu>-2\}\setminus\{(0,-1)\},\quad\cA_\sa=\rr\times]-2,+\infty[.
        \end{aligned}
        \eeq
\item
Hyperbolic  hypergeometric Hamiltonian of the second kind,
$L^2(]0,\infty[)$:
\beq
\begin{aligned}\label{second2-}
K_{\kappa,\mu}^\h&:=-\partial_u^2+\kappa\frac{\cosh u}{\sinh
  u}+\left(\frac{\mu^2}{4}-\frac14\right) 
\frac{1}{\sinh^2
                   u},\\
  \cA_\un&=\cc\times\{\Re\,\mu\geq2\},
\quad \cA_\hol=\cc\times\{\Re\, \mu>-2\}\setminus\{(0,-1)\},\quad\cA_\sa=\rr\times]-2,+\infty[.\end{aligned}
\eeq
\item DeSitterian hypergeometric Hamiltonian of the second kind,
  $L^2(\rr)$,
  \beq
  \begin{aligned}
K_{\kappa,\mu}^{\dS}&:=-\partial_u^2-\kappa\frac{\sinh w}{\cosh
  w}-\left(\frac{\mu^2}{4}-\frac14\right) 
    \frac{1}{\cosh^2 w},\label{second3-}\\
    \cA_\un&=\cA_\hol=\cc\times\cc ,\quad\cA_\sa=\rr\times\rr.\end{aligned}
    \eeq
\end{enumerate}

\end{enumerate}

Gegenbauer Hamiltonians are special cases of hypergeometric
Hamiltonians of both the first and second type. In fact, we have the
following coincidencies:
\begin{align}
  L_\alpha^\s=&L_{\alpha,\alpha}^\s=K_{0,2\alpha}^\s;\label{geg.1}\\
  L_\alpha^\h=&L_{\alpha,\alpha}^\h=K_{0,2\alpha}^\h;\label{geg.2}\\
  L_\alpha^\dS=&L_{\alpha,\alpha}^\dS=K_{0,2\alpha}^\dS.
                 \end{align}

Let us also list the following identities that we prove:
                              \begin{align}
                                K_{\tau,-1}^\s&=K_{\tau,1}^\s,\quad
                                                \tau\neq0, \label{pih1} \\
                                K_{\kappa,-1}^\h&=K_{\kappa,1}^\h.\quad
                                                \kappa\neq0. \label{pih2} 
                                                \end{align}
                                                These identities
                                                imply that $(0,-1)$ are
                                                singularities of
 the functions 
               $(\tau,\mu)\mapsto K_{\tau,\mu}^\s$ and 
                              $(\kappa,\mu)\mapsto K_{\kappa,\mu}^\h$.

Going back to    \eqref{geg.1} and  \eqref{geg.2}, note that
                 \begin{align}   L_\alpha^\s&=K_{0,2\alpha}^\s,\\
                 L_\alpha^\h&=K_{0,2\alpha}^\h,\end{align}
               are the identities for holomorphic functions only for
$               \alpha\neq-\frac12$, because of the above mentioned singularity.
                              Thus the identities
                            \begin{align}   L_{-\frac12}^\s&=K_{0,-1}^\s,\\
                 L_{-\frac12}^\h&=K_{0,-1}^\h.\end{align}
                 should be used as  {\em definitions} of $K_{0,-1}^\s$
                 and $K_{0,-1}^\h$.

\subsection{Transmutations of Green functions}

 Green functions of distinct  Hamiltonians from the above list are
linked by identities, which we find quite curious.
We call them {\em transmutation identities}, since the spectral parameter
undergoes a change into a coupling constant and the other way around.
They
follow from various identities satisfied by the hypergeometric
function and are similar to the transmutation identities for
Hamiltonians related to the confluent equation described in \cite{DL}.

{\red  In the literature the name ``transmutation'' is sometimes
  used for identities of the form $A=VBV^{-1}$, e.g. in \cite{car,ss}.
Note that our transmutation identities are not of this form, and in
fact they seem to  represent a novel contribution.}

Here is the list of transmutation identities  considered
in our paper.  In each case, first we indicate the change of variables involved in a
given transmutation. Then we describe two versions of the identity for Green functions.

\bep
 {\bf Gegenbauer spherical --- Gegenbauer deSitterian:}
    \begin{align}
      ]0,\pi[\ni r\mapsto q\in\rr,&\quad
      \cot r=\sinh q;\\
\sin^\frac12 r\frac{1}{\left(L_{\lambda}^\s-\alpha^2\right)}\left(r,r^\prime \right)\sin^\frac12 r^\prime&=\frac{1}{\left(L_{\alpha}^\dS-\lambda^2\right)}\left(q,q^\prime\right),\\
  \frac{1}{\left(L_{\lambda}^\s-\alpha^2\right)}\left(r,r^\prime \right)&=\cosh^{\frac12}q\frac{1}{\left(L_{\alpha}^\dS-\lambda^2\right)}\left(q,q^\prime\right)\cosh^{\frac12}q'.
\end{align}\label{trans1}\eep
\bep{\bf  1st kind spherical ---  1st kind hyperbolic:}
  \begin{align}
    ]0,\pi[\ni r\mapsto
    q\in\rr_+,&\quad\tan\frac{r}{2}=\sinh\frac{q}{2};\\
     \Big(\cos\frac{r}{2}\Big)^{-\frac12}~ \frac{1}{\left(L^\s_{\alpha,\beta}+\frac{\mu^2}{4}\right)}\left(r,r^\prime\right) ~\Big(\cos\frac{r^\prime}{2}\Big)^{-\frac12}
&=
         \frac{1}{\left(L^\h_{\alpha,\mu}+\frac{\beta^2}{4}\right)}\left(q,q^\prime\right),\\
         \frac{1}{\left(L^\s_{\alpha,\beta}+\frac{\mu^2}{4}\right)}\left(r,r^\prime\right)
&= \Big(\cosh\frac{q}{2}\Big)^{-\frac12}~\frac{1}{\left(L^\h_{\alpha,\mu}+\frac{\beta^2}{4}\right)}\left(q,q^\prime\right)  ~\Big(\cosh\frac{q^\prime}{2}\Big)^{-\frac12}.
  \end{align}\label{trans2}\eep
\bep{\bf 2nd kind hyperbolic ---
 2nd kind deSitterian}
\begin{align} \rr_+\ni u\mapsto w\in\rr,&\quad
                                          \e^{2u}=1+\e^{2w};\\
 (1-\e^{-2u})^{-\frac12}\frac{1}{\left(K^\h_{\nu-\frac{\beta^2}{2},\mu}+\nu+\frac{\beta^2}{2}
  \right)}(u,u^\prime)&(1-\e^{-2u'})^{-\frac12}=\frac{1}{\left(K^\dS_{\nu+\frac{\mu^2}{2},\beta}+\nu-\frac{\mu^2}{2}\right)}(w,w^\prime),\\
   \frac{1}{\left(K^\h_{\nu-\frac{\beta^2}{2},\mu}+\nu+\frac{\beta^2}{2} \right)}(u,u^\prime)=(1+\e^{-2 w})^{-\frac12}&\frac{1}{\left(K^\dS_{\nu+\frac{\mu^2}{2},\beta}+\nu-\frac{\mu^2}{2}\right)}(w,w^\prime)~(1+\e^{-2 w^\prime})^{-\frac12}.
\end{align}\label{trans3}\eep
\bep{\bf 1st kind spherical ---  2nd kind deSitterian}
  \begin{align}
    ]0,\pi[\ni r\mapsto u\in\rr,&\quad\cos r=\tanh u;\\
\sin^{\frac12}r~
    \frac{1}{\left(L^\s_{\alpha,\beta}+\frac{\mu^2}4\right)}\left(r,r^\prime 
    \right)~\sin^{\frac12}r^\prime&=
  \frac{1}{\left(K_{\kappa,\mu}^\dS+\delta\right)}\left(u,u^\prime 
                                      \right),\\
    \frac{1}{\left(L^\s_{\alpha,\beta}+\frac{\mu^2}4\right)}\left(r,r^\prime 
    \right)&=
  \cosh^{\frac12}u\frac{1}{\left(K_{\kappa,\mu}^\dS+\delta\right)}\left(u,u^\prime 
                                      \right)\cosh^{\frac12}u',\\
    \delta=\frac{\alpha^2+\beta^2}{2},&\quad\kappa=\frac{\alpha^2-\beta^2}{2}. \end{align}\label{trans4}\eep
  \bep{\bf  1st kind hyperbolic ---  2nd kind hyperbolic}
  \begin{align}
    \rr_+\ni r\mapsto u\in\rr_+,&\quad \cosh r=\coth u; \\
 \sinh^\frac12 
    r~\frac{1}{\left(L^\h_{\alpha,\beta}+\frac{\mu^2}{4}\right)}\left(r,r^\prime\right)
    ~\sinh^\frac12
    r^\prime&=\frac{1}{\left(K^\h_{\kappa,\mu}+\delta\right)}\left(u,
                     u^\prime \right),
    \\
 \frac{1}{\left(L^\h_{\alpha,\beta}+\frac{\mu^2}{4}\right)}\left(r,r^\prime\right) &=\sinh^{\frac12}u\frac{1}{\left(K^\h_{\kappa,\mu}+\delta\right)}\left(u, u^\prime \right)\sinh^{\frac12}u' ,\\\delta=\frac{\alpha^2+\beta^2}{2},&\quad\kappa=\frac{\alpha^2-\beta^2}{2}.
\end{align}\label{trans5}\eep
\bep{\bf  1st kind deSitterian ---  2nd kind spherical}
  \begin{align}
    \rr\ni r\mapsto u\in]0,\pi[,&\quad\sinh r=-\cot u;\\
  \cosh^{\frac12} r 
    ~\frac{1}{\left(L^\dS_{\alpha,\beta}+\frac{\mu^2}{4}\right)}(r,r^\prime)~\cosh^{\frac12}
    r^\prime&=
              \frac{1}{\left(K^\s_{\tau,\mu}+\delta\right)}\left(u,u^\prime\right) 
              ,\\
      \frac{1}{\left(L^\dS_{\alpha,\beta}+\frac{\mu^2}{4}\right)}(r,r^\prime)&=
\sin^{\frac12} u              \frac{1}{\left(K^\s_{\tau,\mu}+\delta\right)}\left(u,u^\prime\right) \sin^{\frac12}u'
              , \\
    \delta=\frac{\alpha^2+\beta^2}{2},&\quad\tau=\frac{\i(\alpha^2-\beta^2)}{2}. \end{align}
\label{trans6}\eep

\subsection{Geometric applications}

The main original application of hypergeometric Hamiltonians was
Quantum Mechanics, as we describe in Subsection \ref{Historic
  remarks}.
However, probably the most important context where hypergeometric
Hamiltonians appear is geometry, more precisely, the
theory of symmetric spaces and Lie groups. In fact, when we separate
variables for invariant differential operators, e.g. (pseudo-)Laplacians, on symmetric
(pseudo-)Riemannian spaces  we often obtain  some forms of 
hypergeometric Hamiltonians.

This fact plays an important role in Quantum
Field Theory on curved spacetimes, where Green functions of
the d'Alembertian  on  deSitter and anti-deSitter spaces appear
naturally;
see e.g. \cite{DeGa}.

This geometric interpretation is especially striking for Gegenbauer
Hamiltonians. In the following list we show how they arise after
separation of variables of various $d$-dimensional (pseudo-)Laplacians
and restriction to $d-1$-dimensional spherical harmonics of degree
$l$. In all cases $\alpha=\frac{d}{2}-1+l$:

\begin{enumerate} \item $\Delta_d^\s$, Laplacian on  unit sphere
  $\SS^d$, reduces to  spherical Gegenbauer Hamiltonian $L_\alpha^\s$:
\begin{align}
  (\sin
                                                   r)^{\frac{d-1}{2}}\big(-\Delta_d^\s\big)
                                                   (\sin
                                                   r)^{-\frac{d-1}{2}}+\big(\tfrac{d-1}{2}\big)^2&=-\partial_r^2+\frac{\big(\tfrac{d-2}{2}\big)^2-\tfrac14-\Delta_{d-1}^\s}{\sin^2  r}.
                    \end{align}                                                                              
\item $\Delta_d^\h$,  Laplacian on 
  hyperbolic space $\hh^d$, reduces to hyperbolic  Gegenbauer
  Hamiltonian $L_\alpha^\h$:
  \begin{align}
(\sinh r)^{\frac{d-1}{2}}\big(-\Delta_d^\h\big) (\sinh
  r)^{-\frac{d-1}{2}}-\big(\tfrac{d-1}{2}\big)^2=&-\partial_r^2+\frac{\big(\tfrac{d-2}{2}\big)^2-\tfrac14-\Delta_{d-1}^\s}{\sinh^2
                                                   r}.
    \end{align}\item$\Box_d^\dS$,  d'Alembertian on 
  de Sitter space $\dS^d$,
  reduces to deSitterian  Gegenbauer Hamiltonian $L_\alpha^\dS$:
\begin{align}
(\cosh r)^{\frac{d-1}{2}}\Box_d^\dS (\cosh
  r)^{-\frac{d-1}{2}}-\big(\tfrac{d-1}{2}\big)^2=&-\partial_r^2-\frac{\big(\tfrac{d-2}{2}\big)^2-\tfrac14-\Delta_{d-1}^\s}{\cosh^2
                        r}
  .\label{des1}\end{align}
\end{enumerate}

All three types of hypergeometric Hamiltonians of the 1st kind  have   natural
geometric interpretations as well. In the following list we show how they arise
from separation of variables in a (pseudo-)Laplacian on a
(pseudo-)sphere in ``double spherical coordinates''. In all three
examples the coordinates in the ambient pseudo-Euclidean space are
partitioned into two groups.

In the first two cases, these groups are
of dimension $p$ and $q$, and then spherical coordinates are
considered within each group. After restriction to products of
spherical harmonics of degree $l$ and $j$ a
hypergeometric Hamiltonian arises,
expressed in  the relative variable. We have $\alpha=\frac{p}{2}-1+l$, $\beta=\frac{q}{2}-1+j$.

The third case is somewhat
different: $p=q$ is the  dimension of holomorphic and
antiholomorphic (complex) spherical coordinates. The spherical
harmonics are not the usual ones: they are holomorphic and
antiholomorphic harmonics  on 
the complex $p-1$-dimensional sphere.

\begin{enumerate}
\item  $\Delta_{p+q-1}^\s$, Laplacian on  unit sphere 
  $\SS^{p+q-1}$, reduces to spherical hypergeometric Hamiltonian of 1st 
  kind $L_{\alpha,\beta}^\s$:
\begin{align} \notag
                &  \left(\sin\frac{r}{2}\right)^{\frac{p-1}{2}}  \left(\cos\frac{r}{2}\right)^{\frac{q-1}{2}}
 \left(- \Delta_{p+q-1}^\s\right)  \left(\sin\frac{r}{2}\right)^{-\frac{p-1}{2}}
\left(\cos\frac{r}{2}\right)^{-\frac{q-1}{2}}+\left(\tfrac{p+q-2}{2}\right)^2
  \\=&4\left(
                              - \partial_r^2
+\frac{\left(\tfrac{p-2}{2}\right)^2-\tfrac14-\Delta_{p-1}^\s}{4\sin^2
                                  \frac{r}{2}}                                  +\frac{\left(\tfrac{q-2}{2}\right)^2-\tfrac14-\Delta_{q-1}^\s}{4\cos^2
                                  \frac{r}{2}}\right).\label{lapla1-}
\end{align}

\item $\Delta_{p-1,q}$, pseudo-Laplacian on  the hyperboloid 
  $\hh^{p-1,q}$, reduces to hyperbolic hypergeometric Hamiltonian of 1st 
  kind $L_{\alpha,\beta}^\h$:
  \begin{align}  \notag
                &\left(\cosh\frac{r}{2}\right)^{\frac{p-1}{2}}  \left(\sinh\frac{r}{2}\right)^{\frac{q-1}{2}}
 \left(-\Delta_{p-1,q}\right) \left(\cosh\frac{r}{2}\right)^{-\frac{p-1}{2}}
\left(\sinh\frac{r}{2}\right)^{-\frac{q-1}{2}} -\left(\tfrac{p+q-2}{2}\right)^2
  \\
 =&4\left(-                               \partial_r^2
-\frac{\left(\tfrac{p-2}{2}\right)^2-\tfrac14-\Delta_{p-1}^\s}{4\cosh^2
                                  \frac{r}{2}}                                  +\frac{\left(\tfrac{q-2}{2}\right)^2-\tfrac14-\Delta_{q-1}^\s}{4\sinh^2
                                  \frac{r}{2}}\right).\end{align}
                            \item $\Delta_{p-1,p}$, pseudo-Laplacian on  the hyperboloid 
  $\hh^{p-1,p}$, reduces to deSitterian hypergeometric Hamiltonian of 1st 
  kind $L_{\alpha,\beta}^\dS$:
                              \begin{align}\notag
&\left(\cosh r\right)^{\frac{p-1}{2}}\left(\Delta_{p-1,p}\right)\left( \cosh r\right)^{-\frac{p-1}{2}}-(p-1)^2\\ \label{geo3}
  =&4\left(-\partial_r^2-\frac{\left(-\Delta_{p-1}^{\s,\cc}+\left(\frac{p-1}{2}\right)^2-\frac14\right)}{2(1+\i\sinh r)}-\frac{\left(-\bar{\Delta_{p-1}^{\s,\cc}}+\left(\frac{p-1}{2}\right)^2-\frac14\right)}{2(1-\i\sinh r)}\right).
                              \end{align}
                            \end{enumerate}

Unfortunately, we have not found a direct geometric interpretation of
hypergeometric Hamiltonians of the second kind.

\begin{remark} We use the geometric interpretation of Gegenbauer
  Hamiltonians as the main justification for our names of types:
  spherical, hyperbolic and deSitterian. For coherence, we extend these names to
  hypergeometric Hamiltonians of both kinds. The names ``spherical''
  and ``hyperbolic'' seem quite non-controversial, and are used in the
  literature in similar contexts. The name ``deSitterian'' is our
  invention. {\red Its justification is somewhat weaker and based 
  on the geometric interpretation of the Gegenbauer Hamiltonian, see
  \eqref{des1}.  The  
    natural geometric interpretation of
   ``deSitterian  hypergeometric Hamiltonians of the first kind'',
   which we described in \eqref{geo3}
involves the hyperboloid $\hh^{p-1,p}$, which is a deSitter space only
for $p=2$.}
  \end{remark}
  
  \subsection{Boundary conditions}

For some potentials the operator  $L$ initially defined on
$C_\mathrm{c}^\infty]a,b[$ possesses many closed realizations. These
realizations $L_\bullet$  differ only by the behavior of
elements of their domain near the endpoints---in other words, they
differ by {\em boundary conditions}.  The need for boundary conditions
 depends on the behavior of the
potential $V$ near these endpoints.

Let us
consider e.g. the right endpoint $b$. There are two possibilities:
\begin{enumerate} \item One does not need to impose boundary conditon at
  $b$. This will be denoted $\nu_b(L)=0.$
\item There is a 1-parameter family of boundary conditions at
  $b$. This is denoted $\nu_b(L)=2$.
\end{enumerate}

Analogous definitions are valid for the other endpoint $a$.

 The 9 families considered in this paper 
  illustrate Hamiltonians with
various kinds of behaviors of the potential near endpoints.
Let us list the behaviors encountered among these 9 families.
We restrict ourselves to the right endpoint $b$, an analogous
list applies to the left endpoint $a$. Our description is somewhat
informal; for rigorous statements we refer to \cite{DL}.
\begin{enumerate}
{\red
  \item {\bf Short range potential.}
    $b=+\infty$ and $V(x)$ is integrable near $+\infty$ or constant plus integrable.
    Then $\nu_b(L)=0.$ Moreover, eigenfunctions in  $\cD(L_\bullet)$ with
    eigenvalue $-k^2+V(+\infty)$, $\Re(k)>0$ behave as $\e^{-kx}$ with $\Re(k)>0$.
    }
  \item {\bf Bessel type.}  $b$  is finite and
    \beq V(x)\sim
  \Big  (m^2-\frac14\Big)\frac1{(x-b)^2}.\eeq
    Then $\nu_b(L)=2$ iff
     $|\Re\,m|<1$, otherwise $\nu_b(L)=0$. The behavior of elements
      of $\cD(L_\bullet)$ near $b$ is $|x-b|^{\frac12+m}$ for
      $\Re\,m\geq1$,  a linear combination of $|x-b|^{\frac12+m}$ and
$      |x-b|^{\frac12-m}$ for 
$0\leq \Re\,m<1$, $m\neq0$, and a linear combination of $|x-b|^{\frac12}$ and
$|x-b|^{\frac12}\ln|x-b|$ for $m=0$.

The families of operators 
considered in this paper will always have only ``homogeneous'' or ``basic'' 
boundary conditions given by $|x-b|^{\frac12+m}$, with $\Re\, m>-1$. 
Thus in particular for $|\Re\,m|<1$, $m\neq0$, each differential 
expression will have two closed realizations. We will not consider
mixed boundary conditions,  which are discussed  e.g. in. \cite{GTV,DeGe,DeRi}.

{\red Note that $|x-b|^{\frac12+m}$  for $m=\frac12$ is $|x-b|$, and
 for
  $m=-\frac12$ is $|x-b|^0$. Therefore, $m=\frac12$ corresponds to the Dirichlet b.c. and 
$m=-\frac12$ to the Neumann b.c. We discuss separately these cases in Sect.
\ref{The Laplacian on an interval, halfline and line}.}

\item {\bf Whittaker type.}
 $b$  is finite and
    \beq V(x)\sim
  \left (m^2-\frac14\right)\frac1{(x-b)^2}-\frac{\beta}{|x-b|}.\eeq
The conditions are essentially the same as in the Bessel type, with one
difference.
For $\frac12\leq \Re\, m<1$, $(\beta,m)\neq\left(0,\frac12\right)$
the behavior of functions in the domain of closed
realizations of $L$ are linear combinations of $|x-b|^{\frac12+m}$ and
$|x-b|^{\frac12-m}\left(1-\frac{\beta|x-b|}{1-2m}\right)$.

In the families of operators 
considered in this paper we will only consider  ``basic'' 
boundary conditions given by $|x-b|^{\frac12+m}\left(1-\frac{\beta |x-b|}{1-2m}\right)$,
with $\Re\, m>-1$. 
See e.g. \cite{DL,DeRi}
\end{enumerate}

\begin{table}[ht]
\centering
\caption{Boundary behavior of the potentials.}
\label{boundary}

\renewcommand{\arraystretch}{1.2}
{\red
\begin{tabular}{p{4.2cm}p{4cm}p{3.5cm}}
\toprule
Potential & Boundary type at $a$ & Boundary type at $b$ \\
\midrule

Spherical of first kind
& Bessel 
& Bessel \\

Hyperbolic of first kind 
& Bessel 
& Short range potential \\

DeSitterian of first kind 
& Short range potential 
& Short range potential \\

\addlinespace

Spherical of second type 
& Whittaker 
& Whittaker \\

Hyperbolic of second type 
& Whittaker 
& Short range potential \\

DeSitterian of second type 
& Short range potential
& Short range potential \\

\bottomrule
\end{tabular}
}
\end{table}                            
\subsection{Comparison with literature and historic remarks}\label{Historic remarks}

One can roughly divide mathematical literature related to the topic of this paper into 
two parts: algebraic and functional-analytic.

In  algebraic papers one 
considers differential expressions without a Hilbert space setting and
without asking for self-adjointness or closedness.
Functional analytic papers treat  differential operators as
unbounded operators on a certain Hilbert space, usually self-adjoint,
sometimes only closed.

Needless to say,
the algebraic  literature is vast. In fact, the hypergeometric
equation is one of the most classic subjects of mathematics, with
history going back about three centuries. 
From this category one should mention 
\cite{cks,DW} which contain an  algebraic analysis of all 9 families 
of Schr\"odinger operators solvable in terms of the 
hypergeometric function.

1-dimensional Schr\"odinger operators are naturally a special case of
Sturm-Liouville operators, whose history goes back to \cite{liouville}.
There exists large contemporary literature about self-adjoint or closed
realizations of Sturm-Liouville
operators, see e.g.
\cite{GeZin,GTV,DunfordSchwartz,EdmundsEvans}.
In our paper we use mostly
\cite{DeGe}, which is summarized in Section 2 of \cite{DL}.

Each 
Sturm-Liouville operator can appear in many equivalent forms, often with 
distinct names.
By a unitary transformation, called the
Liouville transformation, essentially
each of them can be transformed into a Schr\"odinger
operator, often called its {\em Liouville form} \cite{liouville}.

For instance, in the literature one
often considers the Sturm-Liouville operator called the {\em Jacobi
  operator},
\beq
-(1-x)^{-\alpha}(1+x)^{-\beta}\partial_x(1-x)^{\alpha+1}(1+x)^{\beta+1}\partial_x,\eeq
which acts on the Hilbert space
$L^2\big(]-1,1[,(1-x)^\alpha(1+x)^\beta)$ see e.g. \cite{GLPS,koornwinder}.
Its eigenfunctions are the famous Jacobi polynomials.
By a simple transformation 
the Jacobi operator  is unitarily equivalent to the
trigonometric P\"oschl-Teller Hamiltonian
 (which we call spherical
hypergeometric Hamiltonian of the first kind).

Another common differential operator
is the Legendre operator
\beq
(1+x^2) \partial_x^2-2x \partial_x+\mu(\mu+1) -\frac{\alpha^2}{1-x^2}.
\eeq
Acting on the Hilbert space $L^2\big(]-1,1[,\sqrt{1-x^2}\big)$, used in the
study of spherical harmonics.
It is also equivalent to the trigonometric P\"oschl-Teller Hamiltonian
with $\alpha=\beta$.  Needless to say, spherical harmonics possess a very large literature.

The
trigonometric P\"oschl-Teller Hamiltonian itself is also often studied, see e.g. \cite{FS}.

The Scarf Hamiltonian with $\alpha=\beta$ (which we call the
deSitterian Gegenbauer Hamiltonian) has also a large literature because of its special properties: for certain  values of parameters
it is reflectionless.

An interesting review of various exactly solvable Schr\"odinger
operators is contained in \cite{Everitt}. It does not, however,
contain all 9 families that we consider.

To our knowledge, a complete analysis
of all 9 families interpreted as closed operators,
 including the formulas for the integral
kernels of their resolvents and the computation of their spectra,
seems
to appear for the first time in the literature.
The transmutation
formulas described in our paper are probably new. They 
are analogous to the transmutation formulas for Hamiltonians
related to the confluent equation \cite{DL}.
Another novelty of our paper are the identities 
\eqref{pih1} and \eqref{pih2}. They  are analogous to an identity for the
Whittaker operators described in \cite{DeRi}. 

Let us briefly outline the history of these Hamiltonians in physics. In the physics community, the study of Schr\"odinger operators
exactly solvable in terms of
hypergeometric functions started in the 1930's when physicists studied biatomic
and polyatomic molecules dynamics and the exact solution to the
Schr\"odinger equation. The main reason these physicists considered such Hamiltonians was often due to the limitations of the perturbation method. Therefore, various kinds of exactly solvable Hamiltonians were suggested which fit the experimental data. That was the primary motivation behind the Hamiltonians proposed by Morse \cite{morse}, Rosen-Morse
\cite{rmorse}, Eckart \cite{eckart}, and Manning-Rosen \cite{mrosen}.

The second motivation for this line of research seems to be the search
for exact solutions to the Schrödinger equation. In the work of Eckart \cite{eckart}
and Rosen-Morse \cite{rmorse}, it is clearly noted that these potentials are new
exactly solvable potentials, although their main motivation was
still
the study of polyatomic molecules. It seems that for P\"oschl and
Teller \cite{pt}, the motivation leaned more towards the fact that
their proposed potential was exactly solvable—they even referred to it
as exakt integrierbar (exactly integrable). 

The case of the Scarf Hamiltonian \eqref{scarf-} is a bit different. In the original paper, Scarf did not consider the Hamiltonian in \eqref{scarf}.
We traced this naming to \cite{cks}, where authors refered to it as the hyperbolic Scarf, or Scarf II. However, the origin of this naming is unclear to us.

The uniform study of the Schrödinger equation of hypergeometric type
was started by Bose \cite{bose} and continued by Natanzon
\cite{natanzon}, Ginocchio \cite{ginocchio}, and Milson
\cite{milson}. For a more systematic study and a detailed history of
the topic, we refer to \cite{DW}. In a separate line of research, the
study of these potentials appeared in the factorization method of
Infeld and Hull \cite{hi}, and later in the context of supersymmetric
quantum mechanics and the so-called shape invariance \cite{cks,cotfas}. 

Table \ref{names} presents a comparison of the various names used in the literature and our suggested terminology.        

\begin{table}[ht] \label{names}
\centering
\caption{Names of hypergeometric Hamiltonians appearing in the literature.}
\label{names}

\renewcommand{\arraystretch}{1.2}
{\red
\begin{tabular}{p{3.8cm}p{3.5cm}p{3.2cm}p{3.8cm}}
\toprule
Our suggestion & Name in \cite{DW} & Name in \cite{cks} & Alternative names \\
\midrule

Spherical of 1st kind 
& Trigonometric P\"oschl--Teller 
& Scarf I 
& P\"oschl--Teller of 1st kind, Trigonometric Scarf \\

Hyperbolic of 1st kind 
& Hyperbolic P\"oschl--Teller 
& Scarf II 
& P\"oschl--Teller of 2nd kind \\

DeSitterian of 1st kind 
& Scarf 
& Generalized P\"oschl--Teller 
& Hyperbolic Scarf \\

\addlinespace

Spherical of 2nd kind 
& Rosen--Morse 
& Rosen--Morse I 
& Trigonometric Rosen--Morse \\

Hyperbolic of 2nd kind 
& Eckart 
& Eckart 
& Generalized Morse, Hulth\'en \\

DeSitterian of 2nd kind 
& Manning--Rosen 
& Rosen--Morse II 
& Hyperbolic Rosen--Morse, Woods--Saxon \\
\bottomrule
\end{tabular}
}
\end{table}



The geometric interpretation of hypergeometric Hamiltonians is closely
related to the analysis of hypergeometric equation based on Lie
groups and Lie algebras, which possess large literature
\cite{M1,V,Wa,DHR2}. The interpretation of the deSitterian
hypergeometric Hamiltonian of the 1st kind in terms of complex spheres
in \ref{geo3}
seems to be new.

\subsection{Plan of the paper}

In Sect. \ref{Preliminaries} we recall the definitions of
hypergeometric and Gegenbauer function (the latter, following the
conventions of \cite{DGR}), and we sketch the Liouville method that
allows us to transform a Sturm-Liouville operator into a 1-dimensional
Schr\"odinger operator.

The core of the paper are the sections
\ref{Gegenbauer Hamiltonians},
\ref{Hypergeometric Hamiltonians of the first kind} and
\ref{Hypergeometric Hamiltonians of the second kind}, where the
$3\times3=9$ families of Hamiltonians are introduced and studied. For
each member of each family its spectrum and Green functions are
computed. We also prove various identities that we already listed in
the introduction, including the transmutation identities.
{\red These results are described in $9$ theorems, which have
  similar proofs, based mostly on the abstract theory described in Appendix
\ref{Closed realizations of 1d Schrodinger operators}, and more fully
in \cite{DL,DeGe}. The proof of the first theorem, Thm \ref{geg1}, is
spelled out in more detail, the next ones are more concise.}

Section \ref{The Laplacian on an interval, halfline and line} is
devoted to  1-dimensional Laplacians with various boundary
conditions. We explain  why they are special cases of
hypergeometric Hamiltonians.

 In  Sect. \ref{Major geometric 
  applications}
we  describe how Gegenbauer Hamiltonians and hypergeometric 
Hamiltonians of the first kind arise when we separate variables in 
(pseudo-)Laplacians on some symmetric spaces.

{\red
In Appendix   \ref{Holomorphic families of operators} we recall the
definition of an analytic family of closed operators following
\cite{Kato,DW1}.

In Appendix
\ref{Riemann equation} we give a concise introduction to differential
equations in the complex domain and to the Riemann equation  (also called the
Papperitz equation).
It is very useful to view the Gauss hypergeometric equation
as a special case of the Riemann equation. We introduce the notation for the Riemann
operator, which is directly inspired by the classic notation due to
Papperitz \cite{ww}.
}

In Appendix \ref{Identities for the hypergeometric function}
we collect identities about hypergeometric and Gegenbauer functions
that we use in our paper.

In Appendix \ref{Closed realizations of 1d Schrodinger operators} we
give a concise account of the theory of closed
realizations of 1d Schr\"odinger operators. This account is incomplete
and not fully rigorous---the reader should consult Sect. 2 of
\cite{DL} for a more detailed and rigorous exposition, or
 \cite{DeGe}, where a complete theory with proofs is given. Of course,
 the topic is classic and contained in other texts such as
 \cite{DunfordSchwartz,EdmundsEvans,DeGe}.

In our paper
we try to use notation and conventions that make our formulas,
especially for Green functions, as simple, elegant and symmetric as
possible. This often motivates us to introduce our conventions,
different from the standard ones. In particular, we do not use
the standard conventions for associated Legendre functions, which
could be used for Gegenbauer Hamiltonians.
For readers used to
associated Legendre functions,
in Appendix \ref{Associated Legendre functions vs. Gegenbauer
  functions}
we recall their definitions 
and compare them with the special functions that we use.

\section{Preliminaries}
\label{Preliminaries}
\init
\subsection{Hypergeometric equation}

The {\em hypergeometric equation} is given by the {\em hypergeometric operator}
\beq \F(a,b;c;z,\partial_z):=
z(1-z)\partial_z^2+(c-(a+b+1)z)\partial_z-ab,\label{hyperge}\eeq
where $a,b,c$ are arbitrary complex parameters.
One of solutions of the hypergeometric equation is the famous
hypergeometric function $F(a,b;c;z)$. It is usually convenient to apply to $F(a,b;c;z)$
the so-called {\em Olver's normalization}, which yields the function
\beq\label{olver}
\mathbf{F}(a,b;c;z):=\frac{F(a,b;c;z)}{\Gamma(c)}
=\sum_{j=0}^\infty\frac{(a)_j(b)_j}{\Gamma(c+j)j!}z^j.\eeq



Note that the
  symmetries of the hypergeometric equation are better visible
 if we replace
 $a,b,c$ with $\alpha,\beta ,\mu$:
 {\red
\begin{align}\begin{array}{rrr}\label{newnot}
\alpha=c-1&\ \ \beta =a+b-c,&\ \ \mu=a-b;\\
a=\frac{1+\alpha+\beta +\mu}{2},&\ \ b=\frac{1+\alpha+\beta
  -\mu}{2},&\ \ c=1+\alpha;
\end{array}\end{align}
}
so that the hypergeometric equation has the form
\beq \cF\left({\textstyle\frac{\alpha+\beta+\mu+1}{2}}, 
{\textstyle\frac{\alpha+\beta-\mu+1}{2}}
;1+\alpha;z,\partial_z\right) 
F(z)=0.\label{hypper}
\eeq 

{\red
To understand the properties and  importance of the hypergeometric
equation it is useful to view it as a special case of the Riemann
equation--the most general equation on the Riemann sphere possessing
at most 3 singular points, all of them regular-singular.  
There are many books and survey articles that comprehensively treat
the Riemann and hypergeometric equations, including \cite{ww,Ince,nu,NIST}. Throughout this paper, we follow the
notation established in \cite{DW,DHR1,DHR2}. For the convenience of the
reader in Appendix \ref{Riemann equation} we explain how the hypergeometric
equation arises from the Riemann equation.
}

\subsection{Gegenbauer equation}

The {\em Gegenbauer equation} is  essentially the special case of the hypergeometric 
equation with the symmetry $w\to-w$ and the singular points put at 
$-1,1,\infty$. It is given by the {\em Gegenbauer operator}
\begin{align}
  \cG_{\alpha,\lambda}(w,\partial_w):=
(1-w^2)\partial_w^2-2(1+\alpha )w\partial_w
+\lambda ^2-\big(\alpha +\tfrac{1}{2}\big)^2.\label{gege0}
\end{align}
Here is its relationship to the Riemann operator {\red(defined in \eqref{ff2})}:
\beq
\cG_{\alpha,\lambda}(w,\partial_w)=(1-w^2)\P\left(\begin{array}{cccc}
-1&1&\infty&\\
0&0&\alpha+\lambda+\frac12&w,\partial_w\\
-\alpha&-\alpha&\alpha-\lambda+\frac12&\end{array}\right).
\eeq
Following \cite{DGR}, we introduce two special solutions of the
Gegenbauer equation

\begin{align}\label{geg1}
  \mathbf{S}_{\alpha, \lambda}(w) &=
  \frac{1}{\Gamma(\alpha+1)}F\left(\frac12+\alpha+\lambda, 
  \frac12+\alpha-\lambda;\alpha+1;\frac{1-w}{2}\right),\\\label{geg2}
    \mathbf{Z}_{\alpha, \lambda}(w) &=\frac{(w\pm1)^{-\frac12-\alpha-\lambda}}{\Gamma(\lambda+1)}F\left(\frac12+\alpha+\lambda, 
  \frac12+\lambda;2\lambda+1;\frac2{1\pm w}\right).
  \end{align}
\subsection{Liouville transformation}

An operator of the form
\beq -\partial_r^2+V(r)\eeq
will be called a {\em (1-dimensional) Schr\"odinger operator}.

Let us briefly describe how to  transform
a 2nd order equation
\beq\label{sturm.}
\left(p(z)\partial_z^2+q(z)\partial_z+r(z)\right)u(z){\red =0},\eeq
into an eigenvalue equation of a certain
Schr\"odinger operator.
Consider the operator
\beq\label{sturm}
p(z)\partial_z^2+q(z)\partial_z+r(z),\eeq
that defines the equation \eqref{sturm.}.
We first  multiply \eqref{sturm} from the left by a function $f$, from the
right by a function $g$, obtaining
\beq\label{sturm0}
f(z)\big(p(z)\partial_z^2+q(z)\partial_z+r(z)\big)g(z).\eeq
We choose $f,g$ in such a way, that \eqref{sturm0} has the form
\beq
-t(z)\partial_z^2-\frac12 t'(z)\partial_z+v(z),\eeq
for some function $z\mapsto t(z)$.
Then we change the variable $z$ into $r$, such that
\beq\left(\frac{\d z}{\d r}\right)^2=t(z).\eeq
We obtain 
\beq-\partial_r^2+v\big(z(r)\big).\label{sturm1}\eeq
More details can be found in \cite{DW}.

Operators of the form \eqref{sturm} are often called {\em
  Sturm-Liouville operators} and the transformation that leads from
\eqref{sturm} to \eqref{sturm1}---a  {\em Liouville transformation}.

For
brevity, we will
usually use the term {\em Hamiltonian} instead of {\em (one-dimesional) Schr\"odinger
  operator}. Thus e.g. the {\em hypergeometric operator}  means the
operator \eqref{hyperge}, whereas 
{\em hypergeometric Hamiltonians} will be various Schr\"odinger
operators obtained by transforming the hypergeometric equation.

\section{Gegenbauer Hamiltonians}
\label{Gegenbauer Hamiltonians}
\init
Let us transform the Gegenbauer operator \eqref{gege0}  as follows: 
\begin{eqnarray}
&&-(1-w^2)^{\frac{{\alpha}}{2}+\frac{1}{4}}
\cG_{\alpha,\lambda}(w,\partial_w)  
(1-w^2)^{-\frac{\alpha}{2}-\frac{1}{4}}
\nonumber\\[2ex]
&=&-(1-w^2)\P\left(\begin{array}{cccc}
-1&1&\infty&\\
\frac{\alpha}{2}+\frac{1}{4}&\frac{\alpha}{2}+\frac{1}{4}&\lambda&w,\partial_w\\
-\frac{\alpha}{2}+\frac{1}{4}&-\frac{\alpha}{2}+\frac{1}{4}&-\lambda
&\end{array}\right)\nonumber\\[4ex]
&=&-(1-w^2)\partial_w^2+w\partial_w
+\left(\alpha^2-\frac14\right)\frac{1}{1-w^2}
    -\lambda^2\label{geg}. \end{eqnarray}
  Thus if we set
  \beq L_{\alpha}: =-(1-w^2)\partial_w^2+w\partial_w
+\left(\alpha^2-\frac14\right)\frac{1}{1-w^2},
\label{geg0}\eeq
and $G(w)$ solves  the Gegenbauer equation, then 
\beq 
\big(L_\alpha-\lambda^2\big)(1-w^2)^{\frac{\alpha}{2}+\frac{1}{4}}
G(w)=0.\eeq

{\red We have interpreted
 the Gegenbauer equation 
as the
eigenequation of the operator $L_{\alpha}$ with the
eigenvalue
$\lambda^2$  acting on functions on the complex plane. We would
  like to reinterpret it as an operator on functions of the real
  variable $r\in]a,b[$ for some $-\infty\leq a<b\leq+\infty$.
To this end we need to insert the intervals $]a,b[$ into the complex
plane.
We will consider three  intervals; in each case their
endpoints are among the singular points of the
Gegenbauer equation (that is, $1,-1,\infty$).
For  each of these intervals we perform the Liouville transformation, which  
yields a 1-dimensional Hamiltonian on $C_\mathrm{c}^\infty]a,b[$.
As we will see below, in each of these cases we will obtain a natural holomorphic
family of closed realizations  possessing
elegant formulas for the Green functions.

Here are the
three cases that we consider:}
\begin{enumerate}
\item  $w\in]-1,1[$.  This 
  leads to an operator on $L^2]0,\pi[$,  which we
  call the {\em spherical Gegenbauer Hamiltonian}.
\item $w\in]1,\infty[$.  This  leads to an operator on $L^2(\rr_+)$,
  which we call the  {\em  hyperbolic Gegenbauer Hamiltonian}.
\item $w\in\i\rr$. This leads to an operator on $L^2(\rr)$, which
  we call the {\em deSitterian  Gegenbauer Hamiltonian}.
\end{enumerate}
\begin{center}

\begin{tikzpicture}
]
\draw[help lines,-] (-3,0) -- (3,0) coordinate (xaxis);
\draw[dashed,very thick] (0,-3) -- (0,3) coordinate (desitterian);
\draw[dotted, very thick] (1,0) -- (3,0) coordinate (hyperbolic);
\draw[very thick] (-1,0) -- (1,0) coordinate (spherical);


\node[left] at (0,3) {$\i \infty$};
\node[left] at (0,-3) {$-\i \infty$};
\node[below] at (-1,0) {-1};
\draw[fill,black] (-1,0) circle (1.5pt);
\node[below] at (1,0) {1};
\draw[fill,black] (1,0) circle (1.5pt);
\node[below] at (3,0) {$\infty$};
\draw[fill,black] (3,0) circle (1.5pt);
\node[right] at (0,1.5) {DeSitterian};
\node[below] at (-1,-0.3) { Spherical};
\node[above] at (2,0) {Hyperbolic};
\end{tikzpicture}
    \\
    Figure 1: Gegenbauer equation  on the $w$ plane.\\ \red The
    spherical Hamiltonian acts on the interval marked with a thick
    line, the
    hyperbolic
    Hamiltonian---with a dotted line, and the  deSitterian
    Hamiltonian---with a dashed line.
\end{center}




\subsection{Spherical case}

For $r\in]0,\pi[$, in \eqref{geg} set\beq w=\cos r,\ \ \hbox{ which solves }
\ w'=-(1-w^2)^{\frac12}.\label{a1}\eeq
This leads to the Schr\"odinger equation
\beq \left(L_{\alpha}^\s-\lambda^2\right)\phi(r)=0,
\label{posch-g}\eeq
where
\beq
L_\alpha^\s:=-\partial_r^2+
\left(\alpha^2-\frac{1}{4}\right)\frac{1}{\sin^2r}.
\label{hami1}
\eeq
It has the mirror symmetry $r\to\pi-r$.
It is obtained when we separate variables of the Laplacian on 
the sphere in any dimensions, see e.g. Subsect. \ref{Sphere}. 
Hence our name ``spherical''.


Let us define the function on $]0,\pi[$
 \beq
\mathcal{P}^\s_{\alpha, \lambda}(r)~:=~  \Big(\frac{\sin r }{2}\Big)^{\alpha+\frac12}\mathbf{S}_{\alpha, \lambda}(\cos r).
\eeq
It has the following asymptotic behaviour  near $0$:
\beq \label{beh1}
    \mathcal{P}^\s_{\alpha, \lambda}(r) \sim \frac{1}{\Gamma(1+\alpha)}\Big(\frac{r}{2}\Big)^{\frac12+\alpha}.
\eeq

The following four functions solve the eigenequation
\eqref{posch-g}:
\beq 
\mathcal{P}^\s_{\alpha, \lambda}(r),\quad \mathcal{P}^\s_{-\alpha,
  \lambda}(r),\quad \mathcal{P}^\s_{\alpha,\lambda}(\pi-r),\quad \mathcal{P}^\s_{-\alpha,\lambda}(\pi-r).
\eeq
The following symmetries hold 
\begin{eqnarray}
    \mathcal{P}^\s_{\alpha,\lambda}(r) =\mathcal{P}^\s_{\alpha,-\lambda}(r), \qquad \mathcal{P}^\s_{\alpha,\lambda}(\pi-r)=\mathcal{P}^\s_{\alpha,-\lambda}(\pi-r). 
\end{eqnarray}

The following theorem describes the basic family of closed realizations of $L_{\alpha}^\s$ on  $L^2]0,\pi[$. 

\begin{theoreme}
  For $\Re\,\alpha\geq1$ there exists a unique closed operator
  $L_\alpha^\s$ in the sense of
   $L^2]0,\pi[$, which on $C_\mathrm{c}^\infty]0,\pi[$
   is  given by 
\eqref{hami1}. The
family $\alpha\mapsto L_\alpha^\s$ is holomorphic and possesses a
unique holomorphic extension to $\Re\,\alpha>-1$.
It has only discrete spectrum:
\beq\sigma(L_\alpha^\s)=\sigma_\d(L_\alpha^\s)=\left\{\big(k+\alpha\big)^2\ :\ k\in\nn_0+\frac12\right\}.\eeq
Outside of the spectrum its resolvent is
\begin{align} \notag
\frac{1}{(L_\alpha^\s-\lambda^2)}(x,y) =& \Gamma \left(\alpha -\lambda +\frac{1}{2}\right) \Gamma \left(\alpha +\lambda +\frac{1}{2}\right)  \\
&\times\begin{cases}
    \mathcal{P}^\s_{\alpha,\lambda}(x)\mathcal{P}^\s_{\alpha,\lambda}(\pi-y) ,\quad \text{if} \quad 0<x<y<\pi;
    \\
    \mathcal{P}^\s_{\alpha,\lambda}(y) \mathcal{P}^\s_{\alpha,\lambda}(\pi-x) ,\quad \text{if} \quad 0<y<x<\pi.
\end{cases}\label{resgegs}
\end{align}\label{geg1}
\end{theoreme}

\proof 
{\red Treating $\mathcal{P}^\s_{\alpha,\lambda}(r)$ and
$\mathcal{P}^\s_{-\alpha,\lambda}(r)$ as a basis of
solutions of \eqref{posch-g}, we obtain a connection formula found using
\eqref{gegencon}:
\beq 
    \mathcal{P}^\s_{\alpha, \lambda}(\pi -r) = -\frac{\cos \pi \lambda}{\sin \pi  \alpha} \mathcal{P}^\s_{\alpha,  \lambda}(r) + \frac{\pi}{\sin \pi \alpha} \frac{1}{\Gamma(\frac12 + \alpha+ \lambda)\Gamma(\frac12 + \alpha-\lambda)} \mathcal{P}^\s_{-\alpha,\lambda}(r). 
\eeq  
From\beq 
    \mathcal{P}^\s_{-\alpha, \lambda}(r) \sim \frac{1}{\Gamma(1-\alpha)}\Big(\frac{r}{2}\Big)^{\frac12-\alpha},
\eeq 
 and \eqref{beh1} we obtain
\beq\cW(\cP_{-\alpha,\lambda}^\s,
\cP_{\alpha,\lambda}^\s)=\frac{\sin\pi\alpha}{\pi}.\label{wrons}
\eeq
This yields the  Wronskians 
\begin{align}\label{wro..1}
    \W\left(\mathcal{P}^\s_{\alpha, \lambda}(\pi-r),\mathcal{P}^\s_{\alpha, \lambda}(r)  \right) &= \frac{1}{\Gamma \left(\alpha -\lambda +\frac{1}{2}\right) \Gamma \left(\alpha +\lambda +\frac{1}{2}\right)}
    \\\label{wro..2}
    \W\left(\mathcal{P}^\s_{\alpha, \lambda}(\pi-r) ,\mathcal{P}^\s_{-\alpha, \lambda}(r) \right) &= \frac{ \cos (\pi \lambda )}{ \pi } . 
\end{align}
 Thus  \eqref{resgegs} is an integral kernel of the form
  \eqref{eq:k}.  Let us denote it
by $R_\alpha^\s(-\lambda^2;x,y)$. 

Let us fix temporarily $\lambda$. For $\Re\,\alpha>-1$ the integral kernel
$R_\alpha^\s(-\lambda^2;x,y)$ is square
integrable, and hence it defines a Hilbert-Schmidt operator,  which we
denote $R_\alpha^\s(-\lambda^2)$. By Appendix \ref{Resolvent} there exists an
operator $L_\alpha^\s$, which is a closed realization of the left hand
side of \eqref{hami1} such that
$R_\alpha^\s(-\lambda^2)=(L_\alpha^\s+\lambda^2)^{-1}$.

 The Hilbert-Schmidt norm of
  $R_\alpha^\s(-\lambda^2)$ is
  uniformly bounded for 
  $\alpha\in K$ with $K\subset\{\Re\,\alpha>-1\}$ compact.
Besides, $R_\alpha^\s(-\lambda^2;x,y)$ for any $x,y$
depends analytically on $\alpha$. Hence  for any $f,g\in C_\mathrm{c}(]0,\pi[)$
\beq \alpha\mapsto \int \bar{f(x)}R_\alpha^\s(-\lambda^2;x,y)g(y)\d x\d y\eeq is
analytic. Using Thm \ref{crit} we obtain that
$\{\Re\,\alpha>0\}\ni\alpha\mapsto R_\alpha^\s(-\lambda^2)\in B\big( L^2]0,\pi[\big)$
is an analytic family of bounded operators. Hence, by Thm \ref{crit1},
$\{\Re\,\alpha>0\}\ni\alpha\mapsto L_\alpha^\s\in C\big( L^2]0,\pi[\big)$
is an analytic family of closed operators.

For $\Re\,\alpha\geq 1$, $r^{\frac12-\alpha}$ 
is not $L^2$-integrable, and only $r^{\frac12+\alpha}$ is.
Hence for
such $\alpha$ the right hand side of
\eqref{hami1} possesses a unique closed realization, which has to
coincide with the operator $L_\alpha^\s$ that we have just constructed.
In particular, $L_\alpha^\s$
does not depend on
$\lambda$. By the uniqueness of analytic continuation,
$L_\alpha^\s$ does not depend on the choice of $\lambda$ for all $\Re\,\alpha>-1$.

The  singularities of the Gamma function yield the discrete
spectrum of $L_\alpha^\s$.}
\qed


    
\subsection{Hyperbolic case}
 In \eqref{geg}, for $r\in\rr_+$ we set 
 \beq w=\cosh r,\ \ \hbox{ which solves }
\ w'=(w^2-1)^{\frac12}.\label{a2-g}\eeq
This leads to the Schr\"odinger equation
\beq \left(L_{\alpha}^\h+\lambda^2\right)\phi(r)=0,
\label{hyper-g}\eeq
where
\begin{eqnarray}\label{hami2}
  L_{\alpha}^\h:
  =-\partial_r^2+
\left(\alpha^2-\frac{1}{4}\right)\frac{1}{\sinh^2 r}.\end{eqnarray}
It is obtained when we separate variables of the Laplacian on 
the hyperbolic space of any dimensions, see e.g. Subsect. \ref{Hyperbolic space}.
Hence our name ``hyperbolic''.



Let us define functions on $\rr_+$
\begin{align}
    \mathcal{P}^\h_{\alpha, \lambda}(r) &:= \Big(\frac{\sinh r}{2}\Big)^{\alpha+\frac12} ~ \mathbf{S}_{\alpha,\lambda}(\cosh r),
    \\
    \mathcal{Q}^\h_{\alpha, \lambda}(r) &:= \frac{(\sinh r)^{\alpha+\frac12}}{2^\lambda} ~ \mathbf{Z}_{\alpha,\lambda}(\cosh r).
\end{align}
They have the following asymptotic behavior:
\begin{align}
    \mathcal{P}^\h_{\alpha, \lambda}(r) \sim
  \frac{1}{\Gamma(1+\alpha)}\Big(\frac{r}{2}\Big)^{\frac12+\alpha},\quad
  r\sim0;\\
  \cQ^\h_{\alpha, \lambda}(r) \sim
  \frac{1}{\Gamma(1+\lambda)}\e^{-\lambda r},\quad r\to+\infty.
  \end{align}

The following four functions solve the eigenequation \eqref{hyper-g}:
\beq
\mathcal{P}^\h_{\alpha, \lambda}(r), \quad \mathcal{P}^\h_{-\alpha, \lambda}(r), \quad \mathcal{Q}^\h_{\alpha, \lambda}(r),\quad  \mathcal{Q}^\h_{\alpha, -\lambda}(r).
\eeq
The following symmetries are obvious: 
\beq 
\mathcal{P}^\h_{\alpha, \lambda}(r)=\mathcal{P}^\h_{\alpha, -\lambda}(r), \qquad \mathcal{Q}^\h_{\alpha, \lambda}(r) =\mathcal{Q}^\h_{-\alpha, \lambda}(r). 
\eeq 

The following theorem describes the basic family of  closed realizations of $L^\h_\alpha$ on $L^2[0,\infty[$.
\bet
  For $\Re\,\alpha\geq1$ there exists a unique closed operator
  $L_\alpha^\h$ in the sense of
   $L^2(\rr_+)$, which on $C_\mathrm{c}^\infty(\rr_+)$
   is  given by \eqref{hami2}. The
family $\alpha\mapsto L_\alpha^\h$ is holomorphic and possesses a
unique holomorphic extension to $\Re\,\alpha>-1$.
Here is its discrete spectrum and spectrum:
\begin{align}
    \sigma_\d(L_\alpha^\h)&=\Big\{-\Big(\frac12+\alpha\Big)^2\Big\},\quad
                            -1<\Re\,\alpha<-\frac12;\\
        \sigma_\d(L_\alpha^\h)&=\emptyset,\quad
                            -\frac12{\red \leq}\Re\,\alpha;
                            \label{dicr}  \\
  \sigma(L_\alpha^\h)&=[0,+\infty[\cup   \sigma_\d(L_\alpha^\h).
  \end{align}
Outside of the spectrum, for $\Re\,\lambda>0$,  its resolvent is
\begin{align} \notag
    \frac{1}{\left(L^\h_\alpha+\lambda^2 \right)}(x,y) =& 
\frac{                                                   \Gamma \left(\frac12+\alpha +\lambda\right)}{\sqrt\pi}\\
&\times    \begin{cases}
        \mathcal{P}^\h_{\alpha,\lambda}(x)\mathcal{Q}^\h_{\alpha,\lambda}(y) \qquad &\text{if} \quad 0<x<y<\infty;
        \\
        \mathcal{Q}^\h_{\alpha,\lambda}(x)\mathcal{P}^\h_{\alpha,\lambda}(y) \qquad &\text{if} \quad 0<y<x<\infty.
    \end{cases} \label{resgegh}
\end{align}
\eet

\proof
Consider $\mathcal{P}^\h_{\alpha, \lambda}(r)$ and
$\mathcal{P}^\h_{-\alpha, \lambda}(r)$ as a basis of solutions of \eqref{hyper-g}. The connection formula is
\beq
    \mathcal{Q}^\h_{\alpha, \lambda}(r)  =-\frac{\sqrt\pi}{\sin \pi \alpha~\Gamma\left(\frac12 -\alpha+\lambda\right)}\mathcal{P}^\h_{\alpha, \lambda}(r)+\frac{\sqrt\pi}{\sin \pi \alpha~\Gamma\left(\frac12 +\alpha+\lambda\right)}\mathcal{P}^\h_{-\alpha, \lambda}(r).
\eeq
Similarly as in the spherical case,
we obtain
\beq\cW(\cP_{-\alpha,\lambda}^\h, \cP_{\alpha,\lambda}^\h)=\frac{\sin\alpha}{\pi}.\eeq 
This yields the  Wronskians 
\begin{align}   \nonumber 
    \W\left(\mathcal{Q}^\h_{\alpha, \lambda}(r),\mathcal{P}^\h_{\alpha, \lambda}(r)\right) &= \frac{\sqrt\pi}{ \Gamma \left(\frac12+\alpha +\lambda\right)}.
\end{align}
For $\Re\,\alpha>-1$ and $\Re\,\lambda>0$ the functions
$\mathcal{P}^\h_{\alpha,\lambda}(r)$
resp. $\mathcal{Q}^\h_{\alpha,\lambda}(r)$, are square integrable at
the endpoints.  {\red Using the Schur test} we check that under  these conditions the integral
kernel \ref{resgegh}
defines a bounded operator, depending analytically on $\alpha$.

For $\Re\,\alpha\geq1$, $r^{\frac12-\alpha}$ is not square integrable
  near $0$. Therefore, for such $\alpha$ \eqref{hami2} possesses a
  unique closed realization.

Looking for singularities of the Gamma function we find the discrete
spectrum:
\begin{align}
  \sigma_\d(L_\alpha^\h)&=\left\{-\left(n+\frac12+\alpha\right)^2 \bigg\vert n \in \nn_0,\quad \Re\left(n+\frac12+\alpha\right)<0\right\}. 
\label{dicr1}  \end{align}
It is easy to see that this coincides with 
\eqref{dicr}
  \qed

\subsection{DeSitterian case}
 For $r\in\rr$, in \eqref{geg} we set 
 \beq w=-\i\sinh
 r,\ \ \hbox{ which solves }
\ w'=(w^2-1)^{\frac12}.\label{a2b-g}\eeq
This leads to the Schr\"odinger equation
\beq \left(L_{\alpha}^\dS+\lambda^2\right)\phi(r)=0,
\label{hyper5-g}\eeq
where\beq\label{hami3}
L_{\alpha}^\dS:=-\partial_r^2
-\left(\alpha^2-\frac{1}{4}\right)\frac{1}{\cosh^2 r}
.\eeq
It has the mirror symmetry $r\to-r$.
It is obtained when we separate variables in the d'Alembertian on the
deSitter space of any dimension, see Subsect. \ref{DeSitter space}.
Hence our name ``deSitterian''.

The following theorem describes all closed realizations of $L_{\alpha}^\dS $ on $L^2(\rr)$.

For $r\geq0$ we introduce the following function which solves the eigenequation
\beq
\mathcal{Q}^\dS_{\alpha,\lambda}(r):=\e^{-\i\frac\pi2(\frac12+\alpha+\lambda)}\frac{ (\cosh r)^{\alpha+\frac12}}{2^\lambda} ~{\red\mathbf{Z}_{\alpha,\lambda}}(-\i \sinh r).
\eeq
We extend it to $r\leq0$ by analytic continuation.
Here is its asymptotics:
\beq
\cQ_{\alpha,\lambda}^\dS(r)\sim\frac{1}{\Gamma(\lambda+1)}\e^{-\lambda
  r},\quad r\to+\infty.\eeq
  
Thus the following functions solve the eigenequation \eqref{hyper5-g}:
\beq
\mathcal{Q}^\dS_{\alpha,\lambda}(r),\qquad \mathcal{Q}^\dS_{\alpha,\lambda}(-r),\qquad \mathcal{Q}^\dS_{\alpha,-\lambda}(r), \qquad \mathcal{Q}^\dS_{\alpha,-\lambda}(-r).
\eeq

Let us  describe closed realizations of $L_\alpha^\dS$ on $L^2(\rr)$:
\bet For any $\alpha\in\cc$ there exists a unique closed operator
$L_\alpha^\dS$ in the sense of $L^2(\rr)$ that on
$C_\mathrm{c}^\infty(\rr)$ is given by \eqref{hami3}. The function
$\cc\ni\alpha\mapsto
L_\alpha^\dS$ is holomorphic.
It satisfies $L_\alpha^\dS=L_{-\alpha}^\dS$. 
                         Outside of the spectrum, for $\Re\,\lambda>0$, its resolvent is
\begin{align}&\notag
\frac{1}{(L_{\alpha}^{\dS}+\lambda^2)}(x,y)\\=& \frac{\Gamma \left(\frac12-\alpha +\lambda \right) \Gamma \left(\frac12+\alpha +\lambda \right)}{2}
\begin{cases}
    \mathcal{Q}^{\dS}_{\alpha,\lambda}(x) \mathcal{Q}^{\dS}_{\alpha,\lambda}(-y) \qquad -\infty<x<y<\infty;
    \\
    \mathcal{Q}^{\dS}_{\alpha,\lambda}(y) \mathcal{Q}^{\dS}_{\alpha,\lambda}(-x) \qquad -\infty<y<x<\infty. 
\end{cases}\label{resgegds}
\end{align}
To describe the discrete
spectrum and spectrum of $L_\alpha^\dS$, without loss of generality we
can assume that $\Re\,\alpha\geq0$. Then
\begin{align}\label{discr2}
  \sigma_\d(L_{\alpha}^\dS)=
                            &
                              \left\{-\left(\alpha-k\right)^2
                              \quad\bigg\vert \quad k\in \nn_0+\frac12,\quad k< \Re\,\alpha\right\}
                         \\
  \sigma(L_{\alpha}^\dS)=&[0,\infty[\,\cup \,
                          \sigma_\d(L_{\alpha}^\dS).\end{align}
\label{gegdes}
\eet

\proof
We can use the proof of the theorem \ref{desthm}. Note that
\beq
\mathcal{Q}^\dS_{\alpha,\lambda}(r)=
\frac{\Gamma\big(\frac12+\lambda\big)}{\sqrt\pi}
\cQ_{\alpha,\alpha,2\lambda}^\dS(r)
=\frac{\Gamma(1+2\lambda)}{2^{2\lambda}\Gamma(1+\lambda)}
\cQ_{\alpha,\alpha,2\lambda}^\dS(r).
\eeq
Hence, the Wronskians are
\begin{align}\label{wro.1}
\W\left(\mathcal{Q}^\dS_{\alpha,\lambda}(-r),\mathcal{Q}^\dS_{\alpha,\lambda}(r) \right) &= \frac{2}{\Gamma\left(\frac12+\alpha+\lambda\right)\Gamma\left(\frac12-\alpha+\lambda\right)},\\\label{wro.2}
\W\left(\mathcal{Q}^\dS_{\alpha,-\lambda}(-r),\mathcal{Q}^\dS_{\alpha,\lambda}(r) \right) &= \frac{2 \cos(\pi \alpha)}{\pi}.
\end{align}
{\red Using the Schur test }we check that  the integral kernel \eqref{resgegds} defines a bounded
operator.

The singularities of the Gamma function are at
$\lambda=-\frac12+\alpha-n$ and $\lambda=-\frac12-\alpha-n$,
$n\in\nn_0$. This
gives the following discrete
spectrum:
\begin{align} \label{empty}
  \sigma_\d(L_{\alpha}^\dS)=&
                              \left\{-\left(\frac12+\alpha+n\right)^2
                              \quad\bigg\vert \quad n\in \nn_0,\quad \Re\left(\frac12+\alpha+n\right)<0\right\}\\
                            &\cup
                              \left\{-\left(\frac12-\alpha+n\right)^2
                              \quad\bigg\vert \quad n\in \nn_0,\quad
                              \Re\left(\frac12-\alpha+n\right)<0\right\}.\end{align}
For $\Re\alpha\geq0$ the right hand
                          side of \eqref{empty} is empty.
This yields \eqref{discr2}.
\qed

\noindent{\bf Proof of Prop. \ref{trans1}.}
The transformation $  \cot r=\sinh q$ implies
\begin{align}
  \cos r&=\tanh q,\quad \tan r=\sinh q,\\
  \frac{\d q}{\d r}&=-\frac1{\sin r}=\cosh q.\label{tra..}
\end{align}
The Whipple transformation \eqref{whpl1}, \eqref{whpl2} on the interval $w\in]-1,1[$ has two
versions: above this interval and below, that is 
\begin{align} \label{whip0}
    \mathbf{Z}_{\alpha,\lambda}(w\pm\i0) &=
                                     (\pm\i\sqrt{1-w^2})^{-\frac12-\alpha-\lambda}
                                     \mathbf{S}_{\lambda,\alpha}\left(
                                     \frac{w}{\pm\i\sqrt{1-w^2}}
                                     \right).
\end{align}
This yields
\begin{align}\label{tra1}
  \cQ_{\alpha,\lambda}^\dS(q)&=\Big(\frac2{\sin r}\Big)^{\frac12}\cP_{\lambda,\alpha}^\s(r) 
                               ,\\\label{tra2}
    \cQ_{\alpha,\lambda}^\dS(-q)&=\Big(\frac2{\sin
                                  r}\Big)^{\frac12}\cP_{\lambda,\alpha}^\s(\pi-r)
                                  .\end{align}
                    This yields
     \begin{align}                                                                (\sin
      r)^\frac12\frac{1}{\left(L_{\lambda}^\s-\alpha^2\right)}\left(r,r^\prime
       \right)(\sin
       r^\prime)^\frac12&=\frac{1}{\left(L_{\alpha}^\dS-\lambda^2\right)}\left(q,q^\prime\right),\end{align} 
which proves Prop. \ref{trans1}.
                          \qed

\begin{remark}                                Using \eqref{tra..}, \eqref{tra1}
                                and \eqref{tra2} we obtain
                                \begin{align}
                                  \cW\big(
                                  \cQ_{\alpha,\lambda}^\dS(-q),
                                  \cQ_{\alpha,\lambda}^\dS(q)\big)&=2\cW
                                                                     \big(\cP_{\lambda,\alpha}^\s(\pi-r),
                                                                    \cP_{\lambda,\alpha}^\s(r)\big),\\
                                    \cW\big(
                                  \cQ_{\alpha,\lambda}^\dS(-q),
                                  \cQ_{\alpha,-\lambda}^\dS(q)\big)&=2\cW
                                                                     \big(\cP_{\lambda,\alpha}^\s(\pi-r),
                                                                     \cP_{-\lambda,\alpha}^\s(r)\big).
                                                                     \end{align}
                                 Therefore, \eqref{wro..1} and
                                 \eqref{wro..2} imply
                          \eqref{wro.1} and \eqref{wro.2}. This can be
                          used in an alternative proof of Thm \ref{gegdes}. 
\end{remark}

                          \section{Hypergeometric Hamiltonians of the first kind}
                          \label{Hypergeometric Hamiltonians of the first kind}
\init

Let us transform the hypergeometric operator  as follows:
\begin{eqnarray}
&&-z^{\frac{\alpha}{2}+\frac{1}{4}}(1-z)^{\frac{\beta}{2}+\frac{1}{4}}
\cF\left({\textstyle\frac{\alpha+\beta+\mu+1}{2}},  
{\textstyle\frac{\alpha+\beta-\mu+1}{2}}
;1+\alpha;z,\partial_z\right)  
z^{-\frac{\alpha}{2}-\frac{1}{4}}(1-z)^{-\frac{\beta}{2}-\frac{1}{4}}
\nonumber\\[2ex]
&=&-z(1-z)\P\left(\begin{array}{cccc}
0&1&\infty&\\
\frac{\alpha}{2}+\frac{1}{4}&\frac{\beta}{2}+\frac{1}{4}&\frac{\mu}{2}&z,\partial_z\\
-\frac{\alpha}{2}+\frac{1}{4}&-\frac{\beta}{2}+\frac{1}{4}&-\frac{\mu}{2}
                    &\end{array}\right)\nonumber\\
&&-z(1-z)\left(\partial_z^2+\left(\frac{1}{2z}-\frac{1}{2(1-z)}\right) 
\partial_z\right) 
\nonumber\\&&
+\left(\alpha^2-\frac14\right)\frac{1}{4z}
              +\left(\beta^2-\frac14\right)\frac{1}{4(1-z)}-\frac{\mu^2}{4}\label{riem}.  
\end{eqnarray}
Thus if we set
\beq L_{\alpha,\beta}:=
-z(1-z)\left(\partial_z^2+\left(\frac{1}{2z}-\frac{1}{2(1-z)}\right) 
\partial_z\right) 
+\left(\alpha^2-\frac14\right)\frac{1}{4z}
+\left(\beta^2-\frac14\right)\frac{1}{4(1-z)}\label{riem}
\eeq and $F(z)$ solves  the hypergeometric equation \eqref{hypper}, 
then 
\beq 
\Big(L_{\alpha,\beta} -\frac{\mu^2}{4}\Big)
 z^{\frac{\alpha}{2}+\frac{1}{4}}(1-z)^{\frac{\beta}{2}+\frac{1}{4}}
F(z)=0 .\label{riem0}\eeq

{\red The hypergeometric equation has been interpreted as the
eigenequation of the operator $L_{\alpha,\beta}$ with the
eigenvalue
$\frac{\mu^2}{4}$ acting on functions on the complex plane. Let us reinterpret this  operator as acting
on functions on three intervals whose endpoints are singularities of the
hypergeometric equations. We will see again that these choices lead to
simple formulas for Green functions.} In each of these cases we perform the Liouville transformation, which  
yields a 1-dimensional Hamiltonian. We will consider three cases:
\begin{enumerate}
\item  $z\in]0,1[$, which leads to an operator on $L^2]0,\pi[$,  which we
  call the {\em spherical hypergeometric Hamiltonian of the 1st kind};
\item $z\in]-\infty,0[$, which leads to an operator on $L^2(\rr_+)$,
  which we call the {\em  hyperbolic hypergeometric Hamiltonian of the 1st kind}; 
\item $z\in\frac12+\i\rr$, which leads to an operator on $L^2(\rr)$, which
  we call the {\em deSitterian  hypergeometric Hamiltonian of the 1st kind}.
\end{enumerate}

It will be natural to introduce the parameters
\begin{align}
\delta:&=\frac12(\alpha^2+\beta^2),\label{setti}\\
\kappa:=\frac12(\alpha^2-\beta^2),&\qquad
  \tau:=\frac\i2(\alpha^2-\beta^2)=\i\kappa.\label{setti1}\end{align}

\begin{center}
    \begin{tikzpicture}
\draw[help lines,-] (-3,0) -- (3,0) coordinate (xaxis);
\draw[dashed, very thick] (0,-3) -- (0,3) coordinate (desitterian);
\draw[dotted, very thick] (-1,0) -- (-3,0) coordinate (hyperbolic);
\draw[very  thick] (-1,0) -- (1,0) coordinate (spherical);


\node[left] at (0,3) {\red$\frac12 +\i \infty$};
\node[left] at (0,-3) {\red $\frac12-\i \infty$};
\node[below] at (-1,0) {0};
\draw[fill,black] (-1,0) circle (1.5pt);
\node[below] at (1,0) {1};
\draw[fill,black] (1,0) circle (1.5pt);
\node[below] at (3,0) {$\infty$};
\draw[fill,black] (3,0) circle (1.5pt);
\node[right] at (0,1.5) {DeSitterian};
\node[below] at (1,-0.3) { Spherical};
\node[above] at (-2,0) {Hyperbolic};
\node[below] at (-3,0) {$- \infty$};
\end{tikzpicture}
\\
Figure 2: Hypergeometric Hamiltonians of the first kind on the
$z$-plane.\\\red The spherical Hamiltonian  acts on the interval
marked with a thick line,
 the  hyperbolic Hamiltonian---with a dotted line, and deSitterian
 Hamiltonian---with a dashed line.
\end{center}

\subsection{Spherical case}
\label{Spherical case1}

For $r\in]0,\pi[$, set  in \eqref{riem}\beq z=\sin^2\frac{r}{2}=\frac{1-\cos r}{2},\ \ \hbox{ which solves }
\ z'=z^{\frac12}(1-z)^{\frac12}.\label{a2}\eeq
This leads to the Schr\"odinger equation
\beq \left(L_{\alpha,\beta}^\s-\frac{\mu^2}{4}\right)\phi(r)=0,
\label{posch}\eeq
where
\begin{eqnarray}\label{posch1}
L_{\alpha,\beta}^\s&:=&-\partial_r^2+
\left(\alpha^2-\frac{1}{4}\right)\frac{1}{4\sin^2\frac{r}{2}}+
\left(\beta^2-\frac{1}{4}\right)\frac{1}{4\cos^2 \frac{r}{2}}
\\&=&-\partial_r^2+
\left(\delta-\frac{1}{4}\right)\frac{1}{\sin^2 r}+
\kappa\frac{\cos r}{\sin^2 r}
.\nonumber\end{eqnarray}
In \cite{DW} $L_{\alpha,\beta}^\s$ is called the {\em trigonometric P\"oschl-Teller Hamiltonian}

The case $\alpha=\beta$ is especially important and coincides with the
spherical Gegenbauer Hamiltonian:
\beq
L_\alpha^\s:=L_{\alpha,\alpha}^\s.
\eeq

For $r\in]0,\pi[$,
 define the function
\begin{align} \notag
P_{\alpha,\beta,\mu}^\s(r)&:=\Big(\sin\frac{r}{2}\Big)^{\alpha+\12}\Big(\cos\frac{r}{2}\Big)^{\beta+\12}
\mathbf{F}\left(\frac{\alpha+\beta+\mu+1}{2},
\frac{\alpha+\beta-\mu+1}{2};
  1+\alpha;\sin^2\Big( \frac{r}{2}\Big)\right)\\
&=\Big(\sin\frac{r}{2}\Big)^{\alpha+\12}\Big(\cos\frac{r}{2}\Big)^{-\alpha-\mu-\frac12}
\mathbf{F}\left(\frac{\alpha+\beta+\mu+1}{2},
\frac{\alpha-\beta+\mu+1}{2};
  1+\alpha;-\tan^2\Big( \frac{r}{2}\Big)\right)
  .\end{align}
Note that
\beq
P_{\alpha,\beta,\mu}^\s(r)=P_{\alpha,-\beta,\mu}^\s(r)=P_{\alpha,\beta,-\mu}^\s(r).
\eeq
Asymptotically our function behaves like
\beq\label{asym1}
    P^\s_{\alpha, \beta, \mu}(r) \sim \frac{1
    }{\Gamma(1+\alpha)}\Big(\frac{r}{2}\Big)^{\alpha+\frac12},\quad r\sim0.
    \eeq

Now the following functions solve the eigenvalue problem
\eqref{posch}:
\beq
P_{\alpha,\beta,\mu}^\s(r), \quad P_{-\alpha,\beta,\mu}^\s(r),\quad
P_{\beta,\alpha,\mu}^\s(\pi-r), \quad P_{-\beta,\alpha,\mu}^\s(\pi-r).
\eeq

The following theorem describes the basic  holomorphic family of
closed realizations of $L_{\alpha,\beta}^\s$ on  $L^2]0,\pi[$. 

\bet
For $\Re\,\alpha,\Re\,\beta\geq1$ 
there exists a unique closed operator
  $L_{\alpha,\beta}^\s$ in the sense of
   $L^2]0,\pi[$, which on $C_\mathrm{c}^\infty]0,\pi[$
   is  given by 
\eqref{posch1}. The
family $\alpha,\beta\mapsto L_{\alpha,\beta}^\s$ is holomorphic and possesses a
unique holomorphic extension to $\Re\,\alpha,\Re\,\beta>-1$.
It has only discrete spectrum:
\beq\sigma(L_{\alpha,\beta}^\s)=\sigma_\d(L_{\alpha,\beta}^\s)=\left\{\Big(k+\frac{\alpha+\beta}{2}\Big)^2\ :\ k\in\nn_0+\frac12\right\}.\eeq
Outside of the spectrum its resolvent is
\begin{align} \nonumber
\frac{1}{\left(L^\s_{\alpha,\beta}-\frac{\mu^2}{4}\right)}(x,y) =&\Gamma \left(\frac{1+\alpha+\beta+\mu}{2}\right) \Gamma \left(\frac{1+\alpha+\beta-\mu}{2}\right)
\\
&\times \begin{cases}
    P_{\alpha,\beta,\mu}^\s(x)~P_{\beta,\alpha,\mu}^\s(\pi-y) \qquad \text{if}\quad 0<x<y<\pi;
    \\
    P_{\alpha,\beta,\mu}^\s(y)~P_{\beta,\alpha,\mu}^\s(\pi-x) \qquad \text{if}\quad 0<y<x<\pi.
\end{cases}\label{candi}
\end{align}
\eet

\proof
Considering $P^\s_{\alpha, \beta, \mu}(r)$, and $P^\s_{-\alpha,
  \beta, \mu}(r)$ as a basis of solutions of \eqref{posch}, we can
rewrite the connection formula \eqref{con2} as
\beq
    P^\s_{\beta, \alpha, \mu}(\pi- r) = \frac{\pi~P^\s_{\alpha, \beta, \mu}(r)}{\sin (-\pi \alpha )\Gamma\big(\frac{1-\alpha +\beta +\mu}{2}\big) \Gamma\big(\frac{1-\alpha +\beta -\mu}{2}\big)} +\frac{\pi ~P^\s_{-\alpha, \beta, \mu}(r)}{\sin (\pi \alpha)\Gamma\big(\frac{1+\alpha +\beta +\mu}{2}\big) \Gamma\big(\frac{1+\alpha +\beta -\mu}{2}\big)}.
\eeq
Using \eqref{asym1} and arguing as in the proof of \eqref{wrons}, we obtain
\beq    \W\left( P^\s_{-\alpha, \beta,\mu}(r),P^\s_{\alpha, \beta, 
    \mu}(r) \right) =
\frac{\sin\pi\alpha}{\pi}.\label{wro}\eeq    
From the connection formula and \eqref{wro} we obtain the Wronskian
\begin{align}\label{wro-s}
    \W\left( P^\s_{\beta,\alpha, \mu}(\pi-r),P^\s_{\alpha, \beta, \mu}(r) \right) &= \frac{1}{\Gamma\left(\frac{1+\alpha+\beta+\mu}{2}\right) \Gamma\left(\frac{1+\alpha+\beta-\mu}{2}\right)}.
\end{align}
The $L^2$ integrability conditions at the endpoints are
$\Re\,\alpha>-1, \Re\,\beta>-1$, for $P^\s_{\alpha,\beta,\mu}(r)$
and $P^\s_{\beta,\alpha,\mu}(\pi-r)$ respectively. With these
conditions, we can write the candidate for the resolvent
\eqref{candi}. The $L^2$ norm of this integral kernel is finite. Hence it
defines a bounded (even Hilbert-Schmidt) operator.
For $\Re\,\alpha\geq1, \Re\,\beta\geq1$ it is a unique candidate for the resolvent.
\qed

\subsection{Hyperbolic case}\label{Hyperbolic case1}
For $r\in\rr_+$, in \eqref{riem} we set 
 \beq z=-\sinh^2\frac{r}{2}=\frac{1-\cosh r}{2},\ \ \hbox{ which solves }
\ z'=-(-z)^{\frac12}(1-z)^{\frac12}.\label{a22}\eeq
This leads to the Schr\"odinger equation
\beq \left(L_{\alpha,\beta}^\h+\frac{\mu^2}{4}
\right)\phi(r)=0,
\label{hyper}\eeq
where\begin{eqnarray}\label{posch2}
       L_{\alpha,\beta}^\h&:=&-\partial_r^2+
\left(\alpha^2-\frac{1}{4}\right)\frac{1}{4\sinh^2\frac{r}{2}}-
\left(\beta^2-\frac{1}{4}\right)\frac{1}{4\cosh^2 \frac{r}{2}}
\\&=&-\partial_r^2+
\left(\delta-\frac{1}{4}\right)\frac{1}{\sinh^2 r}+
\kappa\frac{\cosh r}{\sinh^2 r}\nonumber
.\end{eqnarray}
In \cite{DW} $L_{\alpha,\beta}^\h$
is called the
{\em hyperbolic P\"oschl-Teller Hamiltonian}.

The case $\alpha=\beta$ is especially important and coincides with the
hyperbolic Gegenbauer operator:
\begin{eqnarray}L_{\alpha}^\h:= L_{\alpha,\alpha}^\h
.\end{eqnarray}

For $r\in\rr_+$,
let us define
\begin{align} \notag
P_{\alpha,\beta,\mu}^\h(r) &:=\Big(\sinh\frac{r}{2}\Big)^{\alpha+\12}\Big(\cosh\frac{r}{2}\Big)^{\beta+\12}
\mathbf{F}\left(\frac{\alpha+\beta+\mu+1}{2}, 
\frac{\alpha+\beta-\mu+1}{2}; 
                             1+\alpha;-\sinh^2 \frac{r}{2}\right)\\
  &=\Big(\sinh\frac{r}{2}\Big)^{\alpha+\12}\Big(\cosh\frac{r}{2}\Big)^{-\alpha-\mu-\frac12}
\mathbf{F}\left(\frac{\alpha+\beta+\mu+1}{2}, 
\frac{\alpha-\beta+\mu+1}{2}; 
                            1+\alpha;\tanh^2 \frac{r}{2}\right),\\ \notag
Q_{\alpha,\beta,\mu}^\h(r) &:=\Big(\sinh\frac{r}{2}\Big)^{-\mu-\beta-\12}\Big(\cosh\frac{r}{2}\Big)^{\beta+\frac12}
\mathbf{F}\left(\frac{\alpha+\beta+\mu+1}{2}, 
\frac{-\alpha+\beta+\mu+1}{2}; 
                             1+\mu;-\sinh^{-2} \frac{r}{2}\right)\\
  &=\Big(\sinh\frac{r}{2}\Big)^{\alpha+\12}\Big(\cosh\frac{r}{2}\Big)^{-\alpha-\mu-\frac12}
\mathbf{F}\left(\frac{\alpha+\beta+\mu+1}{2}, 
\frac{\alpha-\beta+\mu+1}{2}; 
                            1+\mu;\cosh^{-2} \frac{r}{2}\right)
                            .\end{align}
Note that
\begin{align}
  P_{\alpha,\beta,\mu}^\h(r)&=P_{\alpha,-\beta,\mu}^\h(r)=P_{\alpha,\beta,-\mu}^\h(r),\\
  Q_{\alpha,\beta,\mu}^\h(r)&=Q_{\alpha,-\beta,\mu}^\h(r)=Q_{-\alpha,\beta,\mu}^\h(r),
\end{align} 
Asymptotically, \begin{align}
\label{asym3}    P^\h_{\alpha, \beta, \mu}(r) &\sim
                                                \frac{1}{\Gamma(1+\alpha)}\Big(\frac{r}{2}\Big)^{\alpha+\frac12},
                                                \quad r\to0;\\
 Q^\h_{\alpha, \beta, \mu}(r) &\sim\frac{2^\mu
                 }{\Gamma(1+\mu)} \e^{-\frac{\mu}{2}r},\quad r\to+\infty.
                  \end{align}

Now the following functions solve the eigenvalue problem
\eqref{hyper5}:
\beq
P_{\alpha,\beta,\mu}^\h(r), \quad P_{-\alpha,\beta,\mu}^\h(r),\quad
Q_{\alpha,\beta,\mu}^\h (r), \quad Q_{\alpha,\beta,-\mu}^\h(r).\eeq

The following theorem describes the basic  holomorphic family of
closed realizations of  $L_{\alpha,\beta}^\h$ on the Hilbert space $L^2(\rr_+)$.

\bet For $\Re\,\alpha\geq1$ 
there exists a unique closed operator
  $L_{\alpha,\beta}^\h$ in the sense of
   $L^2(\rr_+)$, which on $C_\mathrm{c}^\infty(\rr_+)$
   is  given by 
\eqref{posch2}. The
family $\alpha,\beta\mapsto L_{\alpha,\beta}^\h$ is holomorphic and possesses a
unique holomorphic extension to $\Re\,\alpha>-1$.
Its discrete spectrum and spectrum are
\begin{align} \notag
  \sigma_\d(L_{\alpha,\beta}^\h) =&\left\{-\Big(\frac{\alpha+\beta}{2}+k\Big)^2\quad 
                                  | \quad k\in\nn_0+\frac12,\quad
                                 k<- \Re\, \frac{\alpha+\beta}{2}\right\},\\ 
  \cup&\left\{-\Big(\frac{\alpha-\beta}{2}+k\Big)^2\quad 
                                  | \quad k\in\nn_0+\frac12,\quad k<-\Re\,\frac{\alpha-\beta}{2}\right\},\\ 
  \sigma(L_{\alpha,\beta}^\h)=&[0,\infty[\cup
                                  \sigma_\d(L_{\alpha,\beta}^\h).
                                  \end{align}
Outside of the spectrum, for $\Re\,\mu>0$,  its resolvent is
    \begin{align} \nonumber
    \frac{1}{(L^\h_{\alpha,\beta}+\frac{\mu^2}{4})}(x,y) =& \Gamma\left(\frac{1+\alpha+\beta+\mu}{2}\right) \Gamma\left(\frac{1+\alpha-\beta+\mu}{2}\right)
    \\
    &\times \begin{cases}
        P^\h_{\alpha,\beta,\mu}(x) ~Q^\h_{\alpha,\beta,\mu}(y) \qquad \text{if} \quad 0<x<y<\infty;
        \\
        P^\h_{\alpha,\beta,\mu}(y) ~Q^\h_{\alpha,\beta,\mu}(x) \qquad \text{if} \quad 0<y<x<\infty.
    \end{cases}\label{resoh}
    \end{align}
\eet

\proof
Considering $P^\h_{-\alpha, \beta, \mu}(r)$, and $P^\h_{-\alpha,
  \beta, \mu}(r)$ as a basis of solutions of \eqref{hyper5}, we can
rewrite  connection formula \eqref{con-1} as
\beq
    Q^\h_{\alpha, \beta, \mu}(r) = -\frac{\pi~P^\h_{\alpha, \beta, \mu}(r)}{\sin \pi \alpha~ \Gamma\left(\frac{1-\alpha -\beta +\mu}{2}\right) \Gamma\left(\frac{1-\alpha +\beta +\mu}{2}\right) } +\frac{\pi~P^h_{-\alpha, \beta, \mu}(r)}{\sin (\alpha \pi) \Gamma\left(\frac{1+\alpha +\beta +\mu}{2}\right) \Gamma\left(\frac{1+\alpha -\beta +\mu}{2}\right) }.
\eeq
Using \eqref{asym3} 
and again arguing as in the proof of \eqref{wrons}, we obtain
we obtain
\beq    \W\left( P^\h_{-\alpha, \beta,\mu}(r),P^\h_{\alpha, \beta, 
    \mu}(r) \right) =
\frac{\sin\pi\alpha}{\pi}.\label{wroh}\eeq    
This yields the Wronskian
\begin{align}
    \W\left( Q^\h_{\alpha,\beta, \mu}(r),P^\h_{\alpha, \beta, \mu}(r) \right)&= \frac{1}{\Gamma\left(\frac{1+\alpha+\beta+\mu}{2}\right) \Gamma\left(\frac{1+\alpha-\beta+\mu}{2}\right)}.
\end{align} 
    The $L^2$ integrability conditions at the endpoints are $\Re\,\alpha>-1$ and $
    \Re\,\mu>0$ for $P^\h_{\alpha,\beta,\mu}(r)$ and
    $\cQ^\h_{\alpha,\beta,\mu}(r)$, respectively. Using the Schur
   Test we check  that integral kernel \eqref{resoh} defines a
    bounded operator. We also see that for $\Re\,\alpha>-1, 
    \Re\,\mu>0$ it is a unique candidate for the resolvent. \qed

\noindent{\bf Proof of Prop. \ref{trans2}.}
The transformation
\beq \tan\frac{r}{2}=\sinh\frac{q}{2}.\eeq
implies
\begin{align}
  \sin\frac{r}{2}=\tanh\frac{q}{2},&\quad 
\frac{\d q}{\d r}=\frac{1}{\cos \frac{r}2}=\cosh\frac{q}{2};\\
  P_{\alpha,\beta,\mu}^\s(r)&=\frac1{\big(\cosh\frac{q}{2}\big)^{\frac12} }
                              P_{\alpha,\mu,\beta}^\h(q),\\
    P_{\beta,\alpha,\mu}^\s(\pi-r)&=\frac1{\big(\cosh\frac{q}{2}\big)^{\frac12} }
  Q_{\alpha,\mu,\beta}^\h(q).
\end{align}
We obtain the transmutation relation
\begin{align}         \frac{1}{\left(L^\s_{\alpha,\beta}+\frac{\mu^2}{4}\right)}\left(r,r^\prime\right)
&= \Big(\cosh\frac{q}{2}\Big)^{-\frac12}~\frac{1}{\left(L^\h_{\alpha,\mu}+\frac{\beta^2}{4}\right)}\left(q,q^\prime\right)  ~\Big(\cosh\frac{q^\prime}{2}\Big)^{-\frac12}.
  \end{align}

\qed

    \subsection{DeSitterian case}
\label{DeSitterian case1}
 For $r\in\rr$,  in \eqref{riem} we set
 \beq z=\frac12-\i\cosh\frac{ r}{2}\sinh \frac{r}{2}=\frac{1-\i\sinh
 r}{2},\ \ \hbox{ which solves }
\ z'=(-z)^{\frac12}(1-z)^{\frac12}.\label{a2b}\eeq
This leads to the Schr\"odinger equation
\beq \left(L_{\alpha,\beta}^\dS+\frac{\mu^2}{4}\right)\phi(r)=0,
\label{hyper5}\eeq
where\begin{eqnarray}\label{scarf}L_{\alpha,\beta}^\dS
&:=&-\partial_r^2-
\left(\alpha^2-\frac{1}{4}\right)\frac{1}{\cosh^2 r}\left(\frac12+\frac{\i\sinh
     r}{2}\right)-
\left(\beta^2-\frac{1}{4}\right)\frac{1}{\cosh^2 r}\left(\frac12-\frac{\i\sinh
     r}{2}\right)
\\       &=&-\partial_r^2
-\left(\delta-\frac{1}{4}\right)\frac{1}{\cosh^2 r}-
\tau\frac{\sinh r}{\cosh^2 r}.\label{a2c}
\end{eqnarray}
This Hamiltonian was proposed and solved by F. Scarf \cite{scarf}
and in \cite{DW} it is called the {\em Scarf Hamiltonian}.

 The case $\alpha=\beta$ is especially important and coincides with
 the deSitterian Gegenbauer Hamiltonian:
 \beq
 L_{\alpha}^\dS:= L_{\alpha,\alpha}^\dS.\eeq
 

Define for $r\geq0$
\begin{align}\notag
                Q_{\alpha,\beta,\mu}^{\dS}(r) 
              :=&
\Big(\frac{\i{+} \sinh r}{2}\Big)^{-\frac\beta2-\frac\mu2-\frac14}\Big(\frac{-\i+ \sinh r}{2}\Big)^{\frac\beta2+\frac14}
\\&\times\mathbf{F}\left(\frac{\alpha+\beta+\mu+1}{2}, 
\frac{-\alpha+\beta+\mu+1}{2}; 
                               1+\mu;\frac2{1-\i \sinh r}\right)\label{re1}\\
  \notag=&
\Big(\frac{\i{+} \sinh r}{2}\Big)^{\frac\alpha2+\frac14}\Big(\frac{-\i+ \sinh r}{2}\Big)^{-\frac\alpha2-\frac\mu2-\frac14}\\
&\times\mathbf{F}\left(\frac{\alpha+\beta+\mu+1}{2}, 
\frac{\alpha-\beta+\mu+1}{2}; 
    1+\mu;\frac2{1+\i  \sinh r}\right).\label{re2}\end{align}
    
We extend $r\to Q_{\alpha,\beta,\mu}^\dS(r)$ to $r<0$ by analyticity. It
satisfies
\beq Q_{\alpha,\beta,\mu}^\dS(r)
\sim \frac{2^\mu\e^{-\frac{\mu}{2}r}}{\Gamma(1+\mu)},\quad r\to+\infty.\eeq

Note that
\begin{align}
  Q_{\alpha,\beta,\mu}^\dS(r)&=Q_{\alpha,-\beta,\mu}^\dS(r)=Q_{-\alpha,\beta,\mu}^\dS(r).
\end{align} 
Now the following functions solve the eigenvalue problem
\eqref{hyper5}:
\beq
Q_{\alpha,\beta,\mu}^\dS (r), \quad Q_{\alpha,\beta,-\mu}^\dS(r),\quad
Q_{\beta,\alpha,\mu}^\dS (-r), \quad Q_{\beta,\alpha,-\mu}^\dS(-r).\eeq

The following theorem describes all closed realizations of $L_{\alpha,\beta}^\dS $ on $L^2(\rr)$.

\bet \label{desthm} For any $\alpha,\beta\in\cc$ there exists a unique closed operator
$L_\alpha^\dS$ in the sense of $L^2(\rr)$ that on
$C_\mathrm{c}^\infty(\rr)$ is given by \eqref{scarf}. The function
$\cc\ni(\alpha,\beta)\mapsto
L_\alpha^\dS$ is holomorphic.
We have $L_{\alpha,\beta}^\dS=L_{-\alpha,\beta}^\dS=L_{\alpha,-\beta}^\dS$.

Outside of the spectrum, for $\Re\,\mu>0$, its resolvent is
\begin{align} \notag
\frac{1}{(L^{\dS}_{\alpha ,\beta}+\frac{\mu^2}{4})}(x,y) =&  \frac{\Gamma \left(\frac{1-\alpha -\beta +\mu}{2}\right) \Gamma \left(\frac{1+\alpha -\beta +\mu }{2}\right) \Gamma \left(\frac{1-\alpha +\beta +\mu }{2}\right) \Gamma \left(\frac{1+\alpha +\beta +\mu }{2}\right)}{2 \pi }
\\
  & \times { \begin{cases}
         Q^{\dS}_{\alpha,\beta, \mu}(x)Q^{\dS}_{\beta,\alpha,\mu}(-y)\quad \text{if} \quad -\infty<x<y<\infty;
     \\
          Q^{\dS}_{\beta,\alpha, \mu}(-y)Q^{\dS}_{\alpha,\beta,\mu}(x)\quad \text{if} \quad -\infty<y<x<\infty.
    \end{cases}} \label{resdes}
    \end{align}
    To describe the discrete spectrum of $L_{\alpha,\beta}^\dS$   assume without loss of generality that
$\Re(\alpha+\beta)\geq0$. We also assume that $\Re(\alpha-\beta)\geq0$
(the case $\Re(\alpha-\beta)\leq0$ is analogous). Then
\begin{align}\notag
  \sigma_\d(L_{\alpha.\beta}^\dS)=&\left\{-\left(\frac{\alpha+\beta}{2}-k\right)^2\bigg\vert\quad 
                              k\in \nn_0+\frac12,\quad k<\Re\,\frac{\alpha+\beta}{2}\right\}\\ \label{specto}
 \cup &\left\{-\left(\frac{\alpha-\beta}{2}-k\right)^2\bigg\vert\quad 
                              k\in \nn_0+\frac12,\quad
                              k< \Re\frac{\alpha-\beta}{2}\right\},
    \\
  \sigma(L_{\alpha,\beta}^\dS)=&[0,\infty[\cup \sigma_\d(L_{\alpha,\beta}^\dS).\end{align}

    \eet

    \proof 
In the connection formula \eqref{popo} we
  insert $z=\frac{1-\i s}{2}$  and multiply it by
$\Big(\frac{1-\i s}{2}\Big)^{\frac\alpha2+\frac14}\Big(\frac{1+\i s}{2}\Big)^{\frac\beta2+\frac14}$,
obtaining
    \begin{align}
      \Big(\frac{1-\i s}{2}\Big)^{\frac\alpha2+\frac14}\Big(\frac{1+\i s}{2}\Big)^{\frac\beta2+\frac14}\mathbf{F}_{\alpha,\beta,\mu}\Big(\frac{1-\i s}{2}\Big)=& \frac{\pi \Big(\frac{1-\i s}{2}\Big)^{\frac\alpha2+\frac14}\Big(\frac{1+\i s}{2}\Big)^{\frac\beta2+\frac14}\Big(\frac{\i s-1}{2}\Big)^\frac{-1-\alpha-\beta-\mu}{2} \mathbf{F}_{\mu,\beta,\alpha}(\frac{2}{1-\i s})}{\sin (-\pi \mu) \Gamma \left(\frac{1+\alpha+\beta-\mu}{2}\right)\Gamma \left(\frac{1+\alpha-\beta-\mu}{2}\right)}\notag\\+&\frac{\pi \Big(\frac{1-\i s}{2}\Big)^{\frac\alpha2+\frac14}\Big(\frac{1+\i s}{2}\Big)^{\frac\beta2+\frac14}\Big(\frac{\i s-1}{2}\Big)^\frac{-1-\alpha-\beta+\mu}{2} \mathbf{F}_{-\mu,\beta,\alpha}(\frac{2}{1-\i s})}{\sin (\pi \mu) \Gamma \left(\frac{1+\alpha+\beta+\mu}{2}\right)\Gamma \left(\frac{1+\alpha-\beta+\mu}{2}\right)}.\label{popo1}
\end{align}
We transform separately \eqref{popo1} above and below the real line:
obtaining resp.
    \begin{align}
& \frac{\pi\e^{\i\frac\pi2(-\alpha-\frac12-\frac\mu2)}\Big(\frac{s+\i}{2}\Big)^{-\frac\beta2-\frac14-\frac\mu2}\Big(\frac{s-\i}{2}\Big)^{\frac\beta2+\frac14}\mathbf{F}_{\mu,\beta,\alpha}(\frac{2}{1-\i  s})}{\sin (-\pi \mu) \Gamma\left(\frac{1+\alpha+\beta-\mu}{2}\right)\Gamma\left(\frac{1+\alpha-\beta-\mu}{2}\right)}\notag\\+&\frac{\pi\e^{\i\frac\pi2(-\alpha-\frac12+\frac\mu2)}\Big(\frac{s+\i}{2}\Big)^{-\frac\beta2-\frac14+\frac\mu2}\Big(\frac{s-\i}{2}\Big)^{\frac\beta2+\frac14}\mathbf{F}_{-\mu,\beta,\alpha}(\frac{2}{1-\i s})}{\sin(\pi\mu)
 \Gamma\left(\frac{1+\alpha+\beta+\mu}{2}\right)\Gamma\left(\frac{1+\alpha-\beta+\mu}{2}\right)},\quad s>0\label{popo1a}\\
  & \frac{\pi\e^{\i\frac\pi2(\alpha+\frac12+\frac\mu2)}\Big(\frac{-s-\i}{2}\Big)^{-\frac\beta2-\frac14-\frac\mu2}\Big(\frac{-s+\i}{2}\Big)^{\frac\beta2+\frac14}\mathbf{F}_{\mu,\beta,\alpha}(\frac{2}{1-\i s})}{\sin (-\pi \mu) \Gamma\left(\frac{1+\alpha+\beta-\mu}{2}\right)\Gamma\left(\frac{1+\alpha-\beta-\mu}{2}\right)}\notag\\+&\frac{\pi\e^{\i\frac\pi2(\alpha+\frac12-\frac\mu2)}\Big(\frac{-s-\i}{2}\Big)^{-\frac\beta2-\frac14+\frac\mu2}\Big(\frac{-s+\i}{2}\Big)^{\frac\beta2+\frac14}
\mathbf{F}_{-\mu,\beta,\alpha}(\frac{2}{1-\i  s})}{\sin(\pi\mu)\Gamma
\left(\frac{1+\alpha+\beta+\mu}{2}\right)\Gamma\left(\frac{1+\alpha-\beta+\mu}{2}\right)}.\quad s<0\label{popo1b}    
\end{align}
Inserting $s=\sinh(- r)=-\sinh r$ into \eqref{popo1b}, and using uniqueness of analytic continuation we get
\begin{align} \notag
    p^{\dS}_{\alpha,\beta,\mu}(r):=& \left(\frac{1- \i \sinh r}{2}\right)^{\frac\alpha2+\frac14} \left(\frac{1+ \i \sinh r}{2}\right)^{\frac\beta2+\frac14}\mathbf{F}_{\alpha,\beta,\mu}\left(\frac{1-\i \sinh r}{2}\right)
    \\ \notag
    =& \frac{\pi \e^{\frac{\i \pi}{2}\left(\alpha+\frac\mu2+\frac12  \right)}}{\sin(-\pi \mu) \Gamma\left(\frac{1+\alpha+\beta-\mu}{2}\right)\Gamma\left(\frac{1+\alpha-\beta-\mu}{2}\right)} Q^\dS_{\beta,\alpha,\mu}(-r)
    \\
    &+\frac{\pi \e^{\frac{\i \pi}{2}\left(\alpha-\frac\mu2+\frac12 \right)}}{\sin(\pi \mu) \Gamma\left(\frac{1+\alpha+\beta+\mu}{2}\right)\Gamma\left(\frac{1+\alpha-\beta+\mu}{2}\right)} Q^{\dS}_{\beta,\alpha,-\mu}(-r).
\end{align}
 By replacing $\alpha$ to $-\alpha$ we obtain the second identity
 \begin{align} \notag
    p^{\dS}_{-\alpha,\beta,\mu}(r)
    =& \frac{\pi \e^{\frac{\i \pi}{2}\left(-\alpha+\frac\mu2+\frac12  \right)}}{\sin(-\pi \mu) \Gamma\left(\frac{1-\alpha+\beta-\mu}{2}\right)\Gamma\left(\frac{1-\alpha-\beta-\mu}{2}\right)} Q^\dS_{\beta,\alpha,\mu}(-r)
    \\
    &+\frac{\pi \e^{\frac{\i \pi}{2}\left(-\alpha-\frac\mu2+\frac12 \right)}}{\sin(\pi \mu) \Gamma\left(\frac{1-\alpha+\beta+\mu}{2}\right)\Gamma\left(\frac{1-\alpha-\beta+\mu}{2}\right)} Q^{\dS}_{\beta,\alpha,-\mu}(-r).
\end{align}
We can rewrite them via
\begin{align} \nonumber
    \begin{bmatrix}
        p^\dS_{\alpha,\beta,\mu}(r)\\p^\dS_{-\alpha,\beta,\mu}(r)
    \end{bmatrix}
    =&
  \frac{\pi}{\sin(\pi \mu)} 
  \begin{bmatrix}
  -\frac{ \e^{\frac{\i \pi}{2}\left(\alpha+\frac\mu2+\frac12  \right)}}{ \Gamma\left(\frac{1+\alpha+\beta-\mu}{2}\right)\Gamma\left(\frac{1+\alpha-\beta-\mu}{2}\right)} &\frac{ \e^{\frac{\i \pi}{2}\left(\alpha-\frac\mu2+\frac12 \right)}}{ \Gamma\left(\frac{1+\alpha+\beta+\mu}{2}\right)\Gamma\left(\frac{1+\alpha-\beta+\mu}{2}\right)} 
  \\
        -\frac{ \e^{\frac{\i \pi}{2}\left(-\alpha+\frac\mu2+\frac12  \right)}}{\Gamma\left(\frac{1-\alpha+\beta-\mu}{2}\right)\Gamma\left(\frac{1-\alpha-\beta-\mu}{2}\right)} &\frac{\e^{\frac{\i \pi}{2}\left(-\alpha-\frac\mu2+\frac12 \right)}}{ \Gamma\left(\frac{1-\alpha+\beta+\mu}{2}\right)\Gamma\left(\frac{1-\alpha-\beta+\mu}{2}\right)} 
    \end{bmatrix}
    \\  &\times
    \begin{bmatrix}
        Q^\dS_{\beta,\alpha,\mu}(-r)\\
        Q^{\dS}_{\beta,\alpha,-\mu}(-r)
    \end{bmatrix}.\label{1stcon}
\end{align}
We  evaluate 
\beq
\det{\begin{bmatrix}
  -\frac{ \e^{\frac{\i \pi}{2}\left(\alpha+\frac\mu2+\frac12  \right)}}{ \Gamma\left(\frac{1+\alpha+\beta-\mu}{2}\right)\Gamma\left(\frac{1+\alpha-\beta-\mu}{2}\right)} &\frac{ \e^{\frac{\i \pi}{2}\left(\alpha-\frac\mu2+\frac12 \right)}}{ \Gamma\left(\frac{1+\alpha+\beta+\mu}{2}\right)\Gamma\left(\frac{1+\alpha-\beta+\mu}{2}\right)} 
  \\
        -\frac{ \e^{\frac{\i \pi}{2}\left(-\alpha+\frac\mu2+\frac12  \right)}}{\Gamma\left(\frac{1-\alpha+\beta-\mu}{2}\right)\Gamma\left(\frac{1-\alpha-\beta-\mu}{2}\right)} &\frac{\e^{\frac{\i \pi}{2}\left(-\alpha-\frac\mu2+\frac12 \right)}}{ \Gamma\left(\frac{1-\alpha+\beta+\mu}{2}\right)\Gamma\left(\frac{1-\alpha-\beta+\mu}{2}\right)} 
    \end{bmatrix}}= -\i \frac{\sin(\pi \mu)\sin(\pi \alpha)}{\pi^2}
\eeq
Therefore we have
\begin{align}\nonumber
   \begin{bmatrix}
        Q^\dS_{\beta,\alpha,\mu}(-r)\\
        Q^{\dS}_{\beta,\alpha,-\mu}(-r)
    \end{bmatrix}
    =&
  \frac{-\i  \pi}{\sin(\pi \alpha)} 
  \begin{bmatrix}
  \frac{\e^{\frac{\i \pi}{2}\left(-\alpha-\frac\mu2+\frac12 \right)}}{ \Gamma\left(\frac{1-\alpha+\beta+\mu}{2}\right)\Gamma\left(\frac{1-\alpha-\beta+\mu}{2}\right)}&-\frac{ \e^{\frac{\i \pi}{2}\left(\alpha-\frac\mu2+\frac12 \right)}}{ \Gamma\left(\frac{1+\alpha+\beta+\mu}{2}\right)\Gamma\left(\frac{1+\alpha-\beta+\mu}{2}\right)} 
  \\
        \frac{ \e^{\frac{\i \pi}{2}\left(-\alpha+\frac\mu2+\frac12  \right)}}{\Gamma\left(\frac{1-\alpha+\beta-\mu}{2}\right)\Gamma\left(\frac{1-\alpha-\beta-\mu}{2}\right)} & -\frac{ \e^{\frac{\i \pi}{2}\left(\alpha+\frac\mu2+\frac12  \right)}}{ \Gamma\left(\frac{1+\alpha+\beta-\mu}{2}\right)\Gamma\left(\frac{1+\alpha-\beta-\mu}{2}\right)} .
    \end{bmatrix}
    \\  &\times
    \begin{bmatrix}
        p^\dS_{\alpha,\beta,\mu}(r)\\p^\dS_{-\alpha,\beta,\mu}(r)
    \end{bmatrix} 
\end{align}
In \eqref{1stcon}, we exchange $\alpha$ and $\beta$, and replace $r$ with $-r$, and we obtain 
\begin{align} \nonumber
    \begin{bmatrix}
        p^\dS_{\beta,\alpha,\mu}(-r)\\p^\dS_{-\beta,\alpha,\mu}(-r)
    \end{bmatrix}
    =&
  \frac{\pi}{\sin(\pi \mu)} 
  \begin{bmatrix}
  -\frac{ \e^{\frac{\i \pi}{2}\left(\beta+\frac\mu2+\frac12  \right)}}{ \Gamma\left(\frac{1+\alpha+\beta-\mu}{2}\right)\Gamma\left(\frac{1-\alpha+\beta-\mu}{2}\right)} &\frac{ \e^{\frac{\i \pi}{2}\left(\beta-\frac\mu2+\frac12 \right)}}{ \Gamma\left(\frac{1+\alpha+\beta+\mu}{2}\right)\Gamma\left(\frac{1-\alpha+\beta+\mu}{2}\right)} 
  \\
        -\frac{ \e^{\frac{\i \pi}{2}\left(-\beta+\frac\mu2+\frac12  \right)}}{\Gamma\left(\frac{1+\alpha-\beta-\mu}{2}\right)\Gamma\left(\frac{1-\alpha-\beta-\mu}{2}\right)} &\frac{\e^{\frac{\i \pi}{2}\left(-\beta-\frac\mu2+\frac12 \right)}}{ \Gamma\left(\frac{1+\alpha-\beta+\mu}{2}\right)\Gamma\left(\frac{1-\alpha-\beta+\mu}{2}\right)} 
    \end{bmatrix}
    \\  &\times
    \begin{bmatrix}
        Q^\dS_{\alpha,\beta,\mu}(r)\\
        Q^{\dS}_{\alpha,\beta,-\mu}(r)
    \end{bmatrix}.
\end{align}
Using \eqref{reqdes}, we know that
\begin{align} \notag
    \begin{bmatrix}
        p^\dS_{\alpha,\beta,\mu}(r)\\p^\dS_{-\alpha,\beta,\mu}(r)
    \end{bmatrix} =  &\frac{\pi}{\sin(\pi \beta)}
    \begin{bmatrix}
        -\frac{1}{\Gamma\left(\frac{1+\alpha-\beta-\mu}{2}\right)\Gamma\left(\frac{1+\alpha-\beta+\mu}{2}\right)}&\frac{1}{\Gamma\left(\frac{1+\alpha+\beta-\mu}{2}\right)\Gamma\left(\frac{1+\alpha+\beta+\mu}{2}\right)}
        \\
        -\frac{1}{\Gamma\left(\frac{1-\alpha-\beta-\mu}{2}\right)\Gamma\left(\frac{1-\alpha-\beta+\mu}{2}\right)}&\frac{1}{\Gamma\left(\frac{1-\alpha+\beta-\mu}{2}\right)\Gamma\left(\frac{1-\alpha+\beta+\mu}{2}\right)}
    \end{bmatrix}
    \\
    &\times \begin{bmatrix}
        p^\dS_{\beta,\alpha,\mu}(-r)\\p^\dS_{-\beta,\alpha,\mu}(-r)
    \end{bmatrix}.
\end{align}
Thus we obtained the connection formula
\begin{align}  \nonumber
    \begin{bmatrix}
        Q^\dS_{\beta,\alpha,\mu}(-r)\\
        Q^{\dS}_{\beta,\alpha,-\mu}(-r).
    \end{bmatrix} = 
    &\frac{-\i \pi^3}{\sin(\pi \alpha)\sin(\pi \beta)\sin(\pi \mu)}
    \begin{bmatrix}
  \frac{\e^{\frac{\i \pi}{2}\left(-\alpha-\frac\mu2+\frac12 \right)}}{ \Gamma\left(\frac{1-\alpha+\beta+\mu}{2}\right)\Gamma\left(\frac{1-\alpha-\beta+\mu}{2}\right)}&-\frac{ \e^{\frac{\i \pi}{2}\left(\alpha-\frac\mu2+\frac12 \right)}}{ \Gamma\left(\frac{1+\alpha+\beta+\mu}{2}\right)\Gamma\left(\frac{1+\alpha-\beta+\mu}{2}\right)} 
  \\ 
        \frac{ \e^{\frac{\i \pi}{2}\left(-\alpha+\frac\mu2+\frac12  \right)}}{\Gamma\left(\frac{1-\alpha+\beta-\mu}{2}\right)\Gamma\left(\frac{1-\alpha-\beta-\mu}{2}\right)} & -\frac{ \e^{\frac{\i \pi}{2}\left(\alpha+\frac\mu2+\frac12  \right)}}{ \Gamma\left(\frac{1+\alpha+\beta-\mu}{2}\right)\Gamma\left(\frac{1+\alpha-\beta-\mu}{2}\right)} 
    \end{bmatrix}
    \\ \notag
    &\times \begin{bmatrix}
        -\frac{1}{\Gamma\left(\frac{1+\alpha-\beta-\mu}{2}\right)\Gamma\left(\frac{1+\alpha-\beta+\mu}{2}\right)}&\frac{1}{\Gamma\left(\frac{1+\alpha+\beta-\mu}{2}\right)\Gamma\left(\frac{1+\alpha+\beta+\mu}{2}\right)}
        \\
        -\frac{1}{\Gamma\left(\frac{1-\alpha-\beta-\mu}{2}\right)\Gamma\left(\frac{1-\alpha-\beta+\mu}{2}\right)}&\frac{1}{\Gamma\left(\frac{1-\alpha+\beta-\mu}{2}\right)\Gamma\left(\frac{1-\alpha+\beta+\mu}{2}\right)}
    \end{bmatrix}
    \\
    &\times  \begin{bmatrix}
  -\frac{ \e^{\frac{\i \pi}{2}\left(\beta+\frac\mu2+\frac12  \right)}}{ \Gamma\left(\frac{1+\alpha+\beta-\mu}{2}\right)\Gamma\left(\frac{1-\alpha+\beta-\mu}{2}\right)} &\frac{ \e^{\frac{\i \pi}{2}\left(\beta-\frac\mu2+\frac12 \right)}}{ \Gamma\left(\frac{1+\alpha+\beta+\mu}{2}\right)\Gamma\left(\frac{1-\alpha+\beta+\mu}{2}\right)} 
  \\
        -\frac{ \e^{\frac{\i \pi}{2}\left(-\beta+\frac\mu2+\frac12  \right)}}{\Gamma\left(\frac{1+\alpha-\beta-\mu}{2}\right)\Gamma\left(\frac{1-\alpha-\beta-\mu}{2}\right)} &\frac{\e^{\frac{\i \pi}{2}\left(-\beta-\frac\mu2+\frac12 \right)}}{ \Gamma\left(\frac{1+\alpha-\beta+\mu}{2}\right)\Gamma\left(\frac{1-\alpha-\beta+\mu}{2}\right)} 
    \end{bmatrix} \begin{bmatrix}
        Q^\dS_{\alpha,\beta,\mu}(r)\\
        Q^{\dS}_{\alpha,\beta,-\mu}(r)
    \end{bmatrix}.
\end{align}
The result of multiplication after simplification with the reflection formula for the gamma functions is
\begin{align}\notag
    &\begin{bmatrix}
    -\frac{\e^{-\frac{\i \pi \mu}{2}}\cos(\pi \alpha)+\e^{\frac{\i \pi \mu}{2}}\cos(\pi\beta) }{\pi}&\frac{2 \pi}{\Gamma\left(\frac{1+\alpha+\beta+\mu}{2}\right)\Gamma\left(\frac{1+\alpha-\beta+\mu}{2}\right)\Gamma\left(\frac{1-\alpha+\beta+\mu}{2}\right)\Gamma\left(\frac{1-\alpha-\beta+\mu}{2}\right)}
    \\
    \frac{-2 \pi}{\Gamma\left(\frac{1+\alpha+\beta+\mu}{2}\right)\Gamma\left(\frac{1+\alpha-\beta+\mu}{2}\right)\Gamma\left(\frac{1-\alpha+\beta+\mu}{2}\right)\Gamma\left(\frac{1-\alpha-\beta+\mu}{2}\right)}& \frac{\e^{-\frac{\i \pi \mu}{2}}\cos(\pi \beta)+\e^{\frac{\i \pi \mu}{2}}\cos(\pi\alpha) }{\pi}
    \end{bmatrix}\\
    &\times\begin{bmatrix}
        Q^\dS_{\alpha,\beta,\mu}(r)\\
        Q^{\dS}_{\alpha,\beta,-\mu}(r)
    \end{bmatrix}   \frac{-\pi}{\sin(\pi \mu)}=
    \begin{bmatrix}
        Q^\dS_{\beta,\alpha,\mu}(-r)\\
        Q^{\dS}_{\beta,\alpha,-\mu}(-r)
    \end{bmatrix}.
\end{align}
Using
\beq
\W\big(Q^\dS_{\alpha,\beta,\mu}(r), Q^\dS_{\alpha,\beta,-\mu}(r)\big)=\frac{\sin\pi\mu}{\pi}
\eeq
we obtain
\begin{align}\label{wron1}
  \W\big(Q_{\beta,\alpha,\mu}^\dS(-r),Q_{\alpha,\beta,\mu}^\dS(r)\big)&=\frac{2 \pi }{\Gamma \left(\frac{1-\alpha -\beta +\mu}{2}\right) \Gamma \left(\frac{1+\alpha -\beta +\mu }{2}\right) \Gamma \left(\frac{1-\alpha +\beta +\mu }{2}\right) \Gamma \left(\frac{1+\alpha +\beta +\mu }{2}\right)}
  \\
\W\big(Q_{\beta,\alpha,-\mu}^\dS(-r),Q_{\alpha,\beta,\mu}^\dS(r)\big)&=\frac{\e^{\i\pi\frac{\mu}{2}}\cos\pi\beta+\e^{-\i\pi\frac{\mu}{2}}\cos\pi\alpha}{\pi}.\label{wron2}
\end{align}
The $L^2$ integrable condition at both endpoints is $\Re(\mu)>0$ for
both $Q_{\alpha,\beta,\mu}^\dS(r)$, and
$Q_{\beta,\alpha,\mu}^\dS(-r)$.
  Therefore, 
for $\Re\,\mu>0$, it is a unique candidate for the resolvent.
Using the Schur Test we see that the integral kernel
  \eqref{resdes} defines a bounded operator.

From the singularities of the Gamma function we obtain
\begin{align}\label{emp1}
  \sigma_\d(L_{\alpha}^\dS)=&\left\{-\left(n+\frac{1+\alpha+\beta}{2}\right)^2\bigg\vert\quad 
                              n\in \nn_0,\quad \Re 
                              \left(n+\frac{1+\alpha+\beta}{2}\right)<0 
                              \right\}\\ \label{emp2}
\cup  &\left\{-\left(n+\frac{1+\alpha-\beta}{2}\right)^2\bigg\vert\quad 
                              n\in \nn_0,\quad \Re 
                              \left(n+\frac{1+\alpha-\beta}{2}\right)<0 
    \right\}\\\label{emp3}
 \cup &\left\{-\left(n+\frac{1-\alpha+\beta}{2}\right)^2\bigg\vert\quad 
                              n\in \nn_0,\quad \Re 
                              \left(n+\frac{1-\alpha+\beta}{2}\right)<0 
                              \right\}\\ \label{emp4}
\cup  &\left\{-\left(n+\frac{1-\alpha-\beta}{2}\right)^2\bigg\vert\quad 
                              n\in \nn_0,\quad \Re 
                              \left(n+\frac{1-\alpha-\beta}{2}\right)<0 
        \right\}.\end{align}
     If
      $\Re(\alpha+\beta)\geq0$
      and $\Re(\alpha-\beta)\geq0$, then \eqref{emp1} and
      \eqref{emp2} are empty and we  obtain \eqref{specto}.
\qed


\section{Hypergeometric Hamiltonians of the second kind}
\label{Hypergeometric Hamiltonians of the second kind}
\init

We transform the hypergeometric equation in a different way:
\begin{eqnarray}\nonumber
&&-4z^{1+\frac{\alpha}{2}}(1-z)^{1+\frac{\beta}{2}}
\F({\textstyle\frac{\alpha+\beta+\mu+1}{2}},
{\textstyle\frac{\alpha+\beta-\mu+1}{2}}
;1+\alpha;z,\partial_z)
z^{-\frac{\alpha}{2}}(1-z)^{-\frac{\beta}{2}}
\nonumber\\[3ex]
&=&
-4z^2(z-1)^2\P\left(\begin{array}{cccc}
0&1&\infty&\\
\frac{\alpha}{2}&\frac{\beta}{2}&\frac\mu2+\12&z,\partial_z\\
-\frac{\alpha}{2}&-\frac{\beta}{2}&-\frac\mu2+\12
&\end{array}\right)\nonumber
\\[3ex]&=&
-4z^2(1-z)^2\left(\partial_z^2+\Big(\frac{1}{z}-\frac{1}{1-z}\Big)\partial_z\right) 
+\alpha^2(1-z)+ \beta^2z-(\mu^2-1)z(1-z).\label{riem3}
\end{eqnarray}

We rearrange the terms in \eqref{riem3} containing $\alpha$ and
$\beta$ as follows:
\begin{align} \alpha^2(1-z)+
  \beta^2z&=\delta+\kappa(1-2z) \label{riem3a}\\
  &=\delta+\tau\i(2z-1), \label{riem3b}\end{align}
where $\delta,\kappa,\tau$ are defined in 
\eqref{setti1}.
Thus if we set
\beq K_{\kappa,\mu}:=
-4z^2(1-z)^2\left(\partial_z^2+\Big(\frac{1}{z}-\frac{1}{1-z}\Big)\partial_z\right) 
+\kappa(1-2z)-(\mu^2-1)z(1-z)\eeq
and if $F(z)$ solves the hypergeometric equation 
\eqref{hypper}, 
then 
\beq
\Big(K_{\kappa,\mu}+\delta\Big)  
z^{\frac{\alpha}{2}}(1-z)^{\frac{\beta}{2}}
F(z)=0.\eeq

{\red We have interpreted \eqref{riem3} as the
eigenequation of the operator $K_{\kappa,\mu}$ with 
eigenvalue
$-\delta$.
Again, we will reinterpret this  operator as acting
on functions on an interval whose endpoints are singularities of the
hypergeometric equations.} In each of these cases we perform the Liouville transformation, which  
yields a 1-dimensional Hamiltonian. We will consider three cases:
\begin{enumerate}
\item $z\in\frac12+\i\rr$, which leads to an operator on $L^2]0,\pi[$,
  which we call the {\em spherical hypergeometric Hamiltonian of the 2nd kind};
\item $z\in]-\infty,0[$, which leads to an operator on $L^2]0,\infty[$
  which we call the {\em  hyperbolic hypergeometric Hamiltonian of the 2nd kind}; 
\item  $z\in]0,1[$, which leads to an operator on $L^2(\rr)$, which we call
 {\em the deSitterian  hypergeometric Hamiltonian of the 2nd kind}.
\end{enumerate}

\begin{center}
\begin{tikzpicture}
\draw[help lines,-] (-3,0) -- (3,0) coordinate (xaxis);
\draw[very thick](0,-3) -- (0,3) coordinate (Spherical);
\draw[dotted, very thick]  (-1,0) -- (-3,0) coordinate (hyperbolic);
\draw[dashed,very thick ]  (-1,0) -- (1,0) coordinate (Desitterian);

\node[left] at (0,3) {\red $\frac12+\i \infty$};
\node[left] at (0,-3) {$\red \frac12-\i \infty$};
\node[below] at (-1,0) {0};
\draw[fill,black] (-1,0) circle (1.5pt);
\node[below] at (1,0) {1};
\draw[fill,black] (1,0) circle (1.5pt);
\node[below] at (3,0) {$\infty$};
\draw[fill,black] (3,0) circle (1.5pt);
\node[right] at (0,1.5) {Spherical};
\node[below] at (1,-0.3) {Desitterian};
\node[above] at (-2,0) {Hyperbolic};
\node[below] at (-3,0) {$- \infty$};
\end{tikzpicture}
\\
\textbf{Figure 3:} Hypergeometric Hamiltonian
of the second kind on the $z$-plane.\\\red
The spherical Hamiltonian acts on the interval marked with
a thick line, the hyperbolic---with a dotted line, and the DeSitterian---with a dashed line.
\end{center}
\subsection{Spherical case}

For $u\in ]0,\pi[$,
in \eqref{riem3} and \eqref{riem3b} we set\beq z=\frac{1}{1-\e^{2\i u}},\ \ \hbox{ which solves }
\ z'=2\i z(1-z).\label{w2a}\eeq
This leads to the Schr\"odinger equation
\beq \left(K_{\tau,\mu}^\s-\delta\right)\phi(u)=0,\label{k-s}
\eeq
where
\beq
K_{\tau,\mu}^\s(u):=
-\partial_u^2+\tau\frac{\cos u}{\sin  u}+
\left(\frac{\mu^2}{4}-\frac14\right) 
\frac{1}{\sin^2
u}.\label{second1}\eeq
This Hamiltonian is known as the {\em Rosen-Morse Hamiltonian} (see
\cite{DW}).

In the case $\tau=0$ we have the coincidence: 
\beq L_\alpha^\s=K_{0,2\alpha}^\s.\eeq

We define for $u\in]0,\frac\pi2]$
\begin{align} \notag
&\qq_{\alpha,\beta,\mu}^\s(u)\\ \notag :=&
\left(\frac{\i}{1-\e^{2\i  u}}\right)^{\frac{-1-\beta-\mu}{2}}\left(\frac{-\i}{1-\e^{-2\i  u}}\right)^{\frac\beta2}
\mathbf{F}\left(\frac{\alpha+\beta+\mu+1}{2}, 
\frac{-\alpha+\beta+\mu+1}{2};\mu+1; 
1-\e^{2\i u}\right)\\ 
=&\left(\frac{\i}{1-\e^{2\i  u}}\right)^{\frac\alpha2}\left(\frac{-\i}{1-\e^{-2\i  u}}\right)^{\frac{-1-\alpha-\mu}{2}}\mathbf{F}\left(\frac{\alpha+\beta+\mu+1}{2}, 
\frac{\alpha-\beta+\mu+1}{2};\mu+1; 
1-\e^{-2\i u}\right).
\end{align}
For $u\in[\frac\pi2,\pi[$ it is continued analytically.
It has the asymptotics
\beq
\qq_{\alpha,\beta,\mu}^\s(u)\sim\frac{1}{\Gamma(\mu+1)}(2u)^{\frac\mu2+\frac12},\quad u\sim0.
  \eeq
Note that
\begin{align}
  \qq_{\alpha,\beta,\mu}^\s(u)&=\qq_{-\alpha,\beta,\mu}^\s(u)=\qq_{\alpha,-\beta,\mu}^\s(u),
  \\
  \qq^\s_{\beta,\alpha,\mu}(\pi-u)&=\qq^\s_{\beta,-\alpha,\mu}(\pi-u)=\qq^\s_{-\beta,\alpha,\mu}(\pi-u),
  \\
\Big(\frac2{\cosh r}\Big)^{\frac12}
  Q_{\alpha,\beta,\mu}^\dS(r)&=\qq_{\alpha,\beta,\mu}^\s(u),
    \\
\Big(\frac2{\cosh r}\Big)^{\frac12} Q_{\beta,\alpha,\mu}^\dS(-r)&=\qq_{\beta,\alpha,\mu}^\s(\pi-u),\quad \sinh r=-\cot u. 
\end{align} 
Now the following functions solve the eigenvalue problem \eqref{k-s}:
\beq
\qq_{\alpha,\beta,\mu}^\s (u), \quad \qq_{\alpha,\beta,-\mu}^\s(u),\quad
\qq_{\beta,\alpha,\mu}^\s (\pi-u), \quad \qq_{\beta,\alpha,-\mu}^\s(\pi-u).
\eeq

The following theorem describes the basic closed realization of   $K_{\tau,\mu}^\s$:
\bet
For $\Re\,\mu\geq2$, $\tau\in\cc$, 
there exists a unique closed operator
  $K_{\tau,\mu}^\s$ in the sense of
   $L^2]0,\pi[$, which on $C_\mathrm{c}^\infty]0,\pi[$
   is  given by 
\eqref{second1}. The
family $\tau,\mu\mapsto K_{\tau,\mu}^\s$ is holomorphic and possesses a
unique holomorphic extension to $\Re\,\mu>-2$, except
for a singularity at $(\tau,\mu)=(0,-1)$.
It has only discrete spectrum:
 \begin{align}\label{only}\sigma(K_{\tau,\mu}^\s)=&\bigg\{-\frac{\tau^2}{(2k+\mu)^2}+\Big(k+\frac\mu2\Big)^2\quad \big\vert\quad k \in \nn_0+\frac12\bigg\}.
 \end{align}
 Set
 \begin{align}
   \alpha:=\sqrt{\delta-\i\tau},\quad
   \beta:=\sqrt{\delta+\i\tau}.\end{align}
 (It does not matter which sign of the square root is taken).
 Outside of the spectrum the resolvent of $K_{\tau,\mu}^\s$ is
\begin{align} \notag
\frac{1}{\left(K^\s_{\tau, \mu}-\delta\right)}(x,y)=&  \frac{ \Gamma \left(\frac{1-\alpha -\beta +\mu }{2} \right) \Gamma \left(\frac{1+\alpha -\beta +\mu }{2} \right) \Gamma \left(\frac{1-\alpha +\beta +\mu }{2} \right) \Gamma \left(\frac{1+\alpha +\beta +\mu }{2} \right)}{4\pi }
\\
  & \times { \begin{cases}
         \qq^{\s}_{\alpha,\beta, \mu}(x)\qq^{\s}_{\beta,\alpha,\mu}(\pi-y),\quad \text{if} \quad 0<x<y<\pi;
     \\
          \qq^{\s}_{\alpha,\beta, \mu}(y)\qq^{\s}_{\beta,\alpha,\mu}(\pi-x),\quad \text{if} \quad 0<y<x<\pi.
    \end{cases}} \label{2ress}
    \end{align}
\eet

\proof
The relation $ \sinh r=-\cot u$ implies
\begin{align}
  \frac{2}{1-\i\sinh r}=1-\e^{2\i
  u},&\quad \frac{2}{1+\i\sinh r}=1-\e^{-2\i u},\\
  \frac{\d u}{\d r}= \sin& u=\frac{1}{\cosh r}.
\end{align}
Therefore, using \eqref{wron1} and \eqref{wron2}, we obtain
\begin{align} \notag
    \W\left( \qq^\s_{\beta,\alpha, \mu}(\pi-u),\qq^\s_{\alpha, \beta,
  \mu}(u) \right) &
                    =2\W\left( Q^\dS_{\beta,\alpha, \mu}(-r),Q^\dS_{\alpha, \beta, \mu}(r) \right) \\
&=\frac{4 \pi }{\Gamma \left(\frac{1-\alpha -\beta +\mu}{2}\right) \Gamma \left(\frac{1+\alpha -\beta +\mu }{2}\right) \Gamma \left(\frac{1-\alpha +\beta +\mu }{2}\right) \Gamma \left(\frac{1+\alpha +\beta +\mu }{2}\right)},
  \\ \notag
  \W\big(\qq^\s_{\beta,\alpha,-\mu}(\pi-u),\qq^\s_{\alpha,\beta,\mu}(u)\big)&
                    =2\W\left( Q^\dS_{\beta,\alpha, -\mu}(-r),Q^\dS_{\alpha, \beta, \mu}(r) \right) \\
  &=\frac{2\left(\e^{\i\pi\frac{\mu}{2}}
                                                                \cos\pi\beta+\e^{-\i\pi\frac{\mu}{2}}\cos\pi\alpha\right)}{\pi}.
\end{align}
The $L^2$ integrability condition for $\qq^\s_{\alpha,\beta,\mu}(u)$
at $0$ and $\qq^s_{\beta,\alpha,\mu}(\pi-u)$ at $\pi$ is
$\Re\,\mu>-2$. The $L^2$ norm of this kernel \ref{2ress} is finite. For
$\Re(\mu)\geq 2$ it is a unique candidate for the resolvent.

As a byproduct we obtain a proof of Prop. \ref{trans4} about the 
transmutation identity $L^\dS\to K^\s$.

The singularities of the Gamma functions in \eqref{2ress} are at
\beq 1+\epsilon_1\alpha+\epsilon_2\beta+\mu=-2n,\quad n\in\nn_0,\eeq
where $\epsilon_1,\epsilon_2\in\{1,-1\}$. 
This implies
\beq
\alpha^2=(2n+1+\mu)^2+2\epsilon_2\beta(2n+1+\mu)+\beta^2.\eeq
Hence,
\begin{align}
  \beta&=\frac{\epsilon_2\kappa}{2n+1+\mu}-\epsilon_2\Big(n+\frac\mu2+\frac12\Big),\\
  \alpha&=-\frac{\epsilon_2\kappa}{2n+1+\mu}-\epsilon_2\Big(n+\frac\mu2+\frac12\Big).
\end{align}
This shows
\beq
\delta=\frac{\kappa^2}{(2n+1+\mu)^2}+\Big(n+\frac12+\frac\mu2\Big)^2.\eeq
Replacing $\kappa^2$ with $-\tau^2$ we obtain
  \eqref{only}. \qed

For $\mu\in\zz$ we have an additional identity for the $\qq^\s$
function, which follows directly from \eqref{degen}:
\begin{align}\label{degen1}
  \Big(\frac{\alpha+\beta-\mu+1}{2}\Big)_\mu 
  \Big(\frac{\alpha-\beta-\mu+1}{2}\Big)_\mu 
  &\qq_{\alpha,\beta,\mu}^\s(u)=\qq_{\alpha,\beta,-\mu}^\s(u). 
     \end{align}
     Using this with $\mu=1$ we obtain
     the following unexpected identity:

     \bet For any $\tau\neq0$, we have $K_{\tau,1}^\s=K_{\tau,-1}^\s$.
     \label{wqw}\eet

     \proof
     Setting $\mu=1$ in \eqref{degen1}  we obtain 
     \begin{align}\label{degen1}
\frac{\alpha^2-\beta^2}{4}
  &\qq_{\alpha,\beta,1}^\s(u)=\qq_{\alpha,\beta,-1}^\s(u). 
     \end{align}
     Setting $\mu=1$ and $\mu=-1$ in the prefactor of
     the right hand side of \eqref{2ress} we obtain
     \begin{align}
 \frac{ \Gamma \left(1-\frac{\alpha +\beta}{2} \right) \Gamma 
       \left(1+\frac{\alpha -\beta }{2} \right) \Gamma 
       \left(1-\frac{\alpha -\beta}{2} \right) \Gamma 
       \left(1+\frac{\alpha +\beta}{2} \right)}{4\pi }
      & =\frac{(\alpha^2-\beta^2)\pi}{16\sin\frac\pi2(\alpha+\beta)\sin\frac\pi2(\alpha-\beta) },\\
 \frac{ \Gamma \left(-\frac{\alpha +\beta}{2} \right) \Gamma 
       \left(\frac{\alpha -\beta }{2} \right) \Gamma 
       \left(-\frac{\alpha -\beta}{2} \right) \Gamma 
       \left(\frac{\alpha +\beta}{2} \right)}{4\pi }
       =&\frac{\pi}{(\alpha^2-\beta^2)\sin\frac\pi2(\alpha+\beta)\sin\frac\pi2(\alpha-\beta) }.
       \end{align}
       We thus obtain
       \beq
\frac1{     (
  K_{\tau,1}^\s+\delta)}=\frac1{(K_{\tau,-1}^\s+\delta)}.\eeq
\qed

Note that \eqref{2ress} is ill defined for
$(\tau,\mu)=(0,-1)$.  Moreover, we know that
$\{2<\Re\,\mu\}\ni \mu\mapsto L_{\frac\mu2}^\s$ is analytic, and for $\mu\neq-1$.

We have $K_{0,\mu}^\s=L_{\frac\mu2}^\s$. Therefore, it
is natural to set
\beq K_{0,-1}^\s:=L_{-\frac12}^\s.\eeq
We know that $L_{-\frac12}^\s\neq L_{\frac12}^\s$. Therefore,
Thm \ref{wqw} implies that the point 
$(\tau,\mu)=(0,-1)$ is a singularity of the function
$(\tau,\mu)\mapsto
K_{\tau,\mu}^\s$. See
\cite{DeRi}, where  a similar phenomenon  is described for the Whittaker operator.

\subsection{Hyperbolic case}

For $u\in\rr_+$, in \eqref{riem3} and \eqref{riem3a} we set
\beq z=\frac{1}{1-\e^{2u}},\ \ \hbox{ which solves }
\ z'=2z(z-1).\label{w2}\eeq
This leads to the Schr\"odinger equation
\beq \left(K_{\kappa,\mu}^\h+\delta\right)\phi(u)=0,\label{k-h}
\eeq
where
\beq\label{second2}
K_{\kappa,\mu}^\h:=-\partial_u^2+\kappa\frac{\cosh u}{\sinh
  u}+\left(\frac{\mu^2}{4}-\frac14\right) 
\frac{1}{\sinh^2
u}.\eeq
This Hamiltonian
in \cite{DW}  is called the {\em Eckart Hamiltonian}.

In the case $\kappa=0$ we have the coincidence
\beq L_\alpha^\h=K_{0,2\alpha}^\h.\eeq

We define
\begin{align}  \notag
\pp_{\alpha,\beta,\mu}^\h(u)&:=
\left(\e^{2u}-1\right)^{-\frac{\alpha}{2}}
\left(1-\e^{-2u}\right)^{-\frac\beta2} ~
\mathbf{F}\left(\frac{\alpha+\beta+\mu+1}{2}, 
\frac{\alpha+\beta-\mu+1}{2};\alpha+1; \frac{1}{1-\e^{2u}}\right)
\\
&= \left(\e^{2u}-1\right)^{-\frac{\alpha}{2}}
(1-\e^{-2u})^{\frac{1+\alpha+\mu}{2}}
~
\mathbf{F}\left(\frac{\alpha+\beta+\mu+1}{2}, 
\frac{\alpha-\beta+\mu+1}{2};\alpha+1;\e^{-2u}\right),
\\ \notag
\qq_{\alpha,\beta,\mu}^\h(u) &:=\left(\e^{2u}-1\right)^{\frac{1+\beta+\mu}2}\left(1-\e^{-2u}\right)^{-\frac\beta2}
\mathbf{F}\left(\frac{\alpha+\beta+\mu+1}{2}, 
\frac{-\alpha+\beta+\mu+1}{2};\mu+1; 
                               1-\e^{2u}\right)\\
  &=\left(\e^{2u}-1\right)^{-\frac{\alpha}2}\left(1-\e^{-2u}\right)^{\frac{1+\alpha+\mu}2}
\mathbf{F}\left(\frac{\alpha+\beta+\mu+1}{2}, 
\frac{\alpha-\beta+\mu+1}{2};\mu+1; 
1-\e^{-2u}\right).
\end{align}
Note that 
\begin{align}
    \pp^\h_{\alpha, \beta \mu}(u) \sim 
  \frac{1}{\Gamma(1+\alpha)}\e^{-\alpha u},\quad u\sim+\infty;\\
      \qq^\h_{\alpha, \beta \mu}(u) 
  \sim\frac1{\Gamma(1+\mu)}(2u)^{\frac\mu2+\frac12},\quad u\sim0. 
\end{align}
We have
\begin{align}
  \pp_{\alpha,\beta,\mu}^\h(u)&=\pp_{\alpha,-\beta,\mu}^\h(u)=\pp_{\alpha,\beta,-\mu}^\h(u),\\
  \qq_{\alpha,\beta,\mu}^\h(u)&=\qq_{-\alpha,\beta,\mu}^\h(u)=\qq_{\alpha,-\beta,\mu}^\h(u); \\
  \Big(\frac2{\sinh r}\Big)^{\frac12}\,P_{\alpha,\beta,\mu}^\h(r)&=\pp_{\alpha,\beta,\mu}^\h(u),\\
  \Big(\frac2{\sinh r}\Big)^{\frac12}\,Q_{\alpha,\beta,\mu}^\h(r)&=\qq_{\alpha,\beta,\mu}^\h(u),\qquad 
                                \cosh r=\coth u.
\end{align} 
Now the following functions solve the eigenvalue problem
\eqref{k-h}:
\beq
\pp_{\alpha,\beta,\mu}^\h (u), \quad \pp_{-\alpha,\beta,\mu}^\h(u),\quad
\qq_{\alpha,\beta,\mu}^\h (u), \quad \qq_{\alpha,\beta,-\mu}^\h(u).
\eeq

Let us describe the basic closed realization of $K_{\kappa,\mu}^\h$ on $L^2(\rr_+)$.
\bet For $\kappa\in\cc,\Re\,\mu\geq2$ 
there exists a unique closed operator
  $K_{\kappa,\mu}^\h$ in the sense of
   $L^2(\rr_+)$, which on $C_\mathrm{c}^\infty(\rr_+)$
   is  given by 
\eqref{second2}. The
family $\kappa,\mu\mapsto K_{\kappa,\mu}^\h$ is holomorphic and possesses a
unique holomorphic extension to $\Re\,\mu>-2$, $(\kappa,\mu)\neq(0,1)$.
Its discrete spectrum and spectrum are
\begin{align} 
  \sigma_\d( K_{\kappa,\mu}^\h)=&\bigg\{-\frac{\kappa ^2}{(2k+\mu)^2}-\Big(k+\frac\mu2\Big)^2\quad\bigg\vert\quad
                                  k \in \nn_0+\frac12 
  ,\quad\Re\left(\frac\kappa{2k+\mu}+k+\frac\mu2\right)<0\bigg\},
  \\
  \sigma( K_{\kappa,\mu}^\h)=&[0,+\infty[\cup
                                  \sigma_\d( K_{\kappa,\mu}^\h).
                                  \end{align}
                                  Set
     \begin{align}\alpha:=\sqrt{\delta+\kappa},\quad\Re\,\alpha>0,\quad\beta:=\sqrt{\delta-\kappa}\end{align}
                                  (The choice of the square root for
                                  $\beta$ does not matter).
                                  Outside of its spectrum the
                                  resolvent of $ K_{\kappa,\mu}^\h$ is
\begin{align} \notag
\frac{1}{(K^\h_{\kappa, \mu}+\delta)} (x,y) =&
                                               \frac{\Gamma\left(\frac{1+\alpha+\beta+\mu}{2}\right)\Gamma\left(\frac{1+\alpha-\beta+\mu}{2}\right)} {2}\\
  &\times\begin{cases} \pp^\h_{\alpha, \beta, \mu}(x)~ \qq^\h_{\alpha, \beta,\mu}(y) \qquad \text{if} \quad 0<y<x<\infty,
\\
\pp^\h_{\alpha,\beta,\mu}(y) ~\qq^\h_{\alpha,\beta,\mu}(x) \qquad \text{if} \quad 0<x<y<\infty.\end{cases}\label{2resh}
\end{align} 
\eet

\proof
The relation $                           \cosh r=\coth u$ implies
\begin{align}
  \sinh^2\frac{r}{2}&=\frac1{\e^{2u}-1},\quad
                      \cosh^2\frac{r}{2}=\frac1{\e^{-2u}-1};\\\frac{\d
  u}{\d r}&=\frac1{\sinh r}=\sinh u,\end{align}
Therefore,
\begin{align} \notag
   \W\left(\qq^\h_{\alpha, \beta, \mu},\pp^\h_{\alpha, \beta , 
  \mu}\right)&=
2   \W\left(Q^\h_{\alpha, \beta, \mu},P^\h_{\alpha, \beta , 
  \mu}\right)\\
  &=\frac{2}{\Gamma\left(\frac{1+\alpha+\beta+\mu}{2}\right)\Gamma\left(\frac{1+\alpha-\beta+\mu}{2}\right)}.\end{align} 
The $L^2$ integrability conditions at endpoints are $\Re\,\alpha>0$,
and $\Re\,\mu>-2$ for $\pp^\h_{\alpha,\beta,\mu}(r)$ and
$\qq^\h_{\alpha,\beta,\mu}(r)$, respectively. For such parameters the
integral kernel \eqref{2resh} defines a bounded operator. For
$\Re\,\alpha>0$, and $\Re\,\mu\geq2$, it is the unique candidate for the resolvent.\\

As a byproduct we obtain a proof of Prop. \ref{trans5} about the 
transmutation identity $L^\h\to K^\h$.\\

The determination of the discrete spectrum follows by similar
arguments as for Thm \ref{wqw}.
\qed

For $\mu\in\zz$ we have an additional identity for the $\qq^\h$
function, which follows directly from \eqref{degen}:
\begin{align}\label{degen1}
  \Big(\frac{\alpha+\beta-\mu+1}{2}\Big)_\mu 
  \Big(\frac{\alpha-\beta-\mu+1}{2}\Big)_\mu 
  &\qq_{\alpha,\beta,\mu}^\h(u)=\qq_{\alpha,\beta,-\mu}^\h(u). 
     \end{align}
     Using this with $\mu=1$ we obtain

     \bet \label{wqwh}
     For any $\kappa\neq0$, we have $K_{\kappa,1}^\h=K_{\kappa,-1}^\h$.
     \eet

     \proof
     Setting $\mu=1$ in \eqref{degen1}  we obtain 
     \begin{align}\label{degen1}
\frac{\alpha^2-\beta^2}{4}
  &\qq_{\alpha,\beta,1}^\h(u)=\qq_{\alpha,\beta,-1}^\h(u). 
     \end{align}
     Moreover,
         \begin{align}
  \Gamma \left(1-\frac{\alpha +\beta}{2} \right) \Gamma 
           \left(1+\frac{\alpha -\beta }{2} \right)
           & =\frac{(\alpha^2-\beta^2)}{4}
             \Gamma \left(\frac{\alpha +\beta}{2} \right) \Gamma 
       \left(\frac{\alpha -\beta }{2} \right) ,       \end{align}

We thus obtain
       \beq
\frac1{     \left(
  K_{\tau,1}^\h+\delta\right)}=\frac1{\left(K_{\tau,-1}^\h+\delta \right)}.\eeq
\qed

Similarly as in the spherical case,
 \eqref{2resh} is ill defined for
$(\kappa,\mu)=(0,-1)$.  Moreover, we know that
$\{2<\Re\mu\}\ni \mu\mapsto L_{\frac\mu2}^\h$ is analytic, and for $\mu\neq-1$
we have $K_{0,\mu}^\h=L_{\frac\mu2}^\h$. Therefore, it
is natural to set
\beq K_{0,-1}^\h:=L_{-\frac12}^\h.\eeq

We know that $L_{-\frac12}^\h\neq L_{\frac12}^\h$. Therefore,
Thm \ref{wqwh} implies that the point 
$(\kappa,\mu)=(0,-1)$ is a singularity of the function
$(\kappa,\mu)\mapsto
K_{\kappa,\mu}^\h$.

\subsection{DeSitterian case}
For $u\in\rr$,
in \eqref{riem3} and \eqref{riem3a} we set \beq z=\frac{1}{1+\e^{2u}},\ \ \hbox{ which solves }
\ z'=2z(z-1).\label{w1}\eeq
This leads to the Schr\"odinger equation
\beq \left(K_{\kappa,\mu}^{\dS} +\delta\right)\phi(u)=0,
\label{rosen}\eeq
where
\beq
K_{\kappa,\mu}^{\dS}:=-\partial_u^2-\kappa\frac{\sinh u}{\cosh
  u}-\left(\frac{\mu^2}{4}-\frac14\right) 
\frac{1}{\cosh^2 u}.\label{second3}\eeq
In \cite{DW} it is called the
{\em Manning-Rosen Hamiltonian}.

In the case $\kappa=0$ we have the coincidence: 
\beq 
L_\alpha^\dS=K_{0,2\alpha}^\dS.
\eeq
We define
\begin{align}  \notag
\pp_{\alpha,\beta,\mu}^\dS(u) &:=
\left(1+\e^{2u}\right)^{-\frac{\alpha}{2}}
\left(1+\e^{-2u}\right)^{-\frac\beta2}
\mathbf{F}\left(\frac{\alpha+\beta+\mu+1}{2}, 
\frac{\alpha+\beta-\mu+1}{2};\alpha+1;\frac{1}{1+\e^{2u}}\right)\\
  &=
\left(1+\e^{2u}\right)^{-\frac{\alpha}{2}}
\left(1+\e^{-2u}\right)^{\frac{1+\alpha+\mu}2}
\mathbf{F}\left(\frac{\alpha+\beta+\mu+1}{2}, 
\frac{\alpha-\beta+\mu+1}{2};\alpha+1;-\e^{-2u}\right).
\end{align}
We have the asymptotics
\beq
\pp_{\alpha,\beta,\mu}^\dS(u)\sim \frac{1}{\Gamma(1+\alpha)}\e^{-\alpha
  u},\quad u\sim+\infty.\eeq
Note that
\begin{align}
  \pp_{\alpha,\beta,\mu}^\dS(u)&=\pp_{\alpha,-\beta,\mu}^\dS(u)=\pp_{\alpha,\beta,-\mu}^\dS(u),\\
\Big(\frac2{\sin r}\Big)^{\frac12}\:  P_{\alpha,\beta,\mu}^\s(r)&=\pp_{\alpha,\beta,\mu}^\dS(u),\quad
                              \cos r=\tanh u.\label{tra-3}
\end{align}

Now the following functions solve the eigenvalue problem
\eqref{rosen}:
\beq
\pp_{\alpha,\beta,\mu}^\dS (u), \quad \pp_{-\alpha,\beta,\mu}^\dS(u),\quad      
\pp_{\beta,\alpha,\mu}^\dS (-u), \quad \pp_{-\beta,\alpha,\mu}^{\dS}(-u).
\eeq

We will find all closed realization of
$K_{\kappa,\mu}^{\dS}$ in the sense of $L^2(\rr)$. 
\bet For any $\kappa,\mu\in\cc$ there exists a unique closed operator
$K_{\kappa,\mu}^\dS$ in the sense of $L^2(\rr)$ that on
$C_\mathrm{c}^\infty(\rr)$ is given by \eqref{second3}. The function
$\cc^2\ni(\kappa,\mu)\mapsto
K_{\kappa,\mu}^\dS$ is holomorphic. The discrete
spectrum and spectrum of $K_{\kappa,\mu}^\dS$ are
\begin{align} \notag
  \sigma_\d( K_{\kappa,\mu}^\dS)=&\left\{-\frac{\kappa ^2}{(2k+\mu )^2}-\Big(k+\frac\mu2\Big)^2\quad\bigg\vert\quad
k \in \nn_0+\frac12 ,\quad k<-\left\vert\Re\,\frac{\kappa}{2k+\mu}\right\vert-\frac\mu2\right\}
  \\
  \cup&\left\{-\frac{\kappa ^2}{(2k-\mu)^2}-\Big(k-\frac\mu2\Big)^2\quad\bigg\vert\quad
                                  k \in \nn_0+\frac12 
  ,\quad k<-\left\vert\Re\,\frac{\kappa}{2k-\mu}\right\vert+\frac\mu2 \right\},\\
        \sigma( K_{\kappa,\mu}^\dS)&=[\kappa,+\infty[\,\cup\, [-\kappa,+\infty[\,\cup\,\sigma_\d\left( K_{\kappa,\mu}^\dS\right).
                                  \end{align}
Here, for $z\in\cc$ we use the notation $[z,+\infty[:=\left\{z+t
\quad|\quad t\in[0,+\infty[\right\}$.

Set
\begin{align}
  \alpha:=\sqrt{\delta+\kappa},\quad\Re\,\alpha>0;\quad
  \beta:=\sqrt{\delta-\kappa},\quad\Re\,\beta>0.\end{align}
Outside of its spectrum, the resolvent of $K_{\kappa,\mu}^\dS$ is
\begin{align} \nonumber
\frac{1}{\left(K_{\kappa,\mu}^{\dS}+\delta\right)}(x,y) =&\frac{\Gamma\left(\frac{1+\alpha+\beta+\mu}{2}\right)\Gamma\left(\frac{1+\alpha+\beta-\mu}{2}\right)} {2}
\\
& \times  \begin{cases}
     \pp^{\dS}_{\beta,\alpha,\mu}(x)~\pp^{\dS}_{\alpha,\beta,\mu}(y),  \qquad \text{if} \quad -\infty<x<y<\infty;
    \\
     \pp^{\dS}_{\beta,\alpha,\mu}(y)~\pp^{\dS}_{\alpha,\beta,\mu}(x),  \qquad \text{if} \quad -\infty<y<x<\infty. \label{2resds}
\end{cases}
\end{align}
We have $K_{\kappa,\mu}^\dS=K_{\kappa,-\mu}^\dS$
\eet

\proof
The relation $                     \cos r=\tanh u$ implies
\begin{align}
  \sin^2\frac{r}2&=\frac1{1+\e^{2u}},\quad
                 \cos^2\frac{r}2=\frac1{1+\e^{-2u}};\\
  \frac{\d u}{\d r}&=\frac{1}{\sin r}=\cosh u.
                    \label{tra-3a} \end{align}
This yields  \eqref{tra-3}. 
Using \eqref{tra-3a}  and \eqref{tra-3}, and then \eqref{wro-s}, we obtain
\begin{align} \notag
\W\left(\pp^{\dS}_{\alpha, \beta , \mu}(u), \pp^{\dS}_{ \beta, \alpha
    , \mu}(-u)\right) &=2\W\left(P^{\s}_{\alpha, \beta , \mu}(r), P^{\s}_{ \beta, \alpha , \mu}(\pi-r)\right) \\&
=\frac{2}{\Gamma\left(\frac{1+\alpha+\beta+\mu}{2}\right)\Gamma\left(\frac{1+\alpha+\beta-\mu}{2}\right)}.
\end{align}
The $L^2$ integrability condition at $+\infty$ of
$\pp_{\alpha,\beta,\mu}(u)$ is $\Re(\alpha)>0$, and at $-\infty$ of
$\pp_{\beta,\alpha,\mu}(-u)$ is $\Re(\beta)>0$. For such parameters,  the
  integral kernel \eqref{2resds} defines a
  bounded operator and is a unique candidate for the resolvent.

As a byproduct we obtain a proof of Prop. \ref{trans4} about the 
transmutation identity $L^\s\to L^\dS$.

The determination of the discrete spectrum is similar as in the
  hyperbolic case.
\qed

\noindent{\bf Proof of Prop. \ref{trans3}.}
The change of variables
\beq 1+\e^{2w}=\e^{2u}\eeq
implies
\begin{align}
-\e^{2w}&=1-\e^{2u},\quad -\e^{-2w}=\frac{1}{1-\e^{2u}};\\  \frac{\d w}{\d u}&=\frac1{1-\e^{-2u}}=1+\e^{-2w},\\
   \pp_{\alpha,\beta,\mu}^\dS(w)&=\left(1-\e^{-2u}\right)^{-\frac12} 
                              \pp_{\alpha,\mu,\beta}^\h(u),\\
    \pp_{\beta,\alpha,\mu}^\dS(-w)&=\left(1-\e^{-2u}\right)^{-\frac12} 
  \qq_{\alpha,\mu,\beta}^\h(u).
  \end{align}
We redefine  parameters 
\beq
    \delta':=\frac12(\alpha^2-\mu^2), \qquad  \kappa':=\frac12(\alpha^2+\mu^2).
\eeq
We obtain the transmutation identity 
\beq
\big(1-\e^{-2u}\big)^{-\frac12}  \frac{1}{\left(K^\h_{\kappa,\mu}+\delta \right)}(u,u^\prime) \big(1-\e^{-2u'}\big)^{-\frac12} =\frac{1}{\left(K^\dS_{\kappa',\beta}+\delta'\right)}(w,w^\prime).
\eeq
\qed

\section{The Laplacian on an interval, halfline and line}
\label{The Laplacian on an interval, halfline and line}
\init

The Laplacians on $]0,\pi[$, $\rr_+$ and $\rr$ with the Dirichlet or
Neumann boundary conditions at endpoints belong to the most widely used
operators. Their Green functions can be easily computed in terms of
elementary functions, without using hypergeometric functions.
In this section we will check that  they are 
special cases of hypergeometric Hamiltonians. We will see that this
coincidence is related to various identities for hypergeometric
functions from Appendix \ref{Half integer case}.

\subsection{Laplacian on an interval}

Consider the Laplacian $-\partial_x^2$ on the interval $]0,\pi[$.
The Dirichlet and Neumann boundary condition will be denoted $\D$ and
$\N$ resp.
Putting them at both $0$ and $\pi$ leads to 4 operators on
$L^2]0,\pi[$. They are special cases of the spherical hypergeometric
Hamiltonian of the first kind:
\begin{align}\label{hamo1}
  L_{\D\D}&:=L_{\frac12,\frac12}^\s=L_\frac12^\s,\\
  L_{\N\D}&:=L_{-\frac12,\frac12}^\s,\\
  L_{\D\N}&:=L_{\frac12,-\frac12}^\s,\\
    L_{\N\N}&:=L_{-\frac12,-\frac12}^\s=L_{-\frac12}^\s.\label{hamo4}
  \end{align}

Resolvents of these operators can be computed in terms of elementary
functions. Indeed, the following functions solve
\beq(-\partial_x^2+k^2)\phi(x)=0\eeq and
  satisfy the Dirichlet/Neumann boundary
  conditions at $0$, resp. at $\pi$:
  \begin{align}
    \text{    Dirichlet:}\quad \sinh kx,&\quad \sinh k(\pi-x);\\
    \text{    Neumann:}\quad \cosh kx,&\quad \cosh k(\pi-x).
                                        \end{align}
    They have the
  Wronskians:
\begin{align}
  \cW\big(\sinh k(\pi-x),\sinh kx\big)&=k\sinh\pi k,\\
  \cW\big(\cosh k(\pi-x),\sinh kx\big)&=k\cosh\pi k,\\
  \cW\big(\sinh k(\pi-x),\cosh kx\big)&=k\cosh\pi k,\\
  \cW\big(\cosh k(\pi-x),\cosh kx\big)&=k\sinh\pi k.
\end{align}
By the usual methods,
 we compute the spectra
 of operators \eqref{hamo1}--\eqref{hamo4}, and for $k^2$ outside of the spectra their Green functions :

\begin{align}
 \sigma(L_{\D\D})&=\{n^2\ |\ n\in\nn_0\},\notag\\
                                        \frac{1}{L_{\D\D}+k^2}(x,y)&=
        \frac1{k\sinh \pi k}\begin{cases}\sinh kx\,\sinh k(\pi-y)
,\quad \text{if} \quad 0<x<y<\pi,
    \\\sinh ky\,\sinh k(\pi-x)
,\quad \text{if} \quad 0<y<x<\pi;
\end{cases}\label{dd1}\\
\sigma(L_{\N\D})&=\{(n+\tfrac12)^2\ |\ n\in\nn_0\},\notag\\
\frac1{L_{\N\D}+k^2}(x,y)&=
        \frac1{k\cosh \pi k}\begin{cases}\cosh kx\,\sinh k(\pi-y)
,\quad \text{if} \quad 0<x<y<\pi,
    \\\cosh ky\,\sinh k(\pi-x)
,\quad \text{if} \quad 0<y<x<\pi;
\end{cases}\label{nd1}\\
\sigma(L_{\D\N})&=\{(n+\tfrac12)^2\ |\ n\in\nn_0\},\notag\\
\frac1{L_{\D\N}+k^2}(x,y)&=
        \frac1{k\cosh \pi k}\begin{cases}\sinh kx\,\cosh k(\pi-y)
,\quad \text{if} \quad 0<x<y<\pi,
    \\\sinh ky\,\cosh k(\pi-x)
,\quad \text{if} \quad 0<y<x<\pi;
\end{cases}\label{dn1}\\
\sigma(L_{\N\N})&=\{n^2\ |\ n\in\nn\},\notag\\
\frac1{L_{\N\N}+k^2}(x,y)&=
        \frac1{k\sinh \pi k}\begin{cases}\cosh kx\,\cosh k(\pi-y)
,\quad \text{if} \quad 0<x<y<\pi,
    \\\cosh ky\,\cosh k(\pi-x)
,\quad \text{if} \quad 0<y<x<\pi.
\end{cases}\label{nn1}
\end{align}

Let us check  that \eqref{dd1}--\eqref{nn1}  agree with the more
general formula \eqref{candi} involving the hypergeometric function and the Gamma
function. We identify
$k=\frac{\i \mu}{2}$.  By \eqref{simp1} and  \eqref{simp2} we obtain 
\begin{align}
  P_{\frac12,\pm\frac12,\mu}^\s(x)&=\frac{\sinh kx}{k\sqrt\pi},\\
  P_{-\frac12,\pm\frac12,\mu}^\s(x)&=\frac{\cosh kx}{\sqrt\pi},
\end{align}
Finally, we use
\begin{align}
  \Gamma\left(1+\i k\right)\,\Gamma\left(1-\i k\right)&=\frac{k\pi}{\sinh k \pi},\\
  \Gamma\left(\frac12+\i k \right)\,\Gamma\left(\frac12-\i k\right)&=\frac{\pi}{\cosh\pi k},\\
    \Gamma(\i k)\,\Gamma\left(-\i k\right)&=\frac{\pi}{k\sinh k\pi}.
  \end{align}

\subsection{Laplacian on the halfline}

Consider the Laplacian $-\partial_x^2$ on the half-line $\rr_+$.
Setting the Dirichlet and Neumann boundary
conditions at $0$ we obtain 2 operators on $L^2(\rr_+)$, which
are special cases of the hyperbolic Gegenbauer Hamiltonian:
\begin{align}
  L_{\D}&: =L_{\frac12,\frac12}^\h =L_{\frac12,-\frac12}^\h
          =L_\frac12^\h,\\
  L_{\N}&:  =L_{-\frac12,\frac12}^\h  =L_{-\frac12,-\frac12}^\h
          =L_{-\frac12}^\h.
            \end{align}

Let us compute their resolvents.
  The following functions solve
\beq(-\partial_x^2+k^2)\phi(x)=0\eeq and
  satisfy the Dirichlet/Neumann boundary
  conditions at $0$ and decay at $+\infty$:
  \begin{align}
    \text{    Dirichlet:}\quad& \sinh kx;\\
    \text{    Neumann:}\quad &\cosh kx;\\
    \text{decaying at $+\infty$:}\quad &\e^{-kx},\quad \Re\, k>0.
                                        \end{align}
    They have the
  Wronskians:
\begin{align}
  \cW\big(\e^{-kx},\sinh kx\big)&=k,\\
  \cW\big(\e^{-kx},\cosh kx\big)&=k.
                                        \end{align}
Now for $\Re\, k>0$,
                                        \begin{align}
\left(L_{\D}+k^2\right)^{-1}(x,y)=
        \frac1{k}\begin{cases}\sinh kx~\e^{-ky}
,\quad \text{if} \quad 0<x<y,
    \\\sinh ky~\e^{-kx}
,\quad \text{if} \quad 0<y<x;
\end{cases}\label{hd1}\\
\left(L_{\N}+k^2\right)^{-1}(x,y)=
      \frac1{k}\begin{cases}\cosh kx~\e^{-ky}
,\quad \text{if} \quad 0<x<y,
    \\\cosh ky~\e^{-kx}
,\quad \text{if} \quad 0<y<x.
\end{cases}\label{hd2}\end{align}  

To check that \eqref{hd1} and \eqref{hd2} agree with \eqref{resoh}, identify
$k=\frac{\mu}{2}$. By \eqref{simp3}, \eqref{simp4} and \eqref{simp5},
we have
\begin{align}
  P_{\frac12,\pm\frac12,\pm\mu}^\h(x)&=\frac{\sinh kx}{k\sqrt\pi},\\
  P_{-\frac12,\pm\frac12,\pm\mu}^\h(x)&=\frac{\cosh kx}{\sqrt\pi},\\
  Q_{\pm\frac12,\pm\frac12,\mu}^\h(x)&=\frac{2^{2k}}{\Gamma(1+2k)}\e^{-kx}.  
\end{align}
Finally, use
\begin{align}
\Gamma(k) \Gamma\left(k+\frac12\right)&=\frac{2^{-2k}\sqrt{\pi} }{k}\Gamma(2k+1),
\\
\Gamma\left(k+\frac12\right) \Gamma\left(k+1\right)&=2^{-2k}\sqrt{\pi} \Gamma(2k+1).
\end{align}
\subsection{Laplacian on the line}

Consider the Laplacian $-\partial_x^2$ on the line $\rr$, denoted $L$.
It is a special case of the deSitterian hypergeometric Hamiltonian of
the first kind:
\begin{align}
  L&:=L_{\pm\frac12,\pm\frac12}^\dS=L_{\pm\frac12,\mp\frac12}^\dS
     =L_{\pm\frac12}^\dS.
            \end{align}

It is well known  how to compute its resolvent:
  The following functions solve
\beq(-\partial_x^2+k^2)\phi(x)=0\eeq 
  \begin{align}
    \text{   decaying at $+\infty$:}\quad \e^{-kx},\quad \Re(k)>0,\\
    \text{   decaying at $-\infty$:}\quad \e^{kx},\quad \Re(k)>0.    
                                        \end{align}
    They have the
  Wronskian:
\begin{align}
  \cW\big(\e^{-kx},\e^{kx}\big)&=2k.
                                        \end{align}
                                        Now for $\Re\, k>0$,
                                        \begin{align}
\left(L+k^2\right)^{-1}(x,y)=
      \frac1{2k}\begin{cases}\e^{kx}\e^{-ky}
,\quad \text{if} \quad x<y,
    \\\e^{ky}\e^{-kx},
\quad \text{if} \quad y<x.
\end{cases}\label{ds.}\end{align}  

To see that \eqref{ds.} agrees with \eqref{resdes}, we identify
$k=\frac{\mu}{2}$, use
\begin{align}
  Q_{\pm\frac12,\pm\frac12,\mu}^\dS(x)= Q_{\pm\frac12,\mp\frac12,\mu}^\dS(x)&=\frac{2^{2k}}{\Gamma(2k+1)}\e^{-kx},
\end{align}
which follows from \eqref{simp5},  and
\beq
\Gamma(k)\,\Gamma\left(k+\frac12\right)^2\Gamma\left(k+1\right) = \frac{2^{-4 k} \pi}{k}\Gamma(2k+1)^2 .
\eeq

  \section{Geometric applications}
\label{Major geometric applications}
\init

In this section we show major applications of hypergeometric
Hamiltonians in geometry.
We will obtain these Hamiltonians as the results of separation of
variables of (pseudo-)Laplacians on various (pseudo-)spheres.

Recall that every (pseudo-)Riemannian manifold is equipped with
a certain natural differential operator called the  {\em (pseudo-)Laplacian}.
Suppose we fix coordinates $x=x_1,\dots,x_d$, the (pseudo-)metric is
given by the field of symmetric invertible  matrices $[g_{ij}]$, so
that the the ``line element'' is
\beq \d s^2=\sum_{1\leq i,j\leq d} g_{ij}\d x_i\d x_j.\eeq
Then the  pseudo-Laplacian is given by
\beq 
\Delta=\frac{1}{\sqrt{|\det g|}}\partial_{x_i}g^{ij}\sqrt{|\det g|}\partial_{x_j},\eeq
where $[g^{ij}]$ is the inverse of $[g_{ij}]$.
In the case of the Riemannian signature, $\Delta$
  is  called the {\em Laplacian} (or the {\em Laplace-Beltrami
  operator}). For the Lorentzian signature the usual name is the
 {\em d'Alembertian}.

 {\em (Pseudo-)spheres} in
a {\em (pseudo-)Euclidean   space} $\rr^{p,q}$ inherit a (pseudo-)Riemannian structure from 
  the ambient space. If the ambient space is Euclidean, they are
  called {\em spheres}. Otherwise, they are various kinds of {\em hyperboloids}.

Below we describe a few examples of 
 separation of variables for a
 (pseudo-)Laplacian on a (pseudo-)sphere in appropriate coordinate systems.
 We will see that after an appropriate gauging, subtraction of a constant and restriction to an
 invariant subspace one
obtains various hypergeometric Hamiltonians.
These computations motivate the names ``spherical'', ``hyperbolic'' and
``deSitterian'' that we use in our paper for various types of
hypergeometric Hamiltonians.

The best known among these Laplacians is $\Delta_d^\s$, the Laplacian
on the $d$-dimensional sphere $\SS^d$. As it is well-known, it
has the spectrum \beq\sigma(\Delta_d^\s)=\{-l(l+d-1)\ |\
l\in\nn_0\}.\eeq (This is, incidentally, a consequence of
computations in Subsection \ref{Sphere} and the properties of the
spherical Gegenbauer Hamiltonian).
Eigenfunctions of $\Delta_\d^\s$ with eigenvalue 
$-l(l+d-1)$ will be called {\em $d$-dimensional spherical harmonics of order $l$.}

  \subsection{Sphere}\label{Sphere}
  The unit $d$-dimensional sphere is defined as
  \beq\SS^d:=\{X\in\rr^{d+1}\ |\ X_0^2+X_1^2+\cdots+X_d^2=1\}.\eeq
  We will denote elements of $\SS^{d-1}$ by $\hat X$ and the
  corresponding element of length  by $\d {\hat X}^2$.
  {\red On $\SS^d$ we will use the coordinates $(r,\hat X)$
\begin{align}
  X_0=\cos r,\quad X_i&=\sin r\hat X_i,\quad i=1,\dots,d,\quad  \hat
                        X\in\SS^{d-1},\end{align}
                       or 
                   $(w,\hat X)$ with 
               $\cos r=w,\quad \sin r=\sqrt{1-w^2}$:}
In these coordinates we first 
write the line element, then the Laplacian:
                         \begin{align}
                           \d s^2&=\d r^2+\sin^2r\d\hat X^2\\&=\frac{\d
                             w^2}{1-w^2}+(1-w^2)\d\hat X^2;\\
\Delta_d^\s&=  \partial_r^2+(d-1)\cot
  r\partial_r+\frac{\Delta_{d-1}^\s}{\sin^2r}\\&=(1-w^2)\partial_w^2-d
                                                   w\partial_w+
                                                   \frac{\Delta_{d-1}^\s}{1-w^2}.
                                                 \end{align}
Finally, we perform an appropriate gauging:
\begin{align}                                                  (\sin
                                                   r)^{\frac{d-1}{2}}\big(-\Delta_d^\s\big)
                                                   (\sin
                                                   r)^{-\frac{d-1}{2}}+\big(\tfrac{d-1}{2}\big)^2&=-\partial_r^2+\frac{\big(\tfrac{d-2}{2}\big)^2-\tfrac14-\Delta_{d-1}^\s}{\sin^2
                                                     r}
                                                  \label{lap-s} .\end{align}

Thus  the Laplacian on $\SS^d$ on $(d-1)$-dimensional spherical harmonics  of order $l$ \eqref{lap-s}
reduces to the spherical Gegenbauer Hamiltonian
$L_\alpha^\s$ with $\alpha=(\frac{d}{2}-1+l)$.

\subsection{Hyperbolic space}
\label{Hyperbolic space}
The $d$-dimensional hyperbolic space is defined as
\beq\hh^d:=\{X\in\rr^{d+1}\ |\ -X_0^2+X_1^2+\cdots+X_d^2=-1,\quad
X_0>0\}.\eeq
{\red On $\hh^d$ we will use the following
cordinates: $(r,\hat X)$, where $\hat X\in\SS^{d-1}$:
\begin{align}
  X_0=\cosh r,\quad i=1,\dots,d,\quad X_i=\sinh r\,\hat X_i,\end{align}
 or  $(w,\hat X)$ with $\cosh r=w,$ $\sinh r=\sqrt{w^2-1}$.}
In these coordinates we first 
write the line element, then the Laplacian:
\begin{alignat}{3}                       
\d s^2=&  \d r^2+\sinh^2r\d\hat X^2\\=&\frac{\d
  w^2}{w^2-1}+(w^2-1)\d\hat X^2;\\
\Delta_d^\h&=  \partial_r^2+(d-1)\coth
  r\partial_r+\frac{\Delta_{d-1}^\s}{\sinh^2r}\\=&(w^2-1)\partial_w^2+d
                                                   w\partial_w+
                                                                                 \frac{\Delta_{d-1}^\s}{w^2-1}.                                                  \end{alignat}
Finally, we perform an appropriate gauging:
\begin{align}                                                 
(\sinh r)^{\frac{d-1}{2}}\big(-\Delta_d^\h\big) (\sinh
  r)^{-\frac{d-1}{2}}-\big(\tfrac{d-1}{2}\big)^2=&-\partial_r^2+\frac{\big(\tfrac{d-2}{2}\big)^2-\tfrac14-\Delta_{d-1}^\s}{\sinh^2
                        r}
                      .\end{align}
                                                 Thus the Laplacian on $\hh^d$
                                                 on $d-1$-dimensional spherical harmonics of order $l$ reduces to
the                                                                                                  hyperbolic Gegenbauer 
                                                 Hamiltonian
$L_\alpha^\h$ with $\alpha=\frac{d}{2}-1+l$.

\subsection{DeSitter space}
\label{DeSitter space}
The deSitter space is defined as
\beq\dS^d:=\{X\in\rr^{d+1}\ |\ -X_0^2+X_1^2+\cdots+X_d^2=1\}.\eeq
{\red On $\dS^d$ we will use the  following coordinates: $(t,\hat X)$, where
$\hat X\in \SS^{d-1}$,
\begin{align}
  X_0&=\sinh t,\quad X_i=\cosh t\,\hat X_i,,\quad
       i=1,\dots,d,\end{align}
or $(w,\hat X)$ with
$  \sinh t=w,$ $\cosh t=\sqrt{1+w^2}.$  }
     In these coordinates we first 
write the line element, then the d'Alembertian:
     \begin{alignat}{3} \notag
       \d s^2&= - \d t^2+\cosh^2t\d\hat X^2\\
       =&-\frac{\d w^2}{w^2+1}+(w^2+1)\d\hat X^2\\
 \notag \Box_d^\dS&= - \partial_t^2+(d-1)\tanh
t\partial_t+\frac{\Delta_{d-1}^\s}{\cosh^2t}\\
=&-(w^2+1)\partial_w^2-d
                                                   w\partial_w+
                                                   \frac{\Delta_{d-1}^\s}{w^2+1}.
                                                                                 \end{alignat}
Finally, we perform an appropriate gauging:
\begin{align}                                                 
(\cosh t)^{\frac{d-1}{2}}\Box_d^\dS (\cosh
  t)^{-\frac{d-1}{2}}-\big(\tfrac{d-1}{2}\big)^2=&-\partial_t^2-\frac{\big(\tfrac{d-2}{2}\big)^2-\tfrac14-\Delta_{d-1}^\s}{\cosh^2
                        t}
  .\end{align}
                                                 Thus the d'Alembertian on $\dS^d$
                                                 on $d-1$-dimensional
                                                 spherical harmonics
                                                 of order $l$ 
                                                 reduces to the 
                                                 deSitterian Gegenbauer 
                                                 Hamiltonian
$L_\alpha^\dS$ with $\alpha=\frac{d}{2}-1+l$.

                \subsection{Sphere in double spherical coordinates}

                Consider the unit sphere of dimension $p+q-1$ with
                coordinates partitioned in two groups:
                \beq \SS^{p+q-1}:=\{(X,Y)\in\rr^{p+q}\ |\
                X_1^2+\cdots+X_p^2+
                Y_1^2+\cdots+Y_q^2
                =1\}.\eeq
                We consider also two spheres of dimension $p-1$ and $q-1$:
                \begin{align}\label{pq}
                  \SS^{p-1}=&\{X\in\rr^p\ :\
                  X_1^2+\cdots+X_p^2=1\},\quad
                                                             \SS^{q-1}=\{Y\in\rr^q\ :\
                              Y_1^2+\cdots+Y_q^2=1\}\end{align}
                            $\SS^{p+q-1}$ is parametrized by
                            $(\tau,\hat X,\hat Y)$, with
$0\leq \tau\leq\frac\pi2$, $\hat X\in\SS^{p-1}$,
$\hat Y\in\SS^{q-1}$:
\begin{align}
  X_i=\sin \tau\,\hat X_i,\quad i=1,\dots,p;\qquad 
  Y_j=\cos \tau\,
  \hat Y_j,\quad 
  i=1, \dots, q.
  \end{align}
  Alternatively, one can use coordinates $(w,\hat X)$ where
  \beq
  \sin^2 \tau=w,\quad
  \cos^2 \tau=1-w.\eeq

We compute the line element and the Laplacian:
  \begin{alignat}{3} \notag
\d s^2&=  \frac{\d w^2}{4w(1-w)}+w\d \hat X^2+(1-w)\d \hat Y^2\\&=
\d\tau^2+\sin^2\tau\d
\hat X^2 +\cos^2\tau\d\hat Y^2\\ \notag
\Delta_{p+q-1}^\s&=4w(1-w)\partial_w^2+2\big(p(1-w)-qw\big)\partial_w+\frac{\Delta_{p-1}^\s}{w}+\frac{\Delta_{q-1}^\s}{1-w}\\=&
\partial_\tau^2+\big((p-1)\cot\tau-(q-1)\tan\tau\big)\partial_\tau+\frac{\Delta_{p-1}^\s}{\sin^2\tau}+\frac{\Delta_{q-1}^\s}{\cos^2\tau}.
\end{alignat}
We perform an appropriate gauging:
\begin{align} 
            \notag    &  (\sin\tau)^{\frac{p-1}{2}}  (\cos\tau)^{\frac{q-1}{2}}
 \big(- \Delta_{p+q-1}^\s\big)  (\sin\tau)^{-\frac{p-1}{2}}
  (\cos\tau)^{-\frac{q-1}{2}}+\big(\tfrac{p+q-2}{2}\big)^2
  \\=&
                              - \partial_\tau^2
+\frac{\big(\tfrac{p-2}{2}\big)^2-\tfrac14-\Delta_{p-1}^\s}{\sin^2
                                  \tau}                                  +\frac{\big(\tfrac{q-2}{2}\big)^2-\tfrac14-\Delta_{q-1}^\s}{\cos^2
                                  \tau}.\label{lapla1}
\end{align}
Finally, we make a substitution $\tau=\frac{r}{2}$:
\beq\eqref{lapla1}=4\Bigg(- \partial_r^2
+\frac{\big(\tfrac{p-2}{2}\big)^2-\tfrac14-\Delta_{p-1}^\s}{4\sin^2
                                  \frac{r}{2}}                                  +\frac{\big(\tfrac{q-2}{2}\big)^2-\tfrac14-\Delta_{q-1}^\s}{4\cos^2
                                  \frac{r}{2}}\Bigg).\eeq
Thus on  products of a spherical harmonic of order $j$ and $l$ we obtain
the spherical hypergeometric Hamiltonian of the first kind $L_{\alpha,\beta}^\s $ with
\beq \alpha=\frac p2-1+j,\quad
\beta =\frac q2-1+l.\eeq

                \subsection{Hyperboloid in double spherical coordinates}
\label{Hyperboloid}                
Consider the hyperboloid of signature $p-1,q$ embedded in the
pseudoEuclidean space of signature $(p,q)$:
\beq\hh^{p-1,q}:=\{(X,Y)\in\rr^{p+q}\ |\
-                X_1^2-\cdots-X_p^2+
                Y_1^2+\cdots+Y_q^2
                =-1\}.\eeq
                Let $\SS^{p-1}$ and $\SS^{q-1}$ be as in \eqref{pq}.
                            $\hh^{p-1,q}$ is parametrized by
                                                $(\tau,\hat X,\hat Y)$, with
$0\leq \tau\leq\infty$, $\hat X\in\SS^{p-1}$,
$\hat Y\in\SS^{q-1}$:
\begin{align}
    X_i=\cosh\tau\,\hat X_i,&\quad i=1,\dots,p;\qquad 
                                           Y_j=\sinh\tau\,\hat Y_j,\quad 
                                           i=1,\dots,q
.\end{align}
Alternatively, one can use coordinates $(w,\hat X,\hat Y)$ where

    \beq
  \cosh^2 r=w,\quad
  \sinh^2 r=w-1.\eeq
The line element and the pseudo-Laplacian in these coordinates:
  \begin{alignat}{3} \notag
\d s^2=&
  \frac{\d w^2}{4w(w-1)}-w\d \hat X^2+(w-1)\d \hat Y^2\\=&
                                                        \d\tau^2-\cosh^2\tau\d\hat X
                                                        ^2+\sinh^2\tau\d\hat Y
                                                        ^2\\ \notag
\Delta_{p-1,q}&=4w(w-1)\partial_w^2+2\big(p(w-1)+qw\big)\partial_w-\frac{\Delta_{p-1}^\s}{w}+\frac{\Delta_{q-1}^\s}{w-1}\\=&
\partial_\tau^2+\big((p-1)\tanh\tau+(q-1)\coth\tau\big)\partial_\tau-\frac{\Delta_{p-1}^\s}{\cosh^2\tau}+\frac{\Delta_{q-1}^\s}{\sinh^2\tau}.\end{alignat}
We perform an appropriate gauging:
\begin{align} 
               \notag &(\cosh\tau)^{\frac{p-1}{2}}  (\sinh\tau)^{\frac{q-1}{2}}
 \big(-\Delta_{p-1,q}\big)  (\cosh\tau)^{-\frac{p-1}{2}}
  (\sinh\tau)^{-\frac{q-1}{2}} -\big(\tfrac{p+q-2}{2}\big)^2
  \\
  =&
-                               \partial_\tau^2
-\frac{\big(\tfrac{p-2}{2}\big)^2-\tfrac14-\Delta_{p-1}^\s}{\cosh^2
                                  \tau}                                  +\frac{\big(\tfrac{q-2}{2}\big)^2-\tfrac14-\Delta_{q-1}^\s}{\sinh^2
                                  \tau}.\label{lapla2}
\end{align}
Finally, we substitute $\tau=\frac{r}{2}$:
\beq
\eqref{lapla2}=
4\Bigg(-                               \partial_r^2
-\frac{\big(\tfrac{p-2}{2}\big)^2-\tfrac14-\Delta_{p-1}^\s}{4\cosh^2
                                  \frac{r}{2}}                                  +\frac{\big(\tfrac{q-2}{2}\big)^2-\tfrac14-\Delta_{q-1}^\s}{4\sinh^2
                                  \frac{r}{2}}\Bigg).\eeq

Thus on the product of a spherical harmonic of order $j$ and $l$ we obtain
the hyperbolic hypergeometric Hamiltonian of the first kind $L_{\alpha,\beta}^\h $ with
\beq \alpha=\frac q2-1+l,\quad 
\beta =\frac p2-1+j.\eeq


\subsection{Complex manifolds}

All the manifolds that we used so far were real.
In the next subsection we will need 
a complex (analytic) manifold. They have 
essentially the same formalism as  real manifolds. Let us briefly
sketch its elements. For more details, see \cite{leb}.

Suppose that a complex manifold is
equipped with local complex coordinates $z=(z_1,\dots,z_d)$ and the holomorphic line element
\beq 
\sum_{1\leq i,j\leq d} g_{ij}\d z_i\d z_j,\eeq
where $g_{ij}$ is a complex symmetric invertible matrix.
The corresponding complex Laplacian is defined by essentially the same formula as
in the real case:
\beq 
\Delta=\frac{1}{\sqrt{\det g}}\partial_{z_i}g^{ij}\sqrt{\det g}\partial_{z_j}.\eeq
{\red Note that there is no absolute value, and $\det g$ is always
  non-zero.
The definition of $\Delta $
  does not depend on the choice
 of the branch of the (double-valued)
  square root.}

Suppose that the manifold is equipped with a conjugation, in the
coordinates given  by $z_i\mapsto\bar z_i$.
We then also have the anti-holomorphic line element
\beq 
\sum_{1\leq i,j\leq d}\bar g_{ij}\d \bar z_i\d \bar z_j.\eeq
and the corresponding conjugate Laplacian
\beq 
\bar\Delta=\frac{1}{\sqrt{\det\bar g}}\partial_{\bar z_i}\bar {g^{ij}}\sqrt{\det\bar g}\partial_{\bar z_j}.\eeq

As our first example consider the space $\cc^{d+1}$ equipped with the line
element
\beq\d Z^2=\d Z_0^2+\d Z_1^2+\cdots+\d Z_p^2.\eeq
The corresponding Laplacian is obviously
\beq\Delta=\partial_{Z_0}^2+\cdots +\partial_{Z_p}^2.\eeq

Note that our standard identification of $\cc^{d+1}$ with $\rr^{2(d+1)}$ will
be
\beq Z_i=\frac{1}{\sqrt2}(X_i+\ii Y_i)\label{zeti}\eeq
(and not $Z_i=X_i+\ii Y_i$.
Therefore,
\beq
\partial_{Z_i}=\frac1{\sqrt2}\big(\partial_{X_i}-\i\partial_{Y_i}\big),\eeq
(and not, as  usual,
$\partial_{Z_i}=\frac1{2}\big(\partial_{X_i}-\i\partial_{Y_i}\big)$.)
Clearly, with this definition $\langle\d
Z_i|\partial_{Z_j}\rangle=\delta_{ij}$.

Another example of a complex manifold is the unit sphere \cite{hiu}  
\beq\SS_\cc^d:=\{Z\in\cc^{d+1} \big\vert Z_0^2+ Z_1^2+\cdots+ Z_d^2=1\}.\eeq

Introducing the complex polar coordinanes
\beq
R:=\sqrt{Z_0^2+\cdots+Z_d  ^2},\quad \hat Z_i:=\frac{Z_i}{R},\eeq
{\red where we  arbitrarily select the branch of the square root,}
we have the direct analog of the formula from the real case:
\beq
\d Z^2=\d R^2+R^2\d \hat Z^2,\eeq
where $\d\hat Z^2$ is the complex line element on $\SS_\cc^d$.
The corresponding complex Laplacian is given by the same expressions
as in the real case:
\begin{align}\notag \Delta_{d,\cc}^\s&=
                                \sum_{0\leq i<j\leq
                                d}(Z_i\partial_{Z_j}-Z_j\partial_{Z_i})^2\\
  &=R^2\Delta_{d+1,\cc}-R^2\partial_R^2-d\, R\,\partial_R.
\end{align}

\subsection{Hyperboloid $\hh^{p-1,p}$ in  complex coordinates}

Consider the hyperboloid $\hh^{p-1,p}$ embedded in $\rr^{2p}$, defined
as in Subsection
\ref{Hyperboloid}, with the coordinates $X_i,Y_i\in\rr$, $i=1,\dots,p$.
We identify $\rr^{2p}$ with $\cc^p$ as in \eqref{zeti},
so that we  obtain two representations of $\hh^{p-1,p}$,  a real and a
complex one:
\begin{align} \notag
\hh^{p-1,p}&=\{(X,Y)\in\rr^{2p} \ |\
            -Y_1^2-\cdots-Y_p^2+X_1^2+\cdots+X_p^2=-1\}\\
  &=\{Z\in\cc^{p} \ |\
             Z_1^2+\cdots+Z_p^2+\bar Z_1^2+\cdots+\bar Z _p^2=1\}
\end{align}
The (real) line element on $\rr^{2p}=\cc^p$ can be written as
\begin{align} \notag
  &-\d Y_1^2-\cdots-\d Y_p^2+\d X_1^2+\cdots+\d X_p^2\\ \notag
  =&\d Z_1^2+\cdots+\d Z_p^2+\d \bar Z_1^2+\cdots+\d \bar Z_1^2\\
  =&\d R^2+R^2\d\hat Z^2+\d \bar R^2+\bar R^2\d\bar{\hat Z}^2
\end{align}
Now on $\hh^{p-1,p}$ we have $R^2+\bar R^2=1$. Therefore
$R^2=\frac{1+\i \sinh r}{2}$ for a unique $r\in\rr$. Thus
we can
parametrize $\hh^{p-1,p}$ with $r\in \rr, \hat Z\in\SS_\cc^{p-1}$
\beq Z_i=\sqrt{\frac{1+\i\sinh r}{2}}\hat Z_i,\eeq
where we take the principal branch of square root. The (real) line
element  and the pseudo-Laplacian are
\begin{align}
\d s^2&=-\frac14\d r^2+
\frac{1+\i\sinh r}{2}\d\hat Z^2+\frac{1-\i\sinh r}{2}\d\bar{\hat Z}^2,\\\
\Delta_{p-1,p}&=-4\partial_r^2-4(p-1)\tanh r\partial_r+\frac2{1+\i\sinh 
  r}\Delta_{p-1}^{\s,\cc}+\frac2{1-\i\sinh 
  r}\bar{\Delta_{p-1}^{\s,\cc}}.\end{align}
We perform an appropriate gauging:
\begin{align} \notag
&\big(\cosh r)^{\frac{p-1}{2}}\big(\Delta_{p-1,p}\big)(\cosh
                r)^{-\frac{p-1}{2}}-(p-1)^2\\
  =&4\Bigg(-\partial_r^2-\frac{\big(-\Delta_{p-1}^{\s,\cc}+(\frac{p-1}{2})^2-\frac14\big)}{2(1+\i\sinh r)}-\frac{\big(-\bar{\Delta_{p-1}^{\s,\cc}}+(\frac{p-1}{2})^2-\frac14\big)}{2(1-\i\sinh r)}\Bigg).
\end{align}
Thus on joint eigenvectors of $\Delta_{p-1}^{\s,\cc}$ and
$\bar{\Delta_{p-1}^{\s,\cc}}$ of degree $l$, resp.  $j$
we obtain the deSitterian hypergeometric Hamiltonian of the first kind
$L_{\alpha,\beta}^\dS$ with
\beq \alpha=\frac p2-1+l,\quad 
\beta =\frac p2-1+j.\eeq

\vspace{2em}
{\bf Acknowledgement.}
J.D. was supported by National Science Center (Poland) under the grant UMO-2019/35/B/ST1/01651. P.K supported by the OPUS grant no. 2022/47/B/ST2/03313 “Quantum geometry and
BPS states” funded by the National Science Centre (Poland).

\appendix

{\red

  \section{Holomorphic families of operators}
\label{Holomorphic families of operators}

The concept of a holomorphic family of bounded operators is
well-known. Less known is the concept of a holomorphic family of
closed operators, which we use throughout our paper.
Both are described in the monograph of Kato  \cite{Kato}, Chap. 7, and 
also in
\cite{DW1}. For the convenience of the reader, we give a concise
account of these concepts in this appendix.

In the following, $\Theta$ is an open subset of $\cc$, $\cH_1,\cH_2$ are Hilbert
spaces, 
$B(\cH_1,\cH_2)$ denotes the space of bounded operators from $\cH_1$
to $\cH_2$ and $C(\cH_1,\cH_2)$
the set of closed operators from $\cH_1$ to $\cH_2$.
   $\cD(H)$ will denote the domain of an operator $H$
and $\rho(H):=\cc\backslash\sigma(H)$ its resolvent set.

There exist several equivalent  definitions of a holomorphic
  family of bounded operators.
Here is one
of them:
Consider a function $\Theta\ni z\mapsto H(z)\in B(\cH_1,\cH_2)$
 We say that this is a {\em
  holomorphic family of bounded operators} if for any $f\in\cH_2$, $g\in\cH_1$
the function $ \Theta\ni z\mapsto (f| H(z) g)\in\cc$
is holomorphic.

We have the following practical criterion:
\bet\label{crit} Suppose that $\Theta\ni z\mapsto H(z)\in B(\cH_1,\cH_2)$ is a
  function
  uniformly bounded on compact subsets of $\Theta$. Suppose in
  addition that there exist dense subspaces $\cX_1\subset \cH_1$, $\cX_2\subset \cH_2$
  such that for any $f\in\cX_1$, $g\in\cX_2$,
  the function 
  $ \Theta\ni z\mapsto (f| H(z) g)\in\cc$ is holomorphic.
Then $\Theta\ni z\mapsto H(z)$ is a holomorphic family of bounded operators.
  \eet

There exists also a natural concept of a holomorphic family of closed
operators, due to Kato, it is however more subtle and less known. Let
us describe its definition.

Consider a function
$\Theta\ni z\mapsto H(z)\in C(\cH_1,\cH_2)$.  We say that it is a {\em
  holomorphic family of closed operators} if for each $z_0\in\Theta$
there exists a neighborhood $\Theta_0$ of $z_0$, a Hilbert space
$\cK$ and a holomorphic family of bounded operators $\Theta_0\ni
z\mapsto A(z)\in B(\cK,\cH_1)$ that map bijectively
$\cK$ onto $\cD(H(z))$ and
\begin{equation*}
\Theta_0\ni z\mapsto H(z)A(z)\in B(\cK,\cH)\label{holo2}
\end{equation*}
is a holomorphic family of bounded operators.

We have the following practical criterion, which works for operators
having non-empty resolvent sets:

\bet\label{crit1}
Consider a function $\Theta\ni z\mapsto H(z)\in C(\cH)$. Suppose in
addition that for any $z\in\Theta$ the resolvent set of $H(z)$
is nonempty. If for any $z_0\in\Theta$ there exists
$\lambda\in\cc$ and a neighborhood $\Theta_0$ of $z_0$ such that
$\lambda\in\rho( H(z))$ for $z\in\Theta_0$ and $z\mapsto
(H(z)-\lambda)^{-1}\in B(\cH)$ is holomorphic on $\Theta_0$,
then $z\mapsto H(z)$ is a holomorphic family of
closed operators.
\eet

\proof We set $\cK:=\cH$ and $A(z):=(H(z)-\lambda)^{-1}$, so that
$ H(z)A(z)=1 +\lambda A(z).$
\qed





\section{Riemann equation}
\label{Riemann equation}

\subsection{Regular-singular points}

Consider an ordinary
2nd order differential equation  with meromorphic coefficients:
\beq\left(b(z)\partial_z^2+c(z)\partial_z+d(z)\right)u(z)=0.\label{fro0}\eeq
If $\frac{c(z)}{b(z)}$ or $\frac{d(z)}{b(z)}$ are singular at $z_0\in\cc$,
then we say that $z_0$ is a singular point of \eqref{fro0}. Following
\cite{ww}, we distinguish a special class of singular points:
\begin{definition}
We say that \eqref{fro0}
has a {\em regular-singular point}  at $z_0$, 
if $\frac{c(z)}{b(z)}$ has
a pole of at most 1st order  at $z_0$, and  $\frac{d(z)}{b(z)}$ has a pole of at most
2nd order  at $z_0$.
\end{definition}

 We can rewrite \eqref{fro0}
as
\beq \left(p(z)(z-z_0)^2\partial_z^2+q(z)(z-z_0)\partial_z+r(z)\right)u(z)=0.\label{fro1}\eeq
If $p(z_0)\neq0$, then $z_0$ is regular-singular iff $p,q,r$ are analytic at $0$.
Define  the {\em indicial polynomial of $z_0$}
  \beq P_{z_0}(\lambda):=\lambda(\lambda-1)p(z_0)+\lambda q(z_0)+r(z_0).\eeq
The roots of $P_{z_0}$ are called {\em indices} of the regular-singular
point $z_0$.

The importance of regular-singular points and their indices
stems from the Frobenius
Method, described in the following theorem \cite{ww,Ince}:

\bet\label{frob}
Assume that $p(z_0)\neq0$ and $p,q,r$ are holomorphic in an open connected
simply connected set $\Omega\subset\cc$ containing $z_0$.
  Let $\rho\in\cc$ satisfy
  \beq P_{z_0}(\rho)=0,\quad P_{z_0}(\rho+n)\neq0,\quad n=1,2,\dots.\eeq
Then there exists a unique function $\tilde u(z)$ holomorphic in
  $\Omega$,
such that $u(z):=(z-z_0)^\rho \tilde u(z)$ solves \eqref{fro1} and
$\tilde u(z_0)=1$.
\eet

\subsection{Riemann equation}

Equation with  3 singular points on the Riemann sphere
$\cc\cup\{\infty\}$, all of them regular-singular, is classified by
the positions of its singularities and the values of indices. It is
called the {\em Riemann}, or sometimes the {\em Papperitz  equation}, and is described in the following
well-known theorem \cite{ww,Ince}. For the Riemann operator we use the
notation introduced in \cite{DW}, which is inspired by the notation due
to Papperitz used in \cite{ww}.

\bet \ben \item Suppose that we are given a 2nd order differential equation
 on the Riemann sphere
having at most
 3 singular points $z_1,z_2,z_3$, all of them regular-singular,
with the following indices
\[
z_1:\  \rho_1,\trho_1;\quad
z_2:\  \rho_2,\trho_2;\quad
z_3:\  \rho_3,\trho_3.
\]
Then the following condition is satisfied:
\beq\rho_1+\trho_1+\rho_2+\trho_2+\rho_3+\trho_3=1.\label{ff3}\eeq
 Such an equation, normalized to have coefficient $1$ at the 2nd derivative,
 is for finite $z_1,z_2,z_3$  given by the operator
\begin{align*}&\P\left(\begin{array}{cccc}
z_1&z_2&z_3&\\
\rho_1&\rho_2&\rho_3&z,\partial_z\\
\trho_1&\trho_2&\trho_3&\end{array}\right):=
\partial_z^2-\left(\frac{\rho_1+\trho_1-1}{z-z_1}+
\frac{\rho_2+\trho_2-1}{z-z_2}+
\frac{\rho_3+\trho_3-1}{z-z_3}\right)\partial_z\\[2ex]
&+\frac{\rho_1\trho_1(z_1-z_2)(z_1-z_3)}{(z-z_1)^2(z-z_2)(z-z_3)}
+\frac{\rho_2\trho_2(z_2-z_3)(z_2-z_1)}{(z-z_2)^2(z-z_3)(z-z_1)}
+\frac{\rho_3\trho_3(z_3-z_1)(z_3-z_2)}{(z-z_3)^2(z-z_1)(z-z_2)}
    ,\end{align*}
and for $z_3=\infty$ by the operator
\begin{eqnarray}\nonumber
&&\P\left(\begin{array}{cccc}
z_1&z_2&\infty&\\
\rho_1&\rho_2&\rho_3&z,\partial_z\\
\trho_1&\trho_2&\trho_3&\end{array}\right)
:=
\partial_z^2-\left(\frac{\rho_1+\trho_1-1}{z-z_1}+
\frac{\rho_2+\trho_2-1}{z-z_2}
\right)\partial_z\nonumber\\
&&+\frac{\rho_1\trho_1(z_1-z_2)}{(z-z_1)^2(z-z_2)}
+\frac{\rho_2\trho_2(z_2-z_1)}{(z-z_2)^2(z-z_1)}
+\frac{\rho_3\trho_3}{(z-z_1)(z-z_2)}.
\label{ff2}\end{eqnarray}
\item With the help of homographies (called also M\"obius
  transformations)
  we can move around singularities. More precisely, if $z\mapsto
  w(z)=\frac{az+b}{cz+d}$, where
 we can  assume that
$ad-bc=1$, then
\[\P\left(\begin{array}{cccc}
w(z_1)&w(z_2)&w(z_3)&\\
\rho_1&\rho_2&\rho_3&w,\partial_w\\
\trho_1&\trho_2&\trho_3&\end{array}\right)=
(cz+d)^4\P\left(\begin{array}{cccc}
z_1&z_2&z_3&\\
\rho_1&\rho_2&\rho_3&z,\partial_z\\
\trho_1&\trho_2&\trho_3&\end{array}\right).
\]
\item By gauging with powers we can shift the indices:
\beq\begin{array}{l}
(z-z_1)^{-\lambda}(z-z_2)^\lambda\P\left(\begin{array}{cccc}
z_1&z_2&z_3&\\
\rho_1&\rho_2&\rho_3&z,\partial_z\\
\trho_1&\trho_2&\trho_3&\end{array}\right)
(z-z_1)^{\lambda}(z-z_2)^{-\lambda}
\\[5mm]=\P\left(\begin{array}{cccc}
z_1&z_2&z_3&\\
\rho_1-\lambda&\rho_2+\lambda&\rho_3&z,\partial_z\\
\trho_1-\lambda&\trho_2+\lambda&\trho_3&\end{array}\right),
\end{array}\eeq
\beq\begin{array}{l}
(z-z_1)^{-\lambda}\P\left(\begin{array}{cccc}
z_1&z_2&\infty&\\
\rho_1&\rho_2&\rho_3&z,\partial_z\\
\trho_1&\trho_2&\trho_3&\end{array}\right)
(z-z_1)^{\lambda}
\\[5mm]=\P\left(\begin{array}{cccc}
z_1&z_2&z_3&\\
\rho_1-\lambda&\rho_2&\rho_3+\lambda&z,\partial_z\\
\trho_1-\lambda&\trho_2&\trho_3+\lambda&\end{array}\right).
\end{array}\eeq

\een\label{riema}\eet

\subsection{From Riemann equation to hypergeometric equation}
\label{hipo}

The Riemann equation can be simplified.
First, by Thm \ref{riema} (2), we can assume that  the points $z_1,z_2,z_3$
are any triplet of distinct points on the Riemann sphere. We choose them
to be $0,1,\infty$.
Then, by Thm \ref{riema} (3), we can assume that  $\rho_1,\rho_2$ are
arbitrary numbers. We choose them to be both $0$. The sum of remaining
indices must be $1$. Hence we can assume that
 $0$ has  indices $0$, $1-c$;
$1$ has indices
$0$, $c-a-b$;
and $\infty$ has indices $a$, $b$.
The hypergeometric operator is
\begin{eqnarray}
\F(a,b;c;z,\partial_z)&:=&
z(1-z)\P\left(\begin{array}{cccc}
0&1&\infty&\\
0&0&a&z,\partial_z\\
1-c&c-a-b&b&\end{array}\right)\\
&=&z(1-z)\partial_z^2+(c-(a+b+1)z)\partial_z-ab.\label{hy1}\end{eqnarray}
The hypergeometric function $F(a,b;c;z)$
is the unique solution of
\beq \cF(a,b;c;z,\partial_z)F(z)=0,\quad F(0)=1.\eeq
Thus it is the solution of the hypergeometric equation given by the Frobenius  method (Thm \ref{frob})
for the singularity $z=0$ and  index $0$.

}

\section{Identities for the hypergeometric function}
\label{Identities for the hypergeometric function}
\init
In this section, we review identities for the hypergeometric and
Gegenbauer functions with the
emphasis on the connection formulas and Kummer's table. There are many
reviews available including \cite{ww,NIST}. Our
presentation is close to  \cite{DHR1,DHR2}.

\subsection{Kummer's table}

Recall that the
hypergeometric equation is given by the operator
$\cF(a,b;c;z,\partial_z)$ defined in \eqref{hyperge}, and the
hypergeometric function with Olver's normalization
$\mathbf{F}(a,b;c;z)$ is
defined in \eqref{olver}.
In this section, for brevity and transparency,
we change the notation following  \cite{DHR1,DHR2}, writing
\beq\label{dhr}
\mathbf{F}_{\alpha,\beta,\mu}(z) := \mathbf{F}\left(\frac{1+\alpha+\beta+\mu}{2},\frac{1+\alpha+\beta-\mu}{2};1+\alpha;z\right).
\eeq
It is obvious from the definition that we have the following identity
    \beq\label{trivial}
        \mathbf{F}_{\alpha,\beta,\mu}(z) = \mathbf{F}_{\alpha,\beta,-\mu}(z).
    \eeq
The following  6 functions form a set of standard solutions of the
hypergeometric equation. Each of the solutions can be expressed in 4
ways (actually, $4\times2$  ways if we include the trivial identity \eqref{trivial}). This yields $6\times4=24$ expressions usually called {\em Kummer's table}:
\begin{center}

\beq
\begin{aligned}
\mathbf{F}_{\alpha,\beta,\mu}(z) &= (1-z)^\frac{-1-\alpha-\beta+\mu}{2}\mathbf{F}_{\alpha,-\mu,-\beta}(\frac{z}{z-1})
\\ 
&=(1-z)^\frac{-1-\alpha-\beta-\mu}{2}\mathbf{F}_{\alpha,\mu,\beta}(\frac{z}{z-1})\label{a3}
\\
&= (1-z)^{-\beta} \mathbf{F}_{\alpha,-\beta,-\mu}(z).
\end{aligned}
\eeq
\beq
\begin{aligned}
(-z)^\alpha \mathbf{F}_{-\alpha,\beta,-\mu}(z) &= (-z)^\alpha (1-z)^\frac{-1+\alpha-\beta-\mu}{2}\mathbf{F}_{-\alpha,\mu,-\beta}(\frac{z}{z-1})
\\
&=(-z)^\alpha (1-z)^\frac{-1+\alpha-\beta+\mu}{2}\mathbf{F}_{-\alpha,-\mu,\beta}(\frac{z}{z-1})
\\
&= (-z)^\alpha (1-z)^{-\beta} \mathbf{F}_{-\alpha,-\beta,\mu}(z).
\end{aligned}
\eeq
\beq
\begin{aligned} 
(-z)^\frac{-1-\alpha-\beta+\mu}{2}\mathbf{F}_{-\mu,\beta,-\alpha}(z^{-1}) &= (1-z)^\frac{-1-\alpha-\beta+\mu}{2}\mathbf{F}_{-\mu,\alpha,-\beta}\left(\frac{1}{1-z}\right)
\\
&=(-z)^{-\alpha}(1-z)^\frac{-1+\alpha-\beta-\mu}{2}\mathbf{F}_{-\mu,-\alpha,\beta}\left(\frac{1}{1-z}\right)
\\
&= (-z)^\frac{-1-\alpha-\beta+\mu}{2} (1-z)^{-\beta} \mathbf{F}_{\alpha,-\beta,-\mu}\left(z^{-1}\right).
\end{aligned}
\eeq
\beq
\begin{aligned}
(-z)^\frac{-1-\alpha-\beta-\mu}{2}\mathbf{F}_{\mu,\beta,\alpha}(z^{-1}) &= (1-z)^\frac{-1-\alpha-\beta-\mu}{2}\mathbf{F}_{\mu,-\alpha,-\beta}\left(\frac{1}{1-z}\right)
\\
&=(-z)^{-\alpha}(1-z)^\frac{-1+\alpha-\beta+\mu}{2}\mathbf{F}_{\mu,\alpha,\beta}\left(\frac{1}{1-z}\right)
\\
&= (-z)^\frac{-1-\alpha-\beta-\mu}{2} (1-z)^{-\beta} \mathbf{F}_{-\alpha,-\beta,\mu}\left(z^{-1}\right).
\end{aligned}
\eeq
\beq
\begin{aligned}
\mathbf{F}_{\beta,\alpha,\mu}(1-z) &= (-z)^\frac{-1-\alpha-\beta+\mu}{2}\mathbf{F}_{\beta,-\mu,-\alpha}(1-\frac{1}{z})
\\
&=(-z)^\frac{-1-\alpha-\beta-\mu}{2}\mathbf{F}_{\beta,\mu,\alpha}(1-\frac{1}{z})
\\
&= (-z)^{-\alpha} \mathbf{F}_{\beta,-\alpha,-\mu}(1-z).
\end{aligned}
\eeq
\beq
\begin{aligned}
(1-z)^{-\beta}\mathbf{F}_{-\beta,-\mu,-\alpha}(1-z) &= (-z)^\frac{-1-\alpha-\beta+\mu}{2}\mathbf{F}_{\beta,-\mu,-\alpha}(1-\frac{1}{z}) 
\\
&=(-z)^\frac{-1-\alpha-\beta-\mu}{2}\mathbf{F}_{\beta,\mu,\alpha}(1-\frac{1}{z})
\\
&= (-z)^{-\alpha} \mathbf{F}_{\beta,-\alpha,-\mu}(1-z).
\end{aligned}
\eeq
    
\end{center}

\subsection{Connection formulas}
Here are  connection formulas. For  $z \notin ]-\infty,0] \cup [1,\infty[$:
\begin{align}  \notag&
    \mathbf{F}_{\beta, \alpha,\mu}(1-z)\\ &= \frac{\pi~\mathbf{F}_{\alpha,\beta, \mu}(z)}{\sin(-\pi \alpha) \Gamma\left( \frac{1-\alpha+\beta-\mu}{2}\right)\Gamma\left( \frac{1-\alpha+\beta+\mu}{2}\right)}+\frac{\pi~ z^{-\alpha}~\mathbf{F}_{-\alpha,\beta, -\mu}(z)}{\sin(\pi \alpha) \Gamma\left( \frac{1+\alpha+\beta+\mu}{2}\right)\Gamma\left( \frac{1+\alpha+\beta-\mu}{2}\right)},\label{con2}
    \\ &\notag
    (1-z)^{-\beta}\mathbf{F}_{-\beta, \alpha,- \mu}(1-z)\\ &= \frac{\pi~ \mathbf{F}_{\alpha,\beta, \mu}(z)}{\sin(-\pi \alpha) \Gamma\left( \frac{1-\alpha-\beta-\mu}{2}\right)\Gamma\left( \frac{1-\alpha-\beta-\mu}{2}\right)}+\frac{\pi~ z^{-\alpha}~\mathbf{F}_{-\alpha,\beta, -\mu}(z)}{\sin(\pi \alpha) \Gamma\left( \frac{1+\alpha-\beta-\mu}{2}\right)\Gamma\left( \frac{1+\alpha-\beta+\mu}{2}\right)},   
\end{align}

\begin{align}  \notag
&    \mathbf{F}_{ \alpha,\beta,\mu}(z)\\ &= \frac{\pi~\mathbf{F}_{\beta,\alpha,\mu}(1-z)}{\sin(-\pi \beta) \Gamma\left( \frac{1+\alpha-\beta-\mu}{2}\right)\Gamma\left( \frac{1+\alpha-\beta+\mu}{2}\right)}+\frac{\pi~ (1-z)^{-\beta}~\mathbf{F}_{-\beta,\alpha, -\mu}(1-z)}{\sin(\pi \beta) \Gamma\left( \frac{1+\alpha+\beta+\mu}{2}\right)\Gamma\left( \frac{1+\alpha+\beta-\mu}{2}\right)},
    \\ \notag
  &z^{-\alpha}\mathbf{F}_{-\alpha, \beta,\mu}(z)\\
  &= \frac{\pi~ \mathbf{F}_{\beta,\alpha, \mu}(z)}{\sin(-\pi \beta) \Gamma\left( \frac{1-\alpha-\beta-\mu}{2}\right)\Gamma\left( \frac{1-\alpha-\beta+\mu}{2}\right)}+\frac{\pi~ (1-z)^{-\beta}~\mathbf{F}_{-\beta,\alpha, -\mu}(1-z)}{\sin(\pi \beta) \Gamma\left( \frac{1-\alpha+\beta-\mu}{2}\right)\Gamma\left( \frac{1-\alpha+\beta+\mu}{2}\right)}.  \label{reqdes} 
\end{align}

For $z\notin[0,\infty[$:

\begin{align}&\notag
(-z)^{\frac{-1-\alpha-\beta-\mu}{2}}\mathbf{F}_{\mu,\beta,\alpha}\left(z^{-1}\right)\\ &= \frac{\pi~\mathbf{F}_{\alpha,\beta,\mu}(z)}{\sin (-\pi \alpha) \Gamma \left(\frac{1-\alpha-\beta+\mu}{2}\right)\Gamma \left(\frac{1-\alpha+\beta+\mu}{2}\right)}+\frac{\pi~(-z)^{-\alpha} \mathbf{F}_{-\alpha,\beta,\mu}(z)}{\sin (\pi \alpha) \Gamma \left(\frac{1+\alpha-\beta+\mu}{2}\right)\Gamma \left(\frac{1+\alpha+\beta+\mu}{2}\right)};\label{con-1}
\\ \notag
&(-z)^\frac{-1-\alpha-\beta+\mu}{2}\mathbf{F}_{-\mu,\beta,\alpha}\left(z^{-1}\right)\\&=\frac{ \pi ~\mathbf{F}_{\alpha,\beta,\mu}(z)}{\sin(-\pi \alpha)\Gamma \left(\frac{1-\alpha-\beta-\mu}{2}\right)\Gamma \left(\frac{1-\alpha+\beta-\mu}{2}\right)}+\frac{ \pi ~(-z)^{-\alpha}~\mathbf{F}_{-\alpha,\beta,\mu}(z)}{\sin(\pi \alpha)\Gamma \left(\frac{1+\alpha-\beta-\mu}{2}\right)\Gamma \left(\frac{1+\alpha+\beta-\mu}{2}\right)}.\label{con-2}
\end{align}

   \begin{align}&\notag
\mathbf{F}_{\alpha,\beta,\mu}(z) \\ \label{popo}&=
\frac{\pi(-z)^\frac{-1-\alpha-\beta-\mu}{2}
  \mathbf{F}_{\mu,\beta,\alpha}(z^{-1})}{\sin (-\pi \mu) \Gamma
  \left(\frac{1+\alpha+\beta-\mu}{2}\right)\Gamma
  \left(\frac{1+\alpha-\beta-\mu}{2}\right)}+\frac{\pi(-z)^\frac{-1-\alpha-\beta+\mu}{2}
  \mathbf{F}_{-\mu,\beta,\alpha}(z^{-1})}{\sin (\pi \mu) \Gamma
  \left(\frac{1+\alpha+\beta+\mu}{2}\right)\Gamma
     \left(\frac{1+\alpha-\beta+\mu}{2}\right)},
     \\&\notag
(-z)^{-\alpha}     \mathbf{F}_{-\alpha,\beta,\mu}(z) \\&=
\frac{\pi(-z)^\frac{-1-\alpha-\beta-\mu}{2}
  \mathbf{F}_{\mu,\beta,\alpha}(z^{-1})}{\sin (-\pi \mu) \Gamma
  \left(\frac{1-\alpha+\beta-\mu}{2}\right)\Gamma
  \left(\frac{1-\alpha-\beta-\mu}{2}\right)}+\frac{\pi(-z)^\frac{-1-\alpha-\beta+\mu}{2}
  \mathbf{F}_{-\mu,\beta,\alpha}(z^{-1})}{\sin (\pi \mu) \Gamma
  \left(\frac{1-\alpha+\beta+\mu}{2}\right)\Gamma
  \left(\frac{1-\alpha-\beta+\mu}{2}\right)}.
     \end{align}

\subsection{Degenerate case}

Let $\mu\in\zz$. Then we have the identity
\begin{align} \notag
  \left(\frac{\alpha+\beta-\mu+1}{2}\right)_\mu
  \left(\frac{\alpha-\beta-\mu+1}{2}\right)_\mu
  &F\left(\frac{\alpha+\beta+\mu+1}{2}, 
  \frac{\alpha-\beta+\mu+1}{2};1+\mu;z\right)\\ \label{degen}
  =&z^{-\mu}F\left(\frac{\alpha+\beta-\mu+1}{2}, 
     \frac{\alpha-\beta-\mu+1}{2};1-\mu;z\right).
     \end{align}

\subsection{Gegenbauer functions}

In the remaining part of this section we review some of the relations
for Gegenbauer function, following mostly \cite{DGR}.

Recall that the Gegenbauer equation is defined by the operator
$\cG_{\alpha,\lambda}(w,\partial_w)$ defined in \eqref{gege0}, 
and the two Gegenbauer functions that we use were defined in
\eqref{geg1} and \eqref{geg2}. Let us rewrite their definitions using
the notation introduced in \eqref{dhr}.

The following function satisfies the Gegenbauer equation
and has value $1$ at $1$:
\beq
S_{\alpha,\pm\lambda}(w) = F_{\alpha,\alpha,\lambda}\left( \frac{1-w}{2}\right).
\eeq
It can be easily seen from \eqref{a3} that the solution behaving as
$(\frac{w+1}{2})^{-\alpha}$ at $1$ is
\beq
S_{-\alpha,\pm  \lambda}(w)  =\Big(\frac{1+w}{2}\Big)^{-\alpha}F_{-\alpha,\alpha,2 \lambda }\left(\frac{1-w}{2}\right)
\eeq
Solution behaving as $w^{-\frac12-\lambda-\alpha}$ at $+\infty$ is
\beq
Z_{\alpha,\lambda}(w) = (w\pm1)^{-\frac12-\alpha-\lambda}F_{2 \lambda,\alpha,\alpha}\left(\frac{2}{1\pm w}\right) 
\eeq
It is useful to introduce Olver's normalization
\beq
\mathbf{S}_{\alpha,\lambda}(w) = \frac{S_{\alpha,\lambda}(w)}{\Gamma(1+\alpha)},\qquad \mathbf{Z}_{\alpha,\lambda}(w) = \frac{Z_{\alpha,\lambda}(w)}{\Gamma(1+\lambda)}.
\eeq

We can read their connection formula from the connection formulas of hypergeometric functions \eqref{con-1} \eqref{con2}, for $\Im(w)<0$:
\begin{align} \label{gegencon}
    \mathbf{Z}_{\alpha,\lambda}(w) &= \frac{\sqrt{\pi } 2^{-\alpha +\lambda -\frac{1}{2}} }{\sin(-\pi \alpha)~\Gamma \left(-\alpha +\lambda +\frac{1}{2}\right)}\mathbf{S}_{\alpha,\lambda}(w) + \frac{\sqrt{\pi } 2^{\alpha +\lambda -\frac{1}{2}}(w^2-1)_\bullet^{-\alpha} }{\sin\pi \alpha~\Gamma \left(\alpha +\lambda +\frac{1}{2}\right)}\mathbf{S}_{-\alpha,\lambda}(w),
    \\ \label{gegencon2}
    \mathbf{S}_{\alpha,\lambda}(-w) &=\frac{\cos \pi \lambda}{\sin(-\pi \alpha)} \mathbf{S}_{\alpha,\lambda}(w)+\frac{\pi ~(1-w^2)^{-\alpha}}{\sin\pi \alpha~\Gamma\left(\frac12+\alpha+\lambda\right)\Gamma\left(\frac12+\alpha-\lambda\right)} \mathbf{S}_{-\alpha,\lambda}(w).
\end{align}
Note that the first connection formula is valid when $w \notin]-\infty,1]$, and the second connection formula is valid when $w\notin]-\infty,-1]\cup [1,\infty[$. Here we borrow a notation from \cite{DGR} where
\beq
 (w^2-1)^\alpha_\bullet := (w-1)^\alpha (w+1)^\alpha. 
\eeq
The functions $(w^2-1)^\alpha$ and $(w^2-1)_\bullet^\alpha$ coincide only if $\Re(w)>0$. In general the function $(w^2-1)_\bullet^\alpha$ is holomorphic on $\cc \setminus ]-\infty,1]$ while $(w^2-1)^\alpha$ is holomorphic on $ \cc \setminus \{ [-1,1] \cup \i \rr \} $. 
\subsection{Whipple transformation}
Gegenbauer equation has an extra symmetry compared to hypergeoemetric
symmetry called Whipple transformation. On the level of its standard
solutions it has the following form:
\begin{align} \label{whpl1}
    \mathbf{Z}_{\alpha,\lambda}(w) &= (w^2-1)_\bullet^{-\frac14-\frac\alpha2-\frac\lambda2}\mathbf{S}_{\lambda,\alpha}\left( \frac{w}{\left(w^2-1\right)_\bullet^{\frac12}} \right),
    \\ \label{whpl2}
    \mathbf{S}_{\alpha,\lambda}(w) &= (w^2-1)_\bullet^{-\frac14-\frac\alpha2-\frac\lambda2}\mathbf{Z}_{\lambda,\alpha}\left( \frac{w}{\left(w^2-1\right)_\bullet^\frac12} \right), \quad \Re(w)>0.
\end{align}
\eqref{whpl2} is obtain by inverting \eqref{whpl1} and using the fact that $w \mapsto \frac{w}{(w^2-1)^\frac12_\bullet}$ is an involution if and only if $\Re(w)>0$.

\subsection{Half integer case}
\label{Half integer case}

Gegenbauer functions with $\alpha=\pm\frac12$ have simple expressions
in terms of elementary functions. 
To see this we change variables in 
Gegenbauer operators. For $w\in]-1,1[$, we substitute $w=\cos\phi$:

\begin{align}\label{substi1}
\cG_{-\frac12,\lambda}\left(w,\partial_w\right)&=\partial_\phi^2+\lambda^2;\\
\cG_{\frac12,\lambda}\left(w,\partial_w\right)&=\frac{1}{\sin\phi}\big(\partial_\phi^2+\lambda^2\big)\sin\phi.
\end{align}

For $w\in]1,+\infty[$, we substitute $w=\cosh\theta$:

\begin{align}
\cG_{-\frac12,\lambda}\left(w,\partial_w\right)&=-\partial_\theta^2+\lambda^2;\\
\cG_{\frac12,\lambda}\left(w,\partial_w\right)&=\frac{1}{\sinh\theta}\big(-\partial_\theta^2+\lambda^2\big)\sinh\theta.\label{substi4}
\end{align}

\eqref{substi1}---\eqref{substi4} easily imply the first column of the following identities for
Gegenbauer functions. (The second column simply follows from the
definitions of Gegenbauer functions).
\begin{align}\label{simp1}
  \cos
  \lambda\phi=&S_{-\frac12,\lambda}(\cos\phi)=F\Big(\lambda,-\lambda;\frac12;\sin^2\frac\phi2\Big),\\\label{simp2}
   \frac{ \sin \lambda\phi}{\lambda\sin\phi}=&S_{\frac12,\lambda}(\cos\phi)=F\Big(1+\lambda,1-\lambda;\frac32;\sin^2\frac\phi2\Big),\\\label{simp3}
  \cosh
  \lambda\theta=&S_{-\frac12,\lambda}(\cosh\theta)=F\Big(\lambda,-\lambda;\frac12;\sinh^2\frac\theta2\Big),\\\label{simp4}
   \frac{ \sinh 
  \lambda\theta}{\lambda\sinh\theta}=&S_{\frac12,\lambda}(\cosh\theta)=F\Big(1+\lambda,1-\lambda;\frac32;\sinh^2\frac\theta2\Big),\\\label{simp5}
  2^\lambda\e^{-\lambda\theta}=&Z_{-\frac12,\lambda}(\cosh\theta) =\Big(2\sinh^2\frac\theta2\Big)^{-\lambda}F\Big(\lambda,\lambda+\frac12;1+2\lambda;-\frac1{\sinh^2\frac\theta2}\Big),\\\label{simp6}
   \frac{
  2^\lambda\e^{-\lambda\theta}}{\sinh\theta}=&Z_{\frac12,\lambda}(\cosh\theta)
=\Big(2\sinh^2\frac\theta2\Big)^{-\lambda-1}F\Big(\lambda+\frac12,\lambda+1;1+2\lambda;-\frac1{\sinh^2\frac\theta2}\Big).
\end{align}

The above formulas can be found e.g.  
 in
equations (4.25)--(4.28) of \cite{DGR} (in a slightly different form). They are straightforward
generalizations of well-known formulas for Chebyshev polynomials. In
fact, for $n\in\nn_0$ the usual Chebyshev polynomials are special
cases of   Gegenbauer functions with $\alpha=\pm\frac12$:
\beq
  T_n(w)=S_{-\frac12,n}(w),\quad U_n(w)=(n+1)S_{\frac12,n+1}(w).\eeq

\section{Closed realizations of 1d Schr\"odinger operators}
\label{Closed realizations of 1d Schrodinger operators}
\init

\subsection{Minimal and maximal realization}

The theory of self-adjoint realizations of 1d Schr\"odinger operators
with real potentials is
well-known and discussed in various sources
\cite{GeZin,GTV}.
Somewhat less known is the theory of their closed realizations, which
allows for complex potentials---however it is also a classic subject
covered in various texts
\cite{DunfordSchwartz,EdmundsEvans,DeGe}.
 We will treat \cite{DeGe} as the basic source
for this topic. It is concisely repeated in Sect. 2 of \cite{DL}.

For the convenience of the reader let us summarize some points from
\cite{DeGe,DL}.

We consider an interval $]a,b[$ and an operator $L$
acting on $f\in C_\mathrm{c}^\infty]a,b[$ given by \eqref{oper}, that
is
\beq L f:=\big( -\partial_x^2 + V(x)\big)f.\eeq
A {\em closed realization} of $L$ is a closed operator $L_\bullet$ in the
sense of
$L^2]a,b[$ that
restricted to  $C_\mathrm{c}^\infty]a,b[$ coincides with $L$.

There
 always exists the maximal closed realization, denoted $L^{\max}$ and
 the minimal closed realization, denoted $L^{\min}$.
 $L^{\max}$ restricted to $\cD(L^{\min})$ coincides with $L^{\min}$.

 In many cases $\cD(L^{\min})=\cD(L^{\max})$, and then
 there exists a unique closed realization of $L$.
 
 Sometimes $\cD(L^{\max})$ is larger than $\cD(L^{\min})$, and then
 there exist also closed realizations of $L$, denote them $L_\bullet$ , which satisfy
\beq\cD(L^{\min})\subset\cD(L_\bullet)\subset\cD(L^{\max})\eeq
and $L_\bullet=L^{\max}\Big|_{\cD(L_\bullet)}$.

\subsection{Resolvent}
\label{Resolvent}
Let $L_\bullet$ be a realization of $L$ with separated boundary
conditions. Following \cite{DL,DeGe} we will  now
sketch how to find 
the spectrum of $L_\bullet$, denoted $\sigma(L_\bullet)$, and how to 
compute the integral kernel of 
$\frac1{L_\bullet-z}$ for $z\in\cc$ outside of $\sigma(L_\bullet)$.

First let us recall the definition of the Wronskian of  two complex
functions $\Phi_1,\Phi_2$ on $]a,b[$:
\beq\label{wronskian}\cW(\Phi_1,\Phi_2)(x):=\Phi_1(x)\Phi_2'(x)-\Phi_1'(x)\Phi_2(x).\eeq
It is easily checked that if both $\Phi_1$ and $\Phi_2$
are eigenfunctions of $-\partial_x^2 + V(x)$ with the
same eigenvalue, 
 then $\cW(\Phi_1,\Phi_2)(x)$
 does not depend on $x$, so that we can write
$\cW(\Phi_1,\Phi_2)$.

Consider a closed realization of $L$, denoted $L_\bullet$.
Let $\cD(L_\bullet)\subset L^2]a,b[$ denote the domain of $L_\bullet$.
Suppose $z\in\cc$ and let $\Psi_a(z,\cdot),\Psi_b(z,\cdot)$ be functions in $AC^1]a,b[$
solving the eigenvalue equation
\begin{align}\label{eigen1}
  \big( -\partial_x^2 + V(x)-z\big)\Psi_a(z,x)&=0,\\\label{eigen2}
  \big( -\partial_x^2 + V(x)-z\big)\Psi_b(z,x)&=0,\end{align}
$\Psi_a(z,\cdot)$ is in $\cD(L_\bullet)$ near $a$ and
$\Psi_b(z,\cdot)$ is in $\cD(L_\bullet)$ near $b$.
Set
\beq\label{eigen3}\cW(x):=\cW\big(\Psi_b(z,\cdot),\Psi_a(z,\cdot)\big).\eeq
Define the integral kernel
\begin{align}\label{eq:k}
	R_\bullet(z;x,y):= &\frac{1}{\cW(z)}
	\begin{cases}
		\Psi_a(z,x)\;\! \Psi_b(z,y) & \hbox{ if }  a<x < y<b, \\
		\Psi_a(z,y)\;\! \Psi_b(z,x) & \hbox{ if }  a<y < x<b. 
	\end{cases}
\end{align}
Note that \eqref{eq:k}  does not depend on the choice of
$\Psi_a(z,\cdot)$ and $\Psi_b(z,\cdot)$.

Suppose that \eqref{eq:k} defines a bounded operator, which we denote
$R_\bullet(z)$.
Then \begin{align}
       & z\not\in\sigma(L_\bullet)\quad
       \text{and }\quad
\frac{1}{L_\bullet-z}=R_\bullet(z).\label{eigen4}\end{align}

Conversely, if $z\not\in\sigma(L_\bullet)$, then the functions
  $\Psi_a,\Psi_b$ with the above properties
  exist and the operator $R_\bullet(z)$ is bounded.

{\red Above we described the path from $L_\bullet$ to
  $R_\bullet(z)$. In our constructions we will use the reverse
  direction, from 
 $R_\bullet(z)$ to $L_\bullet$, which is described in the following
 theorem.

 \bet \label{eigen5}
 Suppose that 
 $z\in\cc$ and let $\Psi_a(z,\cdot),\Psi_b(z,\cdot)$ be functions in $AC^1]a,b[$
solving the eigenvalue equations \eqref{eigen1}, resp. \eqref{eigen2}, such that
$\Psi_a(z,\cdot)$ is in $L^2$ near $a$ and
$\Psi_b(z,\cdot)$ is in $L^2$ near $b$.
Define $\cW(x)$ by \eqref{eigen3} and $R_\bullet(z;x,y)$ by \eqref{eq:k}.
Suppose that \eqref{eq:k} defines a bounded operator on $L^2]a,b[$, which we denote
$R_\bullet(z)$. Then there exists a unique operator closed $L_\bullet$ that
satisfies
                         \eqref{eigen4}. \eet
                         }

\section{Associated Legendre functions vs. Gegenbauer functions}
\label{Associated Legendre functions vs. Gegenbauer functions}
\init

In the literature, many authors use a Legendre function instead of
Gegenbauer functions. Here, we briefly discuss the relations between
these functions and Gegenbauer functions. For Legendre function we use
\cite{NIST2} as our reference and \cite{DHR1} as our reference for
Gegenbauer function.

The Legendre differential operator  is
\beq
\cL_\mu^\alpha:=(1+z^2) \partial_z^2-2z \partial_z+\mu(\mu+1) -\frac{\alpha^2}{1-z^2}.
\eeq
It is equivalent to the Gegenbauer operator, in fact
\begin{align} \notag
&(1-w^2)^{\mp\frac\alpha2}\cL_\mu^\alpha (1-w^2)^{\pm\frac\alpha2}\\=&
(1-w^2)\partial_w^2-2(\pm\alpha+1)w\partial_w+(\mu\mp\alpha)
(\mu\pm\alpha+1) =\cG_{\pm\alpha,\mu+\frac12}.
\end{align}

Certain distinguished funnctions annihilated by this operator are
called associated Legendre functions. There various choices  for these
functions, which we quote following \cite{NIST2}:

\noindent\textbf{The associated Legendre function of the first kind} is 
\begin{align} \notag
    \mathbf{P}^\alpha_\mu(z)&= \left( \frac{z+1}{z-1}\right)^\frac{\alpha}{2} \mathbf{F}\left(\mu+1,-\mu;1-\alpha;\frac{1-z}{2}\right)
    \\
    &=\frac{2^\alpha}{\left(z^2-1\right)^\alpha_\bullet}\mathbf{S}_{-\alpha,\mu+\frac12}(z)
\end{align}
\textbf{The Ferrers function of the first  kind} is
\begin{align} \notag
    \mathcal{P}^\alpha_\mu(z)&= \left( \frac{z+1}{1-z}\right)^\frac{\alpha}{2} \mathbf{F}\left(\mu+1,-\mu;1-\alpha;\frac{1-z}{2}\right)
    \\
    &=\frac{2^\alpha}{\left(1-z^2\right)^\alpha_\bullet}\mathbf{S}_{-\alpha,\mu+\frac12}(z)
\end{align}
And \textbf{the associated Legendre function of the second kind} is
\begin{align} \notag
    \mathbf{Q}^\alpha_{\mu}(z)&=\e^{\i\pi\alpha } \frac{\pi^\frac12 \Gamma(\alpha+\mu+1)\left(z^2-1\right)^\frac{\alpha}{2}}{2^{\mu+1}z^{\alpha+\mu+1}}\mathbf{F}\left(\frac{\alpha+\mu}{2}+1,\frac{\alpha+\mu}{2}+\frac12;\frac32+\mu;z^{-2}\right)
    \\
    &=\e^{\i\pi\alpha } \frac{\pi^\frac12 \Gamma(\alpha+\mu+1)\left(z^2-1\right)^\frac{\alpha}{2}}{2^{\mu+1}}\mathbf{Z}_{\alpha,\mu+\frac12}(z).
\end{align}

Note that Legendre functions are represented with upper and lower
indices. One should not confuse them with other, quite analogous functions defined in
the text, whose parameters are lower indices.

The Legendre functions are closely related to the functions $\cP^\s$, $\cP^\h$,
and $\cQ^\h$ that we introduced in the section on Gegenbauer
Hamiltonians, which you can see on the right of the following
comparison:
\begin{align}
\cP_\mu^\alpha(\cos r)&=\Big(\frac2{\sin 
                            r}\Big)^{\frac12}\cP_{-\alpha,\mu+\frac12}^\s(r),\\
  {\bf P}_\mu^\alpha(\cosh r)&=\Big(\frac2{\sinh 
                                   r}\Big)^{\frac12}\cP_{-\alpha,\mu+\frac12}^\h(r),\\
  \mathbf{Q}_\mu^\alpha(\cosh r)&  =\frac{1}{(2\sinh r)^{\frac12}}\e^{\i\pi\alpha } \sqrt\pi \Gamma(\alpha+\mu+1)\cQ_{-\alpha,\mu+\frac12}^\h(r).
  \end{align}

  \noindent
{\bf Conflict of interest.} This manuscript has no conflict of interest.

\end{document}